\documentclass[12pt,doublespacing]{article}

\usepackage{fullpage}
\usepackage{setspace}
\usepackage{authblk}

\usepackage{cite}
\usepackage{hyperref} 

\usepackage{graphicx}

\usepackage{amsmath}
\usepackage{amssymb}
\allowdisplaybreaks

\usepackage{float}
\usepackage{subfig}
\usepackage{tabularx}

\begin{document}
\title{Enhanced Third-Harmonic Generation in a Bound State in the Continuum Assisted Multiband All-Dielectric Metasurface}

\author[1]{Partha Mondal}
\author[2]{Hadeel Alamoudi}
\author[2]{Rodrigo P. Sanchez}
\author[1]{Redha H. Al Ibrahim}
\author[1]{Boon S. Ooi}
\author[2]{Iman Roqan}
\author[1]{Hakan Bagci\vspace{0.5cm}}

\affil[1]{Electrical and Computer Engineering (ECE) Program
\authorcr Computer, Electrical and Mathematical Science and Engineering (CEMSE) Division
\authorcr King Abdullah University of Science and Technology (KAUST)
\authorcr Thuwal, 23955-6900, Saudi Arabia\vspace{0.5cm}}

\affil[2]{Material Science and Engineering (MSE) Program
\authorcr Physical Science and Engineering (PSE) Division
\authorcr King Abdullah University of Science and Technology (KAUST)
\authorcr Thuwal, 23955-6900, Saudi Arabia}

\footnotetext[1]{Corresponding author: Partha Mondal (partha.mondal@kaust.edu.sa)}

\date{}
\maketitle
\newpage

\begin{abstract}
Multiband Fano resonances are demonstrated in the near-infrared (near-IR) using an all-dielectric metasurface whose unit cell consists of four silicon nanoblocks on a glass substrate. An in-plane asymmetry triggers symmetry-protected quasi-bound states in the continuum (QBICs), producing multiple high-$Q$ resonances. Their origin is identified through multipolar decomposition of the scattering cross section and field distributions at the resonances. The strong field localization at these resonances enables efficient multiband third-harmonic (TH) generation in the ultraviolet (UV), with a maximum simulated conversion efficiency of $8.5 \times 10^{-3}$ at a peak pump intensity of $1.6\,\mathrm{GW/cm^{2}}$. The metasurface is fabricated in symmetric and asymmetric configurations, and its linear and nonlinear responses are measured under normal incidence. A TH conversion efficiency of $1.2 \times 10^{-6}$ is obtained at a peak pump intensity of $3.25\,\mathrm{GW/cm^{2}}$. These results establish a route to multiband photonic devices, including multiwavelength lasers, multiband harmonic generation, and single-photon sources for quantum photonics.
\end{abstract}

\newpage\clearpage

\section{Introduction}
Nonlinear optical processes drive a wide range of modern photonic technologies, from tunable ultrafast lasers based on frequency up- and down-conversion~\cite{wu2025advancements} to high-capacity optical communication systems, advanced data storage~\cite{soman2021tutorial}, ultrafast signal processing~\cite{salem2013application}, nonlinear spectroscopy~\cite{de2016techniques}, and the generation of entangled photon pairs for emerging quantum applications~\cite{ma2024engineering}. In conventional bulk materials, these processes typically require high optical intensities, and the extended interaction length imposes stringent phase-matching conditions between the fundamental and generated waves. Optical nanostructures overcome these constraints by strongly localizing electromagnetic fields at the nanoscale, thereby enhancing light–matter interactions. This field confinement relaxes phase-matching constraints and enables efficient nonlinear frequency conversion at substantially reduced power levels. In particular, optical metasurfaces have emerged as a versatile, subwavelength-thick platform for tailoring nonlinear optical responses~\cite{sain2019nonlinear}. For instance, intense field localization of surface plasmon resonances in metallic nanostructures offers an efficient mechanism for second-harmonic (SH) generation and third-harmonic (TH) generation~\cite{rahimi2018nonlinear}. Notably, SH generation is allowed only in noncentrosymmetric media, whereas TH generation is a more general process that also occurs in centrosymmetric media. TH generation has found widespread applications in nonlinear imaging, spectroscopy, lasing, and optical telecommunications~\cite{zheng2023third,konorov2003femtosecond}. However, the inherent Ohmic losses and low damage thresholds of plasmonic materials severely limit their performance in high-efficiency nonlinear devices. In contrast, high-refractive-index dielectric materials with low absorption losses offer a promising alternative for nonlinear metasurfaces~\cite{minovich2015functional, kivshar2018all}.

Silicon (Si) metasurfaces, with their strong intrinsic third-order nonlinearity, have been widely used to achieve enhanced TH generation through intense electromagnetic field localization supported by multipolar resonances. High-$Q$ resonances further amplify light–matter interactions, producing a much stronger nonlinear response~\cite{hail2024third}. Various resonant modes, including anapole modes~\cite{grinblat2016enhanced}, dipolar resonances~\cite{shorokhov2016multifold}, and Fano resonances~\cite{yang2015nonlinear, shorokhov2016multifold, hahnel2023multi}, have been exploited to reinforce field localization and thereby improve TH conversion efficiency. Notably, a marked increase in TH generation in Si metasurfaces has been achieved via multipolar interference between electric quadrupole and magnetic dipole modes, leading to pronounced near-field amplification~\cite{chen2018third}.  

In this context, bound states in the continuum (BICs) have attracted considerable interest over the past decade for their ability to support ultranarrow spectral linewidths arising from extraordinary field localization~\cite{zhou2023bound, zhou2025bound, barreda2025bound}. A BIC is a radiation-free mode embedded within a radiation continuum with an infinite $Q$ factor. In practice, however, material absorption, fabrication-induced imperfections, and external perturbations inevitably limit the $Q$ factor, converting an ideal BIC into a quasi-BIC (QBIC). This transformation lets the otherwise trapped mode radiate weakly, so it can be excited by an incident wave while keeping a very high $Q$, which makes QBICs attractive for practical photonic applications~\cite{liu2019high, 10.1021/acsami.6c02939,Liu:26}. Several mechanisms have been proposed to realize BICs, including symmetry-protected BICs~\cite{li2022multiple}, Fabry–Pérot BICs~\cite{li2023narrowband}, accidental BICs~\cite{sidorenko2021observation}, and Friedrich–Wintgen BICs~\cite{zhang2024flat}. Among these, symmetry-protected BICs are particularly prominent, as field localization is achieved through controlled breaking of structural, material, or polarization symmetry. In particular, in-plane structural deformations within the unit cell break the symmetry, allowing precise tuning of QBIC resonant wavelengths and $Q$ factors~\cite{wang2024dual, huang2024realizing, zhou2025ultrahigh}. Efficient engineering of high-$Q$ QBIC modes enables strong field localization, which directly translates into high TH conversion efficiency~\cite{tonkaev2025unconventional, PhysRevLett.134.023804}. For instance, unit cells composed of tilted elliptical nanoparticle pairs~\cite{moretti2022introducing, moretti2024si} or asymmetric nanobars~\cite{wang2024resonantly, koshelev2019nonlinear} have been employed to realize QBIC-assisted TH generation. Efficient TH generation has also been demonstrated in metasurfaces with single-particle unit cells, where symmetry breaking is introduced via asymmetric geometries, such as cylindrical nanoparticles with off-center holes that disrupt radial symmetry~\cite{xu2019dynamic, sun2025efficient}, kite-shaped nanopillar arrays~\cite{hsiao2024enhancement}, planar chiral structures~\cite{shi2022planar}, and asymmetric square nanoblocks~\cite{liu2019high, oea-2020-0030-Ganxuetao}.

Closely related are Fano resonances, which arise from the interference between a discrete localized state and a continuum of radiative states and provide an efficient route to high-$Q$ resonances in metasurfaces~\cite{limonov2017fano}. They exhibit a characteristically asymmetric spectral profile, which has been exploited in a wide range of photonic applications~\cite{agarwal2024all}. High-$Q$ Fano resonances supported by a variety of geometrical configurations within the unit cell have demonstrated multifold enhancement of nonlinear optical responses~\cite{shorokhov2016multifold, yang2015nonlinear, bijloo2024near, tong2016enhanced}. A strong connection between Fano resonances and QBICs has been established in isolated AlGaAs nanodisks, revealing that both phenomena are governed by similar underlying physical mechanisms~\cite{melik2021fano}. Recent studies have shown that symmetry-protected QBICs offer a robust route for generating high-$Q$ Fano resonances~\cite{zhang2018high, yu2021ultra}. To achieve such resonances with large modulation depths, diverse dielectric nanostructures with tailored unit-cell geometries have been proposed~\cite{xing2023multiple, limonov2021fano}. 

From an application perspective, the demand for high-$Q$ resonances has evolved from single-band to multiband operation. However, most reported dielectric metasurfaces support Fano resonances within a single spectral band, and only limited efforts have focused on realizing multiple Fano resonances over a broad spectral range~\cite{zhang2018high, li2022modifying, yu2021multiple, Lv:24}. Multiband high-$Q$ Fano resonances, enabled by multimode interactions and multipolar interference, are highly attractive for applications including multiband channel filters and sensors, multiwavelength spectroscopy, multiwavelength lasers, and multiband harmonic generation~\cite{liu2020multipole, yang2021multiple}. Moreover, multiple Fano resonances provide greater flexibility for harmonic generation across distinct spectral bands~\cite{liu2018enhanced, qin2024polarization}, facilitating the development of multichannel nonlinear photonic devices. Despite these advancements, realizing multiband Fano resonances with simultaneously high $Q$ factors and large modulation depths remains a challenge.

This work presents the design, fabrication, and experimental characterization of an all-dielectric metasurface that supports multiple high-$Q$ Fano resonances in the near-infrared (near-IR) spectral region. The metasurface consists of a periodic array of unit cells on a glass substrate, each cell comprising four amorphous silicon (a-Si) nanoblocks. Structural asymmetry is deliberately introduced to excite QBICs, giving rise to additional Fano resonances in the transmission spectrum. A systematic study is conducted to elucidate the dependence of the QBIC modes on geometrical parameters and the degree of asymmetry. Multipolar decomposition analysis is employed to reveal the physical mechanisms underlying the observed resonances. In addition, the spectral features are quantitatively analyzed using the Fano line-shape model, confirming high-$Q$ factors across multiple resonant bands. The linear optical responses of both symmetric and asymmetric metasurfaces are experimentally characterized under normal-incidence excitation, showing good agreement with theoretical predictions. Furthermore, TH generation is experimentally demonstrated at $382.1\,\mathrm{nm}$ in the ultraviolet (UV) region under femtosecond excitation at a fundamental wavelength of $1146.4\,\mathrm{nm}$. A maximum TH conversion efficiency of $1.2\times10^{-6}$ is achieved at a peak pump intensity of $3.25\,\mathrm{GW/cm^{2}}$. The experimental findings are corroborated by full-wave nonlinear simulations. Finally, a nonlinear analysis investigates the metasurface's potential for TH generation at other resonant wavelengths. The results indicate that the additional high-$Q$ resonance modes strongly enhance field localization, enabling efficient TH generation over a broad spectral range in the UV region. The principal novelty of this work lies in the integration of eight QBIC-assisted TH channels, broad near-UV frequency conversion, polarization-selective excitation, mode-resolved multipolar composition, and a single-layer architecture, in contrast to previous studies primarily focused on optimizing TH generation at single or dual QBIC resonances.

\section{Metasurface design and results}
\begin{figure}[ht!]
    \centering
    \subfloat[]{\includegraphics[width=0.65\textwidth]{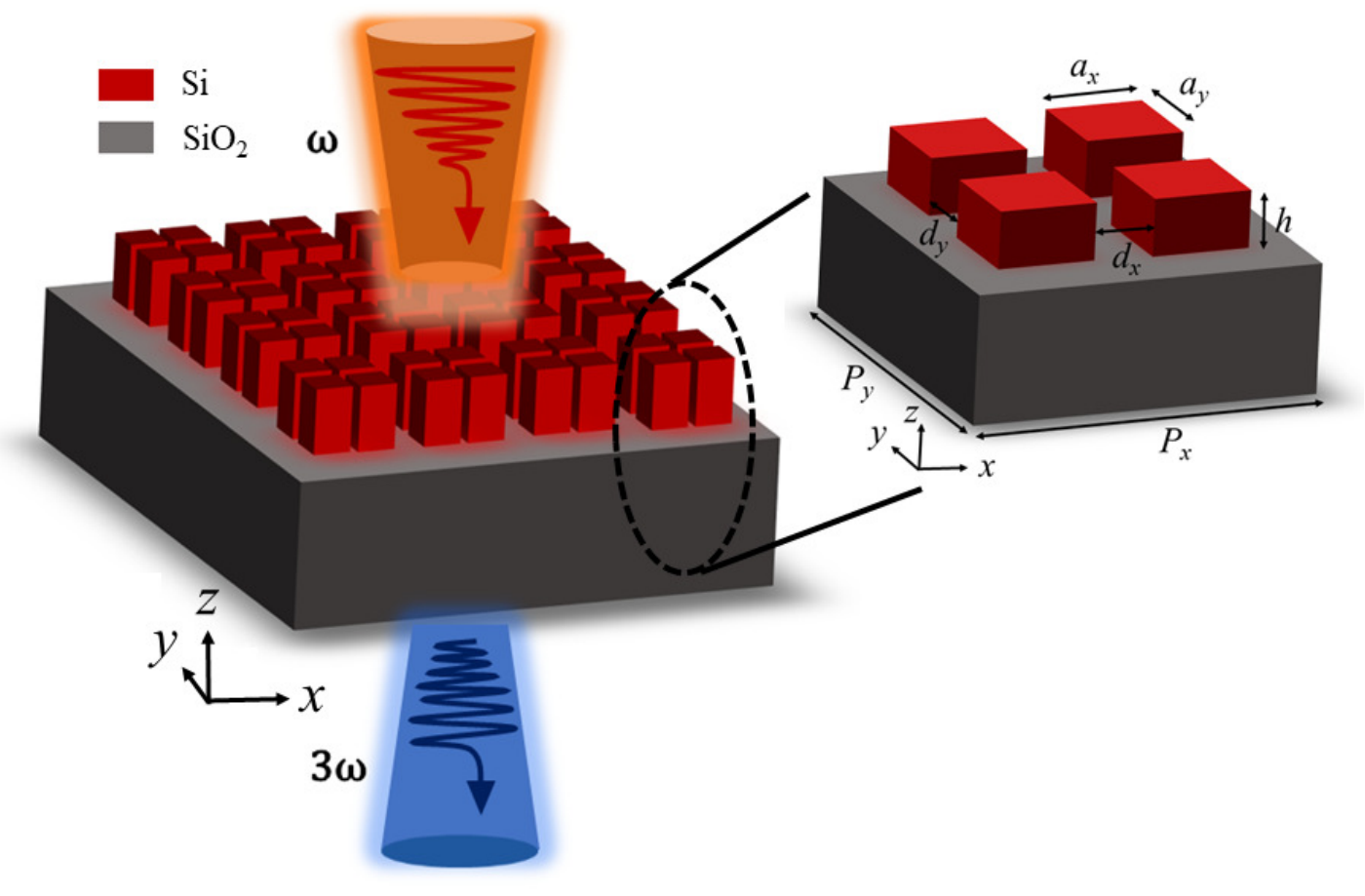}}\hspace{12pt}
    \subfloat[]{\includegraphics[width=0.275\textwidth]{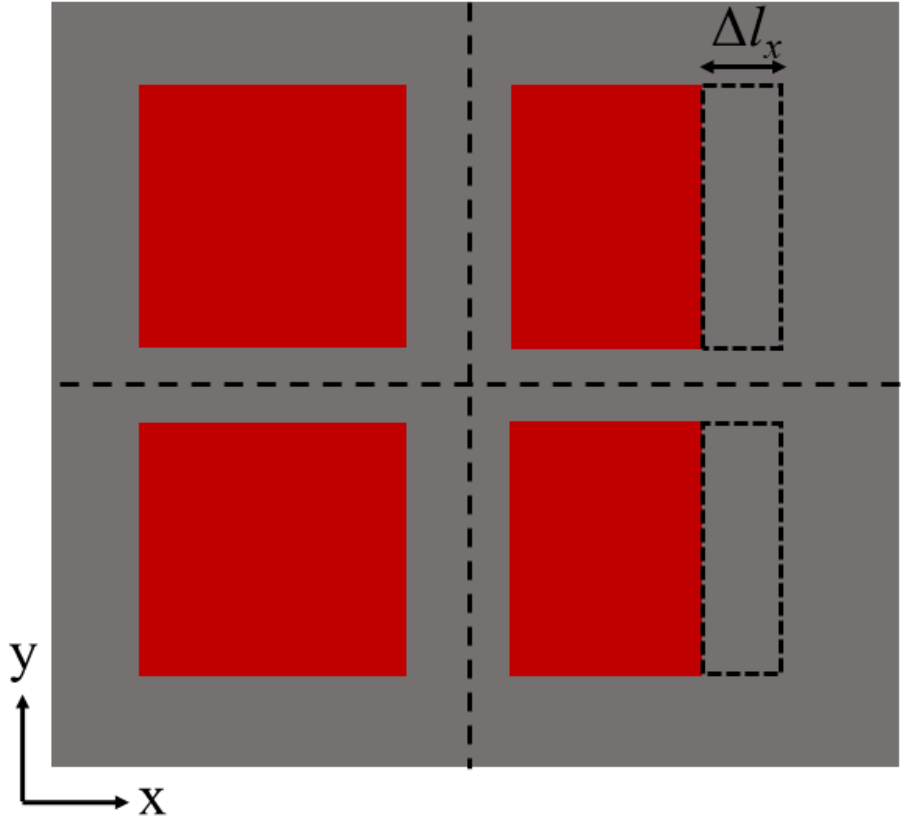}}     
    \caption{Design of the nonlinear metasurface. (a) Schematic of the all-dielectric metasurface for enhanced TH generation, where the unit cell consists of four Si nanoblocks. (b) Top view of the asymmetric unit cell, showing the reduction $\Delta l_x$ in the $x$-dimension of the two nanoblocks on the right.}
    \label{fig1}
\end{figure}
The geometry of the proposed metasurface is illustrated in Figure~\ref{fig1}. The structure consists of Si nanoblocks and a uniform glass substrate that supports them. The isometric view of the metasurface is shown in Figure~\ref{fig1}(a), where the inset illustrates a single unit cell. Each unit cell consists of four Si nanoblocks of height $h$. The periodicities along the $x$- and $y$-directions are denoted by $P_{x}$ and $P_{y}$, respectively. The in-plane dimensions of the nanoblocks are represented by $a_{x}$ (along the $x$-direction) and $a_{y}$ (along the $y$-direction), while the separation distances between adjacent blocks are denoted by $d_{x}$ (along the $x$-direction) and $d_{y}$ (along the $y$-direction). A $y$-polarized plane wave propagating along the $-z$-direction is incident on the metasurface, with broadband excitation in the near-IR part of the spectrum. Simulations are carried out using the finite-difference time-domain (FDTD) method. Periodic boundary conditions are applied along the $x$- and $y$-directions to model the infinite array, while perfectly matched layers (PMLs) are employed along the $z$-direction to absorb the outgoing scattered fields. The wavelength-dependent complex refractive indices of the Si metasurface and glass (SiO$_2$) substrate are taken from the Palik optical-constant database~\cite{palik1985handbook}, with the crystalline Si data used to approximate the optical response of the fabricated a-Si nanoblocks. The Si and glass Palik models are employed over both the fundamental and the corresponding TH wavelength ranges. Therefore, the full-wave simulations account for material dispersion and absorption, including the strong UV absorption of Si at the TH frequency.
 Figure~\ref{fig1}(b) illustrates the top view of the perturbed unit cell, where structural asymmetry is introduced by reducing the dimension of the two nanoblocks on the right along the $x$-direction by $\Delta l_x$. The asymmetry parameter is defined as the ratio of this reduction to the reference nanoblock dimension $a_x$: $\delta = \Delta l_x/a_{x}$. 

\begin{figure}[t!]
    \centering
    \subfloat[]{\includegraphics[width=0.625\columnwidth]{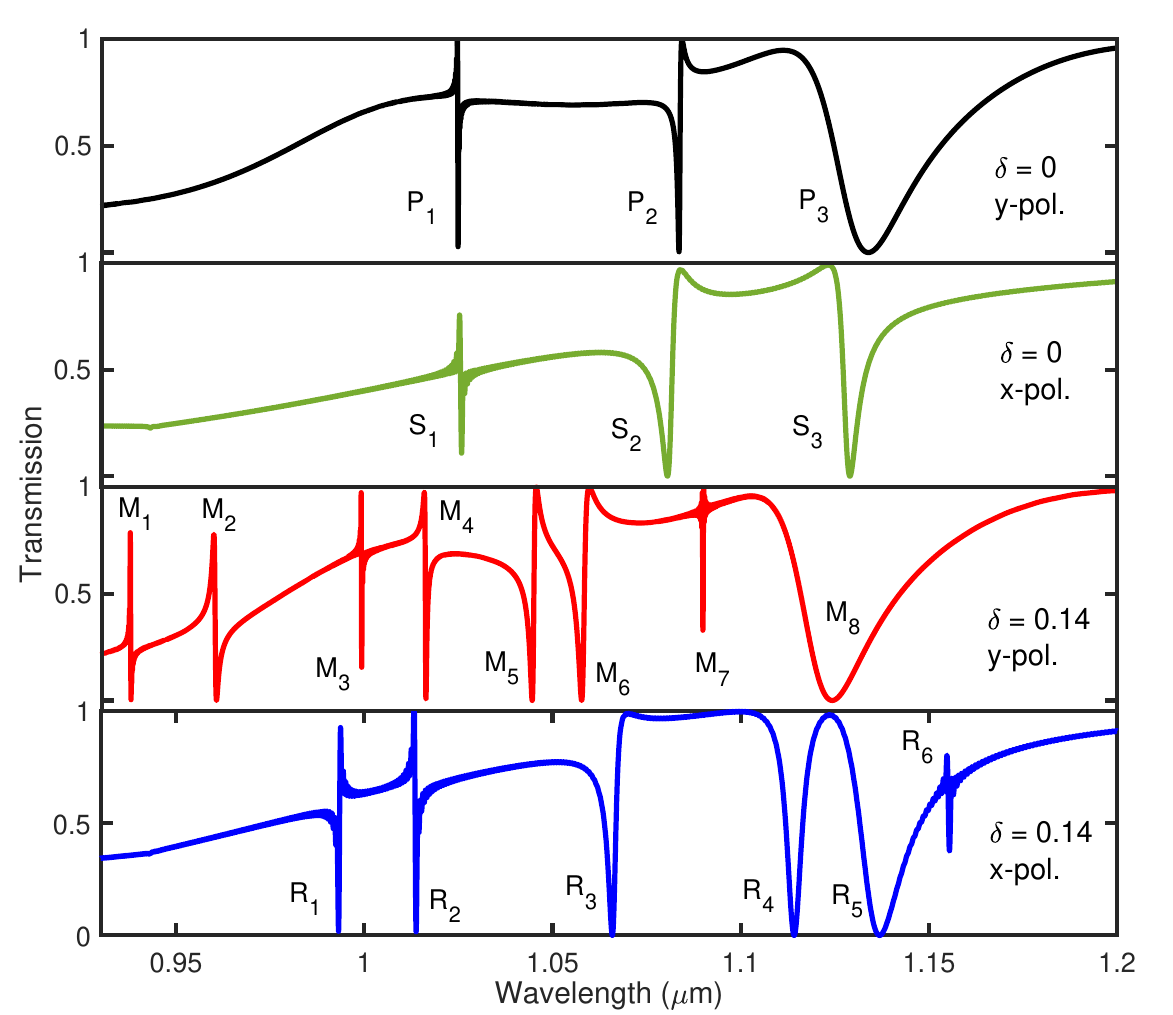}}\hspace{12pt}
    \subfloat[]{\includegraphics[width=0.65\columnwidth]{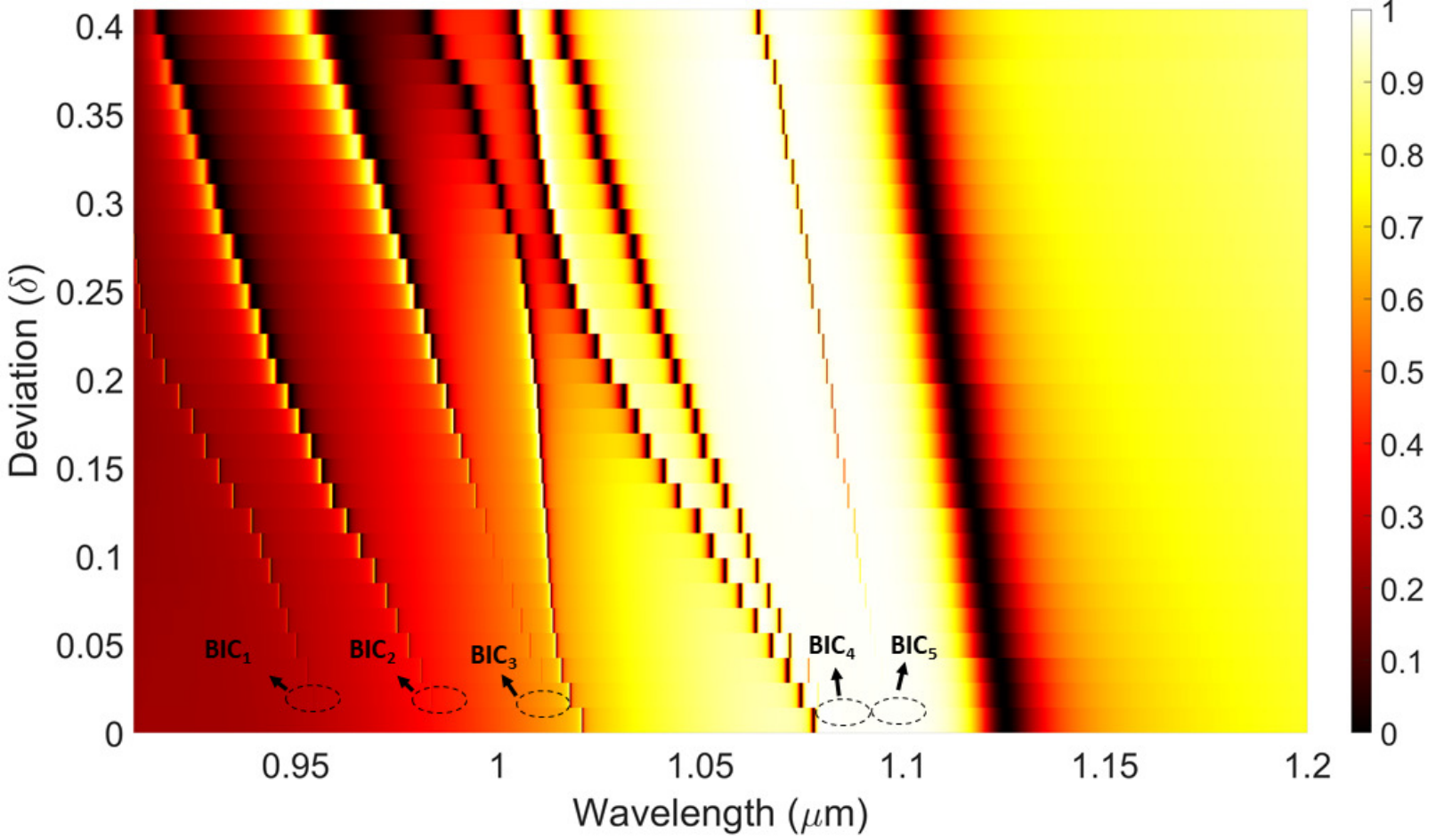}} 
    \caption{Normalized transmission spectra of the metasurface. (a) Transmission spectra of the symmetric ($\delta=0$) and asymmetric ($\delta=0.14$) metasurfaces under $x$- and $y$-polarized excitation. (b) Transmission spectrum as a function of the asymmetry parameter $\delta$ under $y$-polarized excitation.}
    \label{fig2}
\end{figure}

\begin{figure}[!ht]
    \centering
    \subfloat[]{\includegraphics[width=0.33\columnwidth]{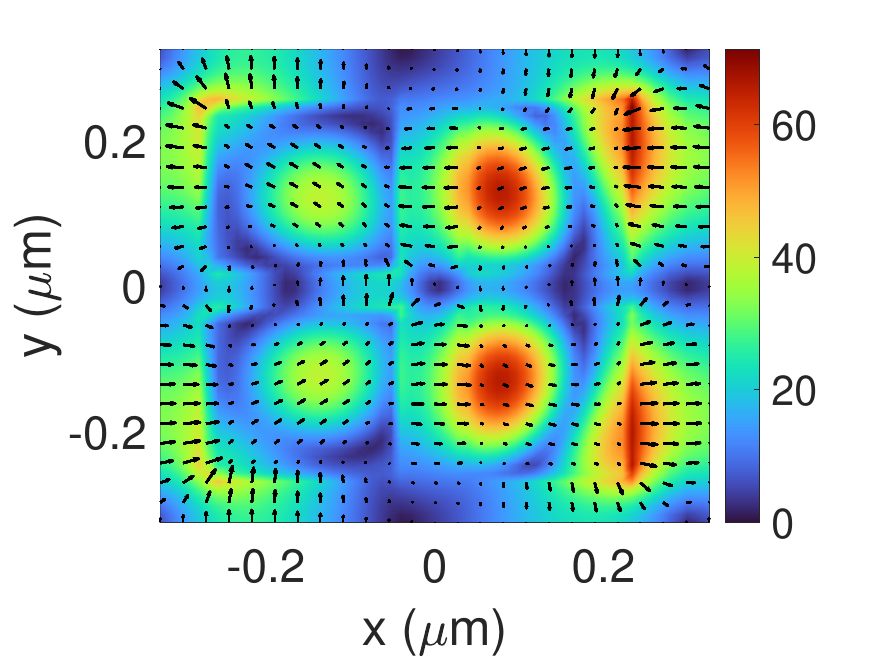}}
    \subfloat[]{\includegraphics[width=0.33\columnwidth]{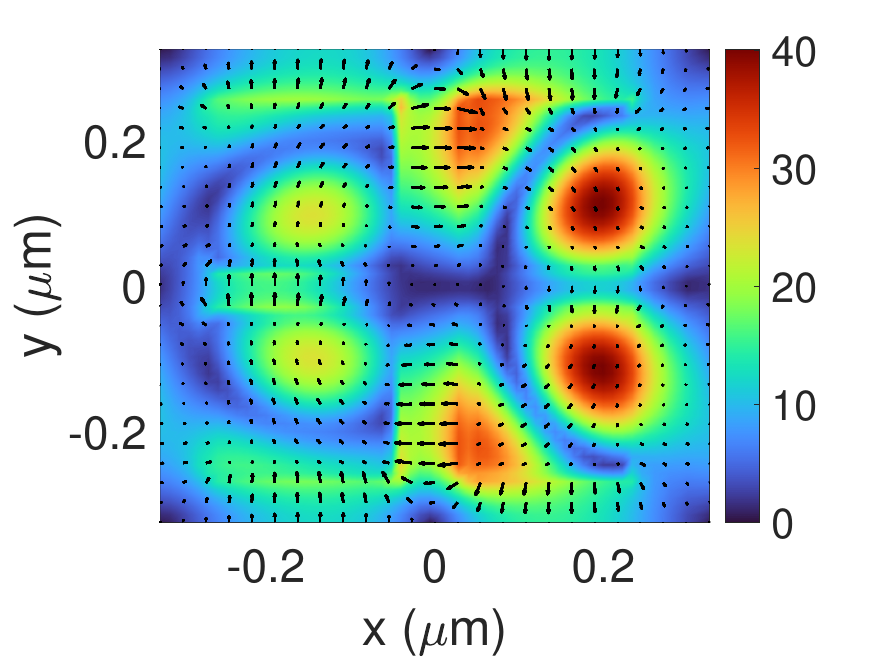}} 
    \subfloat[]{\includegraphics[width=0.33\columnwidth]{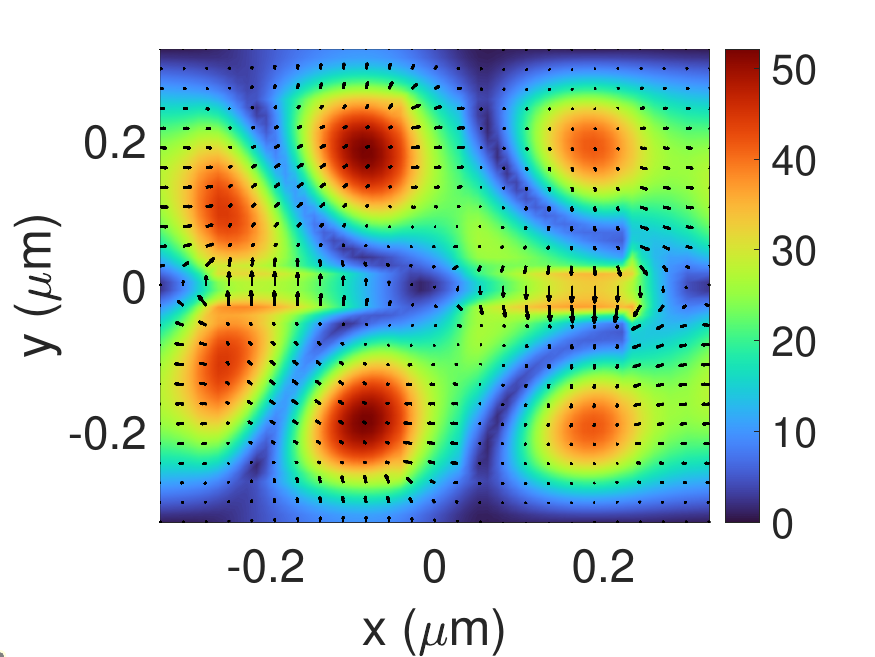}}\\
    \subfloat[]{\includegraphics[width=0.33\columnwidth]{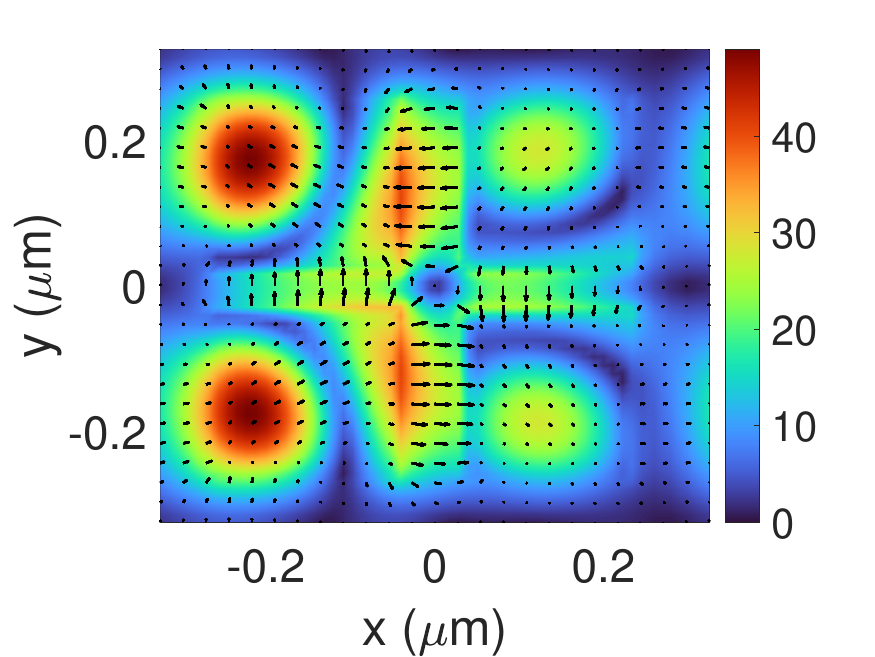}}
    \subfloat[]{\includegraphics[width=0.33\columnwidth]{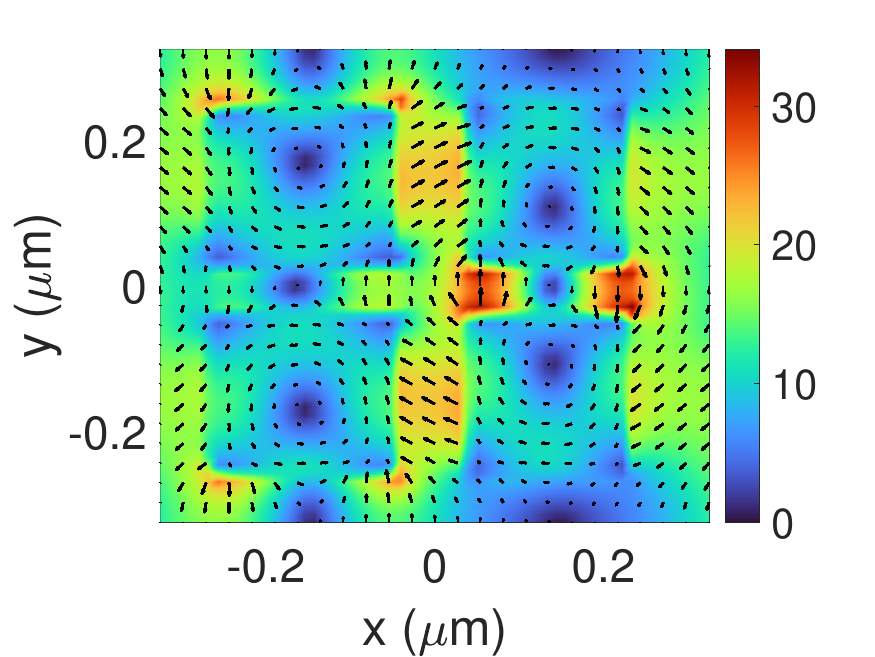}}
    \subfloat[]{\includegraphics[width=0.33\columnwidth]{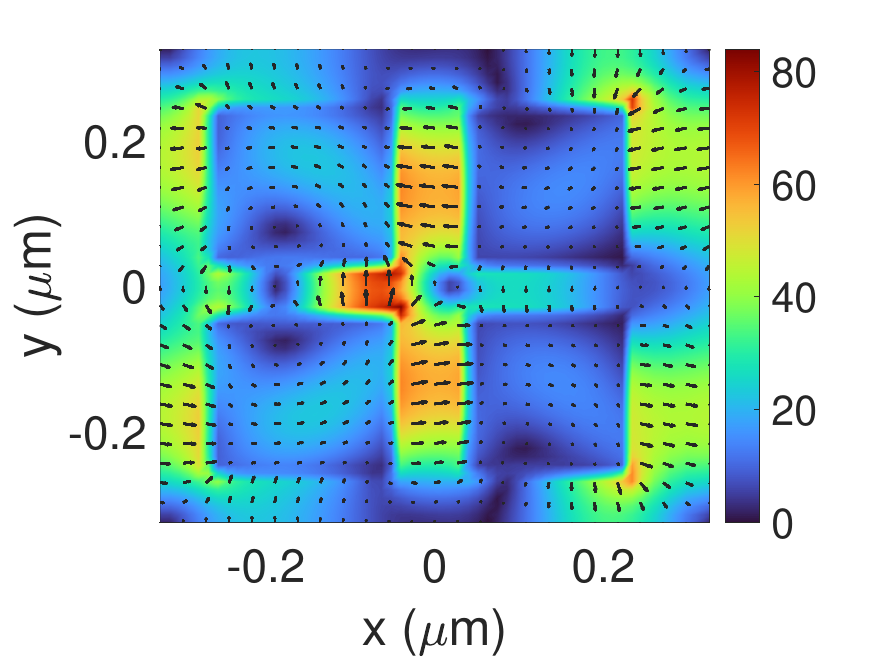}}\\
    \subfloat[]{\includegraphics[width=0.33\columnwidth]{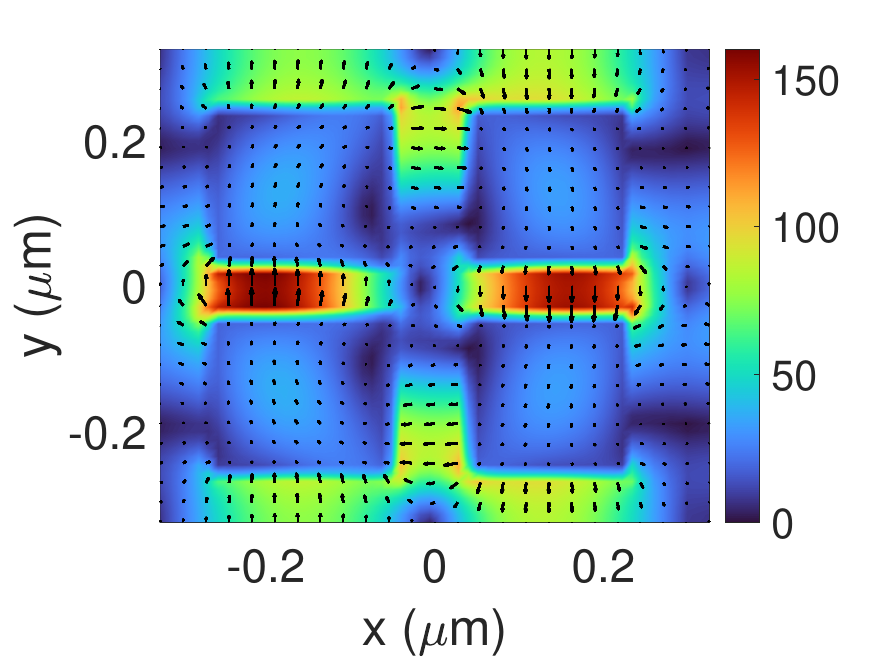}}
    \subfloat[]{\includegraphics[width=0.33\columnwidth]{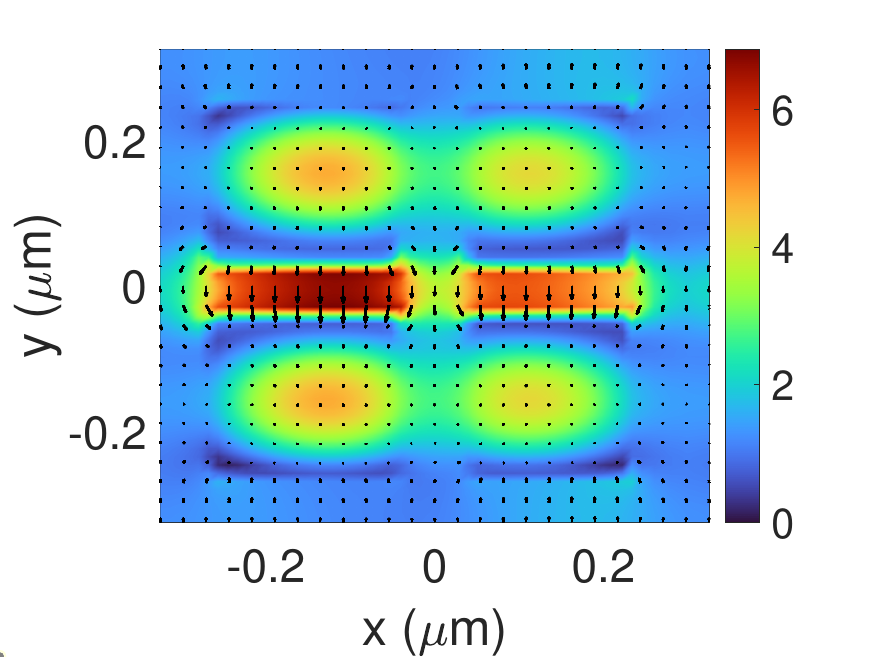}}
    \caption{Simulated electric field distributions in the $xy$-plane at $\delta = 0.14$ for the resonances (a) $M_{1}$, (b) $M_{2}$, (c) $M_{3}$, (d) $M_{4}$, (e) $M_{5}$, (f) $M_{6}$, (g) $M_{7}$, and (h) $M_{8}$.}
    \label{fig_Efield_xy}
\end{figure}

The geometrical parameters of the metasurface are chosen as $P_x=P_y=P=650\, \mathrm{nm}$, $a_{x}=a_{y}= 220\,\mathrm{nm}$, $d_{x} = 85\,\mathrm{nm}$, $d_{y} = 65\,\mathrm{nm}$, and $h = 285\,\mathrm{nm}$. To investigate the polarization dependence of the metasurface, the transmission spectra are calculated under both $x$- and $y$-polarized excitation. The results are shown in Figure~\ref{fig2}(a). For the symmetric metasurface ($\delta = 0$), three distinct resonances are observed at $1024.6\,\mathrm{nm}$ ($P_{1}$), $1083.4\,\mathrm{nm}$ ($P_{2}$), and $1133.5\,\mathrm{nm}$ ($P_{3}$) under $y$-polarized excitation. Nearly identical transmission spectra for the $x$-polarized excitation are observed at $1025.7\,\mathrm{nm}$ ($S_{1}$), $1080.4\,\mathrm{nm}$ ($S_{2}$), and $1128.9\,\mathrm{nm}$ ($S_{3}$). However, introducing a perturbation $\Delta l_x$ ($\delta=0.14$) breaks the in-plane mirror symmetry of the unit cell, resulting in distinctly different polarization-dependent responses. Under $y$-polarized excitation, eight pronounced Fano resonances are obtained at $937.69\,\mathrm{nm}$ ($M_{1}$), $960.4\,\mathrm{nm}$ ($M_{2}$), $999.04\,\mathrm{nm}$ ($M_{3}$), $1016.1\,\mathrm{nm}$ ($M_{4}$), $1044.4\,\mathrm{nm}$ ($M_{5}$), $1057.5\,\mathrm{nm}$ ($M_{6}$), $1089.7\,\mathrm{nm}$ ($M_{7}$), and $1124.2\,\mathrm{nm}$ ($M_{8}$). In contrast, under $x$-polarized excitation six weaker resonances are observed at $993.15\,\mathrm{nm}$ ($R_{1}$), $1013.75\,\mathrm{nm}$ ($R_{2}$), $1065.85\,\mathrm{nm}$ ($R_{3}$), $1114.15\,\mathrm{nm}$ ($R_{4}$), $1136.85\,\mathrm{nm}$ ($R_{5}$), and $1155.45\,\mathrm{nm}$ ($R_{6}$). Because the perturbation $\Delta l_x$ shortens the nanoblocks along $x$, the resulting QBIC modes present a greater material extent along $y$; consequently, the $y$-polarized incident field couples more efficiently to these symmetry-broken modes, producing stronger resonance contrast and field localization together with a larger number of accessible high-$Q$ channels, whereas the $x$-polarized field couples weakly. Therefore, the subsequent multipolar decomposition and nonlinear TH generation analyses are performed under $y$-polarized excitation. The appearance of these additional resonances demonstrates the significance of symmetry breaking in exciting QBICs. For the metasurface considered here, this leads to the formation of multiple Fano resonances within the near-IR spectral range. The $Q$ factor of each resonance is extracted by fitting the transmission spectra with the classical Fano line-shape formula~\cite{wang2024dual}
\begin{equation*}
T(\omega)=T_{0}+A_{0} \frac{\displaystyle \left[  q+\frac{2(\omega-\omega_{0})}{\tau}\right]^{2}}{\displaystyle 1+\left[\frac{2(\omega-\omega_{0})}{\tau}\right]^{2}},
\end{equation*} 
where $T_{0}$ denotes the transmission offset, $A_{0}$ is the coupling constant, $q$ is the Fano asymmetry parameter, $\omega_{0}$ is the resonant frequency, and $\tau$ corresponds to the resonance linewidth. Once these parameters are extracted, the $Q$ factor is obtained using $Q = \omega_{0}/\tau$. For the asymmetric unit cell with $\delta = 0.14$, the $Q$ factors extracted from resonances $M_{i}$, $i=1,2,\ldots,8$ are $8\,731$, $1\,509$, $15\,084$, $2\,942$, $1\,002$, $635$, $10\,167$, and $52$, respectively. The corresponding fitting curves for all resonances are provided in Figure~S1 of the supplementary material. 
Figure~\ref{fig2}(b) illustrates the influence of the asymmetry parameter $\delta$ on the transmission spectra. As $\delta$ increases, additional Fano resonances emerge and sharpen, and their wavelengths exhibit a progressive blueshift. The $Q$ factor is strongly dependent on the asymmetry parameter $\delta$: increasing structural deviation broadens the QBIC resonance dips, leading to a pronounced reduction in the $Q$ factor, which follows an inverse proportionality to $\delta^2$. To gain deeper insight into the physical origin of the resonance modes, the electromagnetic field distributions in the unit cell with $\delta = 0.14$ are analyzed. Figure~\ref{fig_Efield_xy} shows the electric field distributions in the $xy$-plane at the resonance wavelengths. At the shorter wavelengths $(M_{1}$ -- $M_{3})$, the field localization occurs within the individual nanoblocks, whereas at higher wavelengths $(M_{4}$ -- $M_{8})$ the electric field becomes increasingly localized within the narrow gaps between adjacent nanoblocks, reflecting enhanced near-field coupling and hybridization of neighboring multipole modes. Figure~S2 in the supplementary material depicts the electric field distributions in the $xz$-plane at the resonance wavelengths, while Figure~S3 shows the corresponding magnetic field distributions in the $xy$-plane. The results indicate that the multiple resonances arise from the complex coupling and hybridization of multipole modes within the asymmetric unit cell, with dominant electric quadrupole, magnetic quadrupole, and toroidal dipole contributions.

\begin{figure}[ht!]
    \centering
    \subfloat[]{\includegraphics[width=0.475\columnwidth]{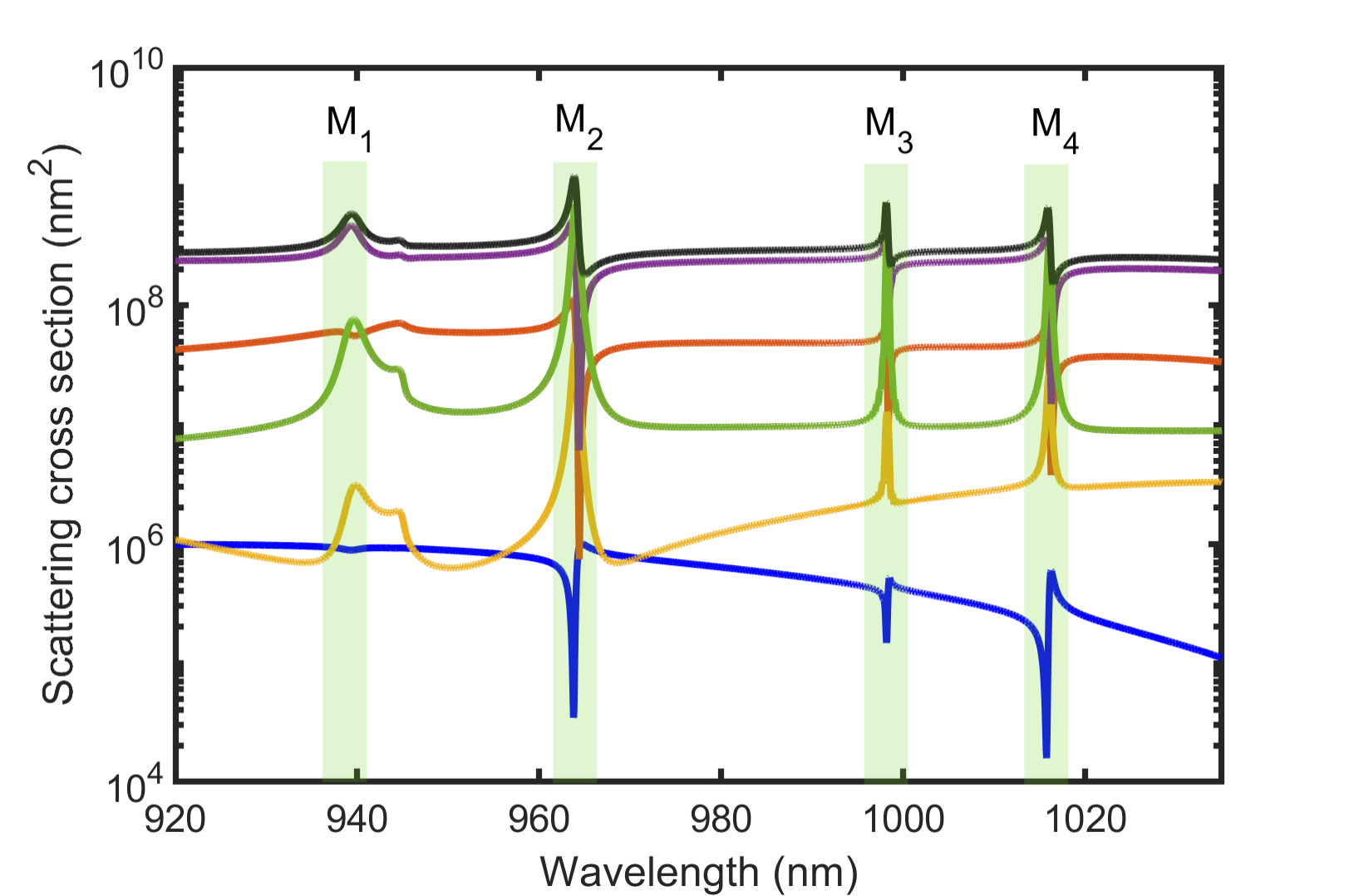}}\hspace{12pt}
    \subfloat[]{\includegraphics[width=0.475\columnwidth]{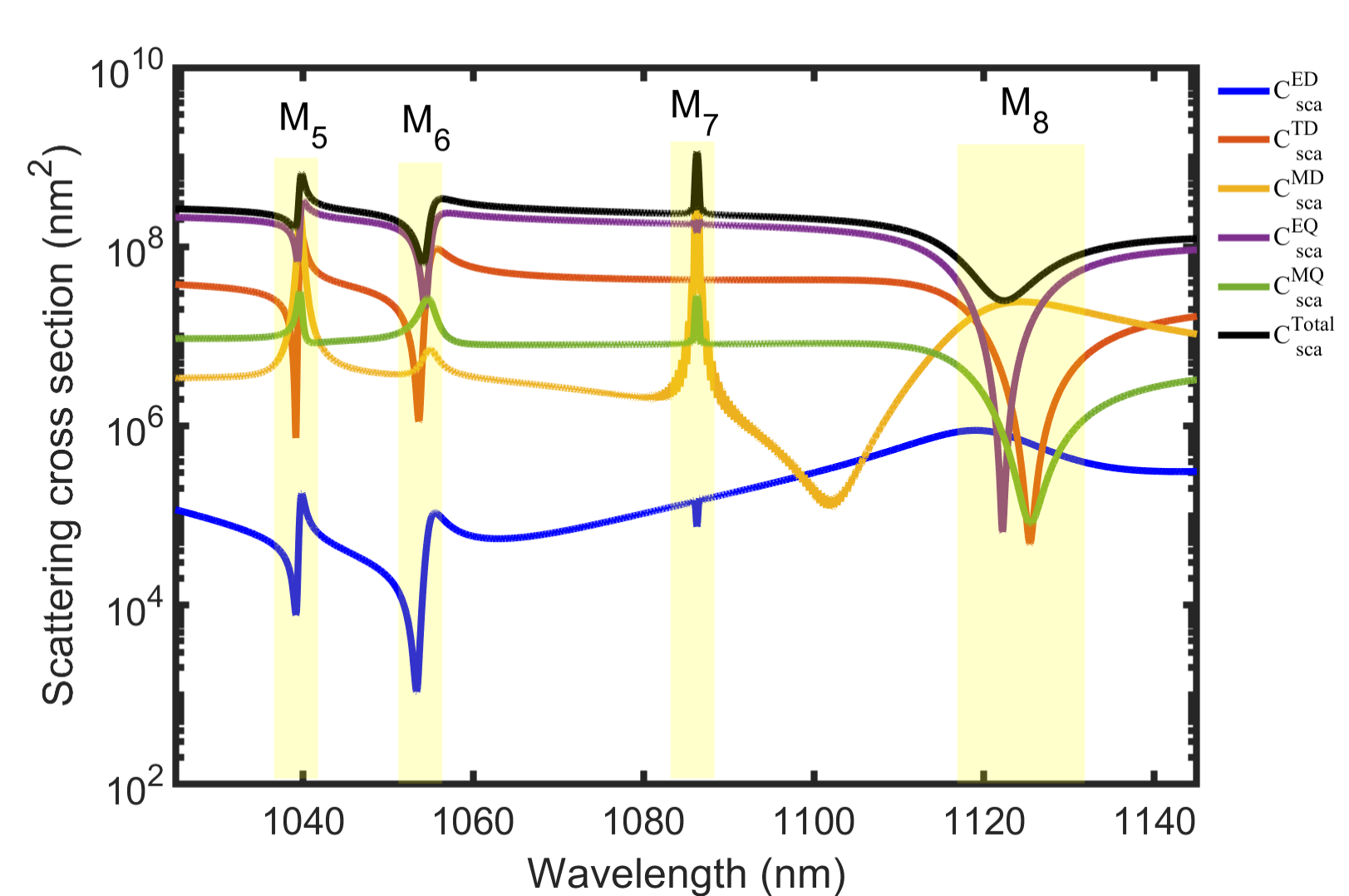}} 
    \caption{Simulated multipolar decomposition of scattering cross section for resonances (a) $M_{1}$ -- $M_{4}$ and (b) $M_{5}$ -- $M_{8}$. $C^{\mathrm{ED}}_{\mathrm{sca}}$, $C^{\mathrm{TD}}_{\mathrm{sca}}$, $C^{\mathrm{MD}}_{\mathrm{sca}}$, $C^{\mathrm{EQ}}_{\mathrm{sca}}$, and $C^{\mathrm{MQ}}_{\mathrm{sca}}$ denote the electric-dipole, toroidal-dipole, magnetic-dipole, electric-quadrupole, and magnetic-quadrupole contributions to the scattering cross section, respectively.}
    \label{fig_multipole}
\end{figure}

To quantify this behavior, a Cartesian multipole decomposition of the scattering cross section is performed at each resonance. This decomposition includes electric dipole, magnetic dipole, toroidal dipole, electric quadrupole, and magnetic quadrupole components, whose theoretical formulations are provided in Section 1 of the supplementary material. Figures~\ref{fig_multipole}(a) and \ref{fig_multipole}(b) show the results for all resonant wavelengths at $\delta = 0.14$. At most resonances, the electric and magnetic quadrupole modes dominate the scattering response. This can be attributed to the presence of four asymmetric nanoblocks within the unit cell. The magnetic and toroidal dipole modes also contribute significantly, whereas the electric dipole contribution remains comparatively weak across all resonances. The field distributions presented in the supplementary material are in good agreement with these results, further validating the underlying physical interpretation.

\begin{figure}[t!]
    \centering
     \subfloat[]{\includegraphics[width=0.45\columnwidth]{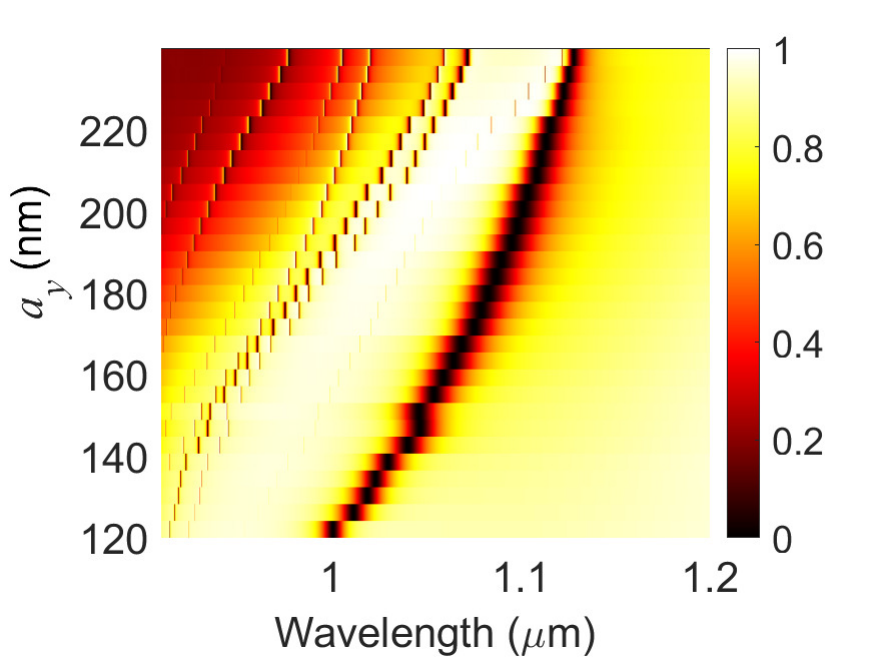}}\hspace{12pt}
    \subfloat[]{\includegraphics[width=0.45\columnwidth]{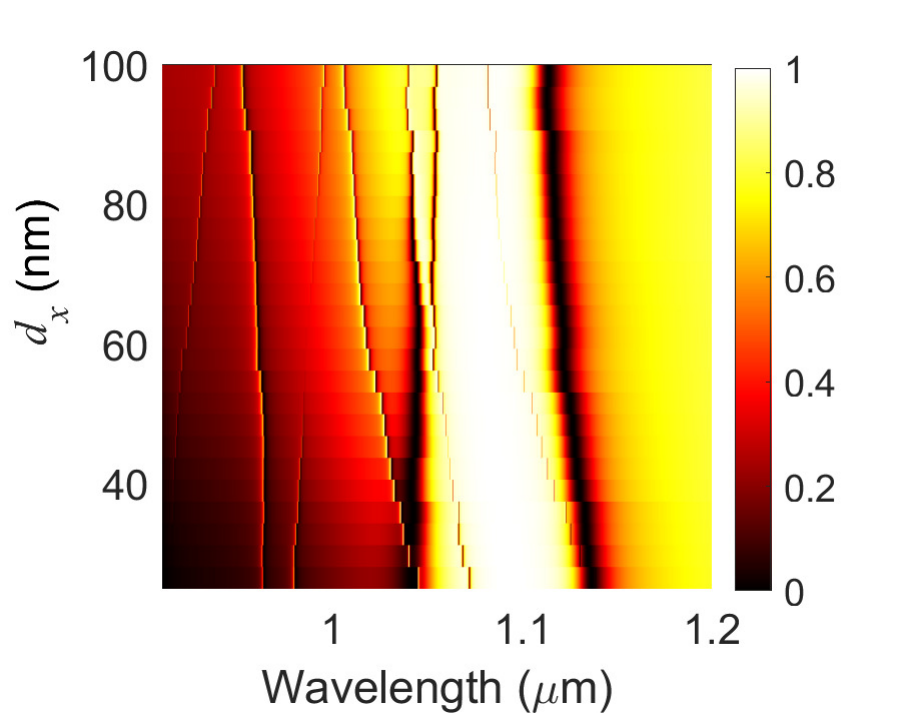}}\\ 
    \subfloat[]{\includegraphics[width=0.45\columnwidth]{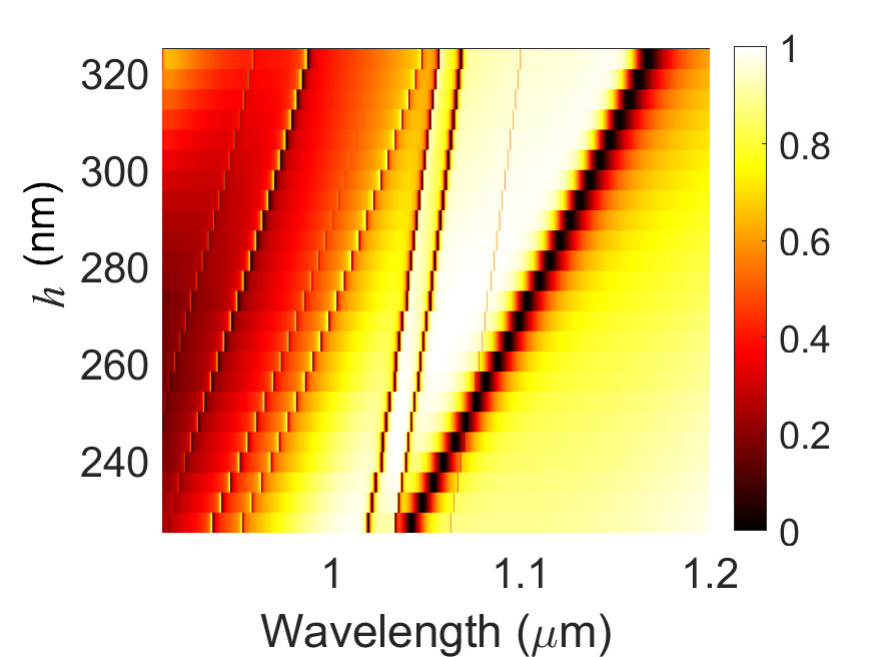}}\hspace{12pt}
        \subfloat[]{\includegraphics[width=0.45\columnwidth]{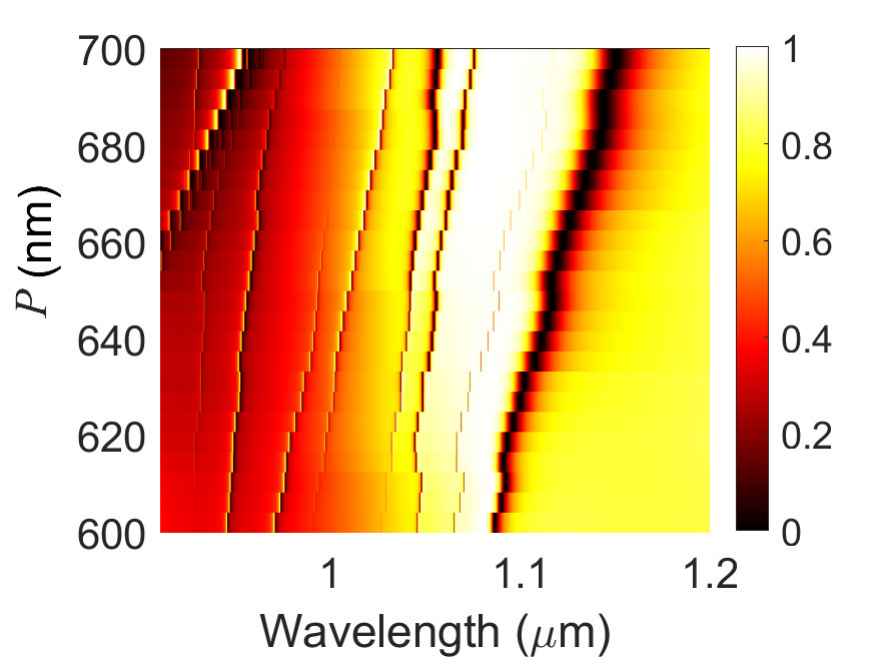}}
   \caption{Normalized transmission spectra for (a) varying the nanoblock dimension $a_{y}$ along the $y$-direction, (b) varying the separation distance $d_{x}$ along the $x$-direction, (c) varying the nanoblock height $h$, and (d) varying the period $P$ of the unit cell.}
    \label{fig_structure_parameter}
\end{figure}

The influence of geometrical parameters on the transmission spectra is systematically investigated at a fixed asymmetry parameter of $\delta = 0.14$. Figure~\ref{fig_structure_parameter}(a) presents the effect of varying the nanoblock dimension $a_y$ along the $y$-direction while holding all other parameters at their design values. Increasing $a_y$ produces a pronounced redshift of all resonances in the transmission spectrum, indicating strong sensitivity of the resonant modes to the in-plane block dimensions. In contrast, increasing the separation distance $d_{x}$ along the $x$-direction leads to a blueshift of the resonance wavelengths, as shown in Figure~\ref{fig_structure_parameter}(b). This behavior can be attributed to the modified near-field coupling between adjacent nanoblocks as the separation distance changes. The impact of the nanoblock height $h$ on the transmission response is illustrated in Figure~\ref{fig_structure_parameter}(c). The results demonstrate that variation in $h$ provides substantial tunability over the spectral positions of the resonances, offering an additional degree of freedom for resonance engineering. Furthermore, a slight redshift of the resonances is observed when the period $P$ is increased from $600\,\mathrm{nm}$ to $700\,\mathrm{nm}$, as shown in Figure~\ref{fig_structure_parameter}(d). This shift arises from changes in lattice coupling and collective diffraction effects associated with the periodic arrangement. Overall, these results confirm that the spectral characteristics of the multiband Fano resonances can be effectively tailored through precise control of the geometrical parameters, providing a versatile platform for engineering high-Q, narrowband resonances for a wide range of applications, including sensing, filtering, nonlinear optics, and other advanced photonic technologies.

\section{Nonlinear simulation}
Next, full-wave nonlinear simulations are performed to evaluate the efficiency of TH generation at each resonant wavelength. The nonlinear simulations are performed under the undepleted-pump approximation, which neglects the modifications to the fundamental field caused by energy transfer to the generated TH field. This approximation is justified because the conversion efficiency remains below $1\%$ and the metasurface is subwavelength in thickness, resulting in limited pump depletion. The wavelength-dependent complex permittivity of Si described above is employed at both the fundamental and TH frequencies, thereby accounting for linear absorption and attenuation of the generated TH field. However, the model neglects two-photon absorption (TPA), free-carrier absorption (FCA), and thermally induced changes in the material properties, so the calculated efficiencies represent upper-bound estimates. The incident $y$-polarized plane wave (the pump in the simulations) serves as a fixed excitation for the nonlinear polarization. The third-order nonlinear susceptibility of amorphous Si is assumed to be dispersionless over the wavelength range from $930$ to $1150\,\mathrm{nm}$. The computational procedure involves the following steps. First, the electric field distribution at the fundamental frequency $(\omega)$ is obtained over the unit cell from the Fourier transform of the time-domain fields computed by FDTD. Subsequently, the nonlinear polarization is evaluated as
\begin{equation*}
P_i^{(3)}(\mathbf{r},3\omega) = \epsilon_{0}\sum_{j,k,l}\chi^{(3)}_{ijkl}\,E_j(\mathbf{r},\omega)\,E_k(\mathbf{r},\omega)\,E_l(\mathbf{r},\omega),
\end{equation*}
where $P_i^{(3)}$ is the $i$-th Cartesian component of the nonlinear polarization, $\epsilon_{0}$ is the vacuum permittivity, $\chi^{(3)}_{ijkl}$ is the third-order susceptibility tensor, and the indices $i,j,k,l \in \{x,y,z\}$. For an isotropic material such as amorphous Si, the tensor reduces to a single independent component, and the nonlinear polarization takes the form
\begin{equation*}
\mathbf{P}^{(3)}(\mathbf{r},3\omega) = \epsilon_{0}\,\chi^{(3)}\,[\mathbf{E}(\mathbf{r},\omega)\cdot\mathbf{E}(\mathbf{r},\omega)]\,\mathbf{E}(\mathbf{r},\omega),
\end{equation*}
where $\chi^{(3)}$ is the independent scalar susceptibility. Finally, the far-field radiation generated by the induced nonlinear polarization is calculated using Green’s function formalism. The total radiated TH power is obtained by integrating the far-field intensity over the full solid angle. In these simulations, $\chi^{(3)}$ of Si is taken as $2\times10^{-19}\,\mathrm{m^2/V^2}$ and is assumed to be dispersionless over the investigated pump-wavelength range, whereas the linear refractive indices remain wavelength dependent.

\begin{figure}[t!]
    \centering
    \subfloat[]{\includegraphics[width=0.33\columnwidth]{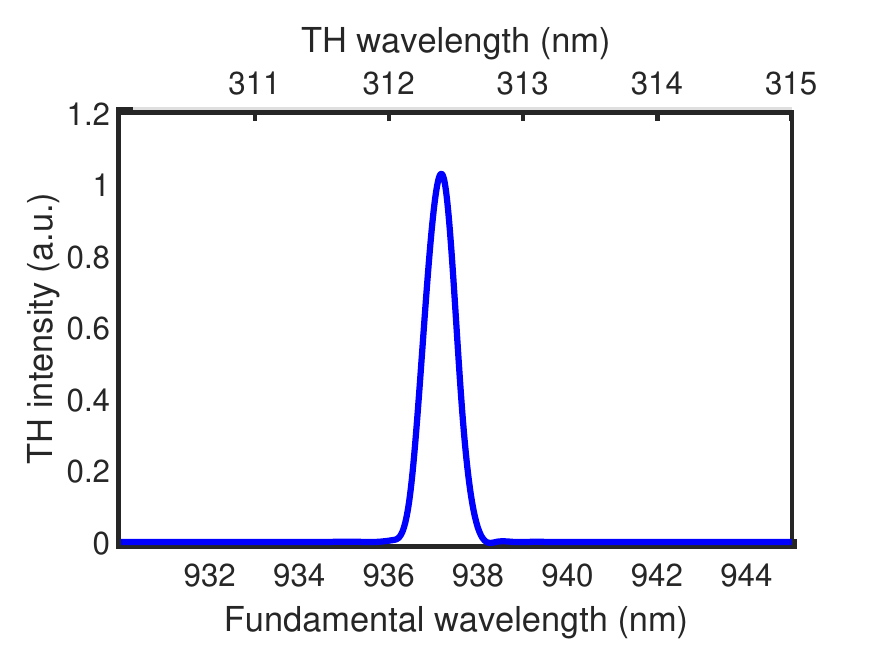}}
    \subfloat[]{\includegraphics[width=0.33\columnwidth]{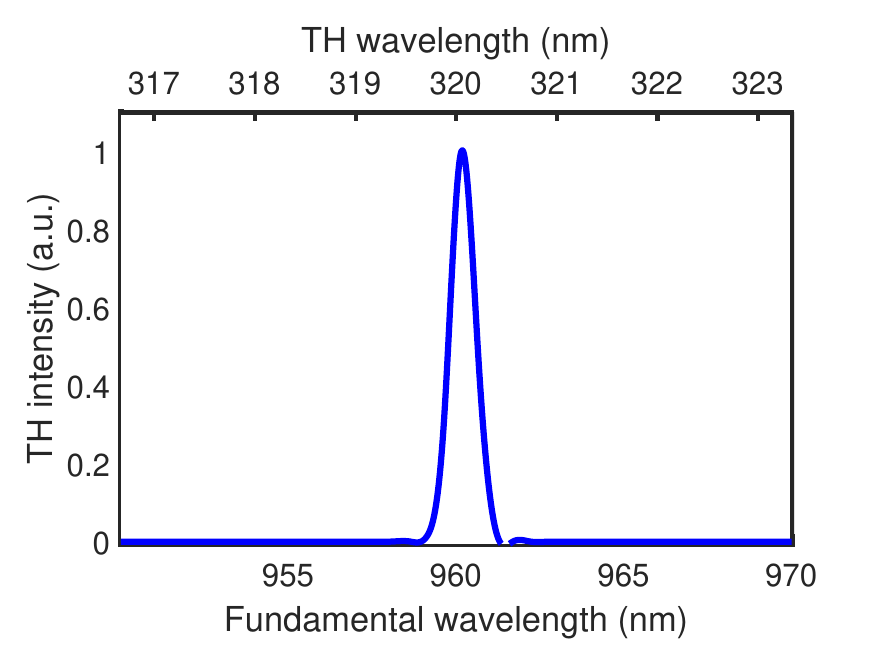}} 
    \subfloat[]{\includegraphics[width=0.33\columnwidth]{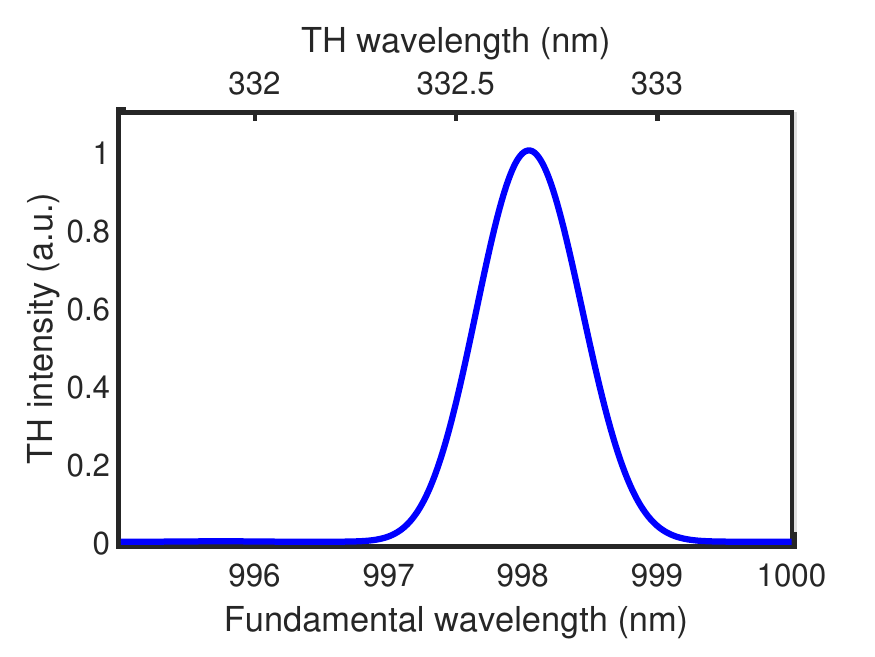}}\vspace{-0.2cm}\\
    \subfloat[]{\includegraphics[width=0.33\columnwidth]{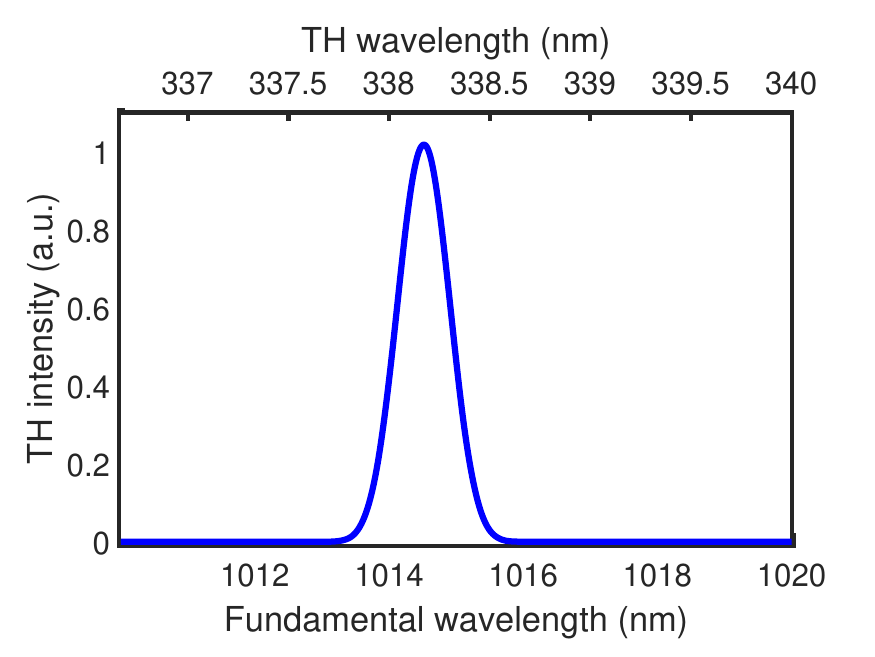}}
    \subfloat[]{\includegraphics[width=0.33\columnwidth]{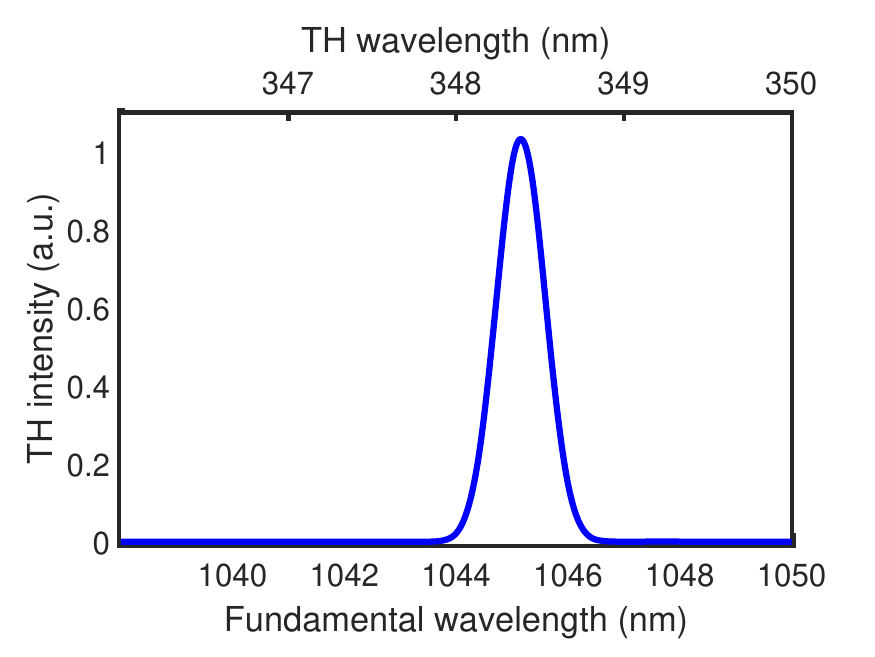}}
    \subfloat[]{\includegraphics[width=0.33\columnwidth]{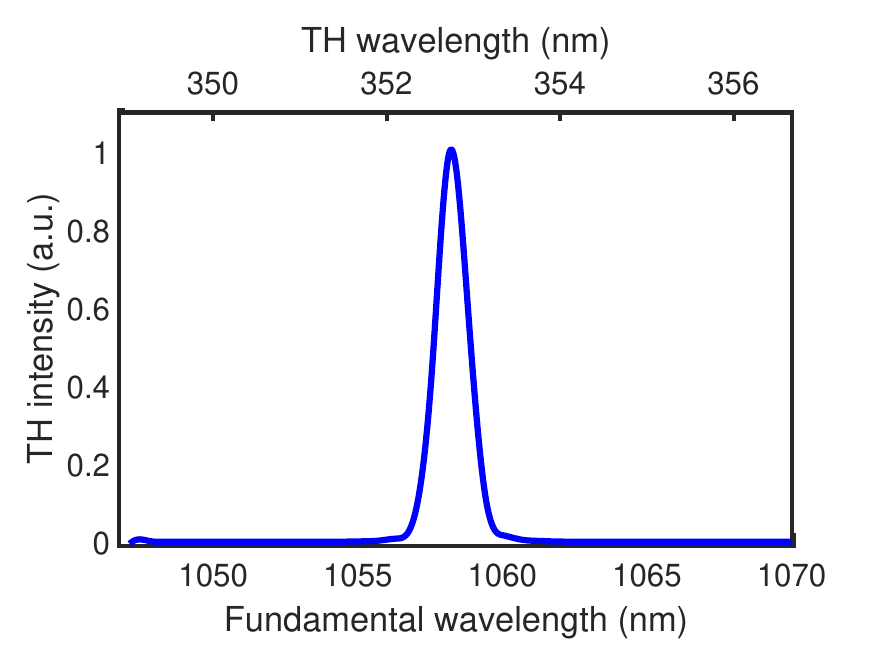}}\vspace{-0.2cm}\\
    \subfloat[]{\includegraphics[width=0.33\columnwidth]{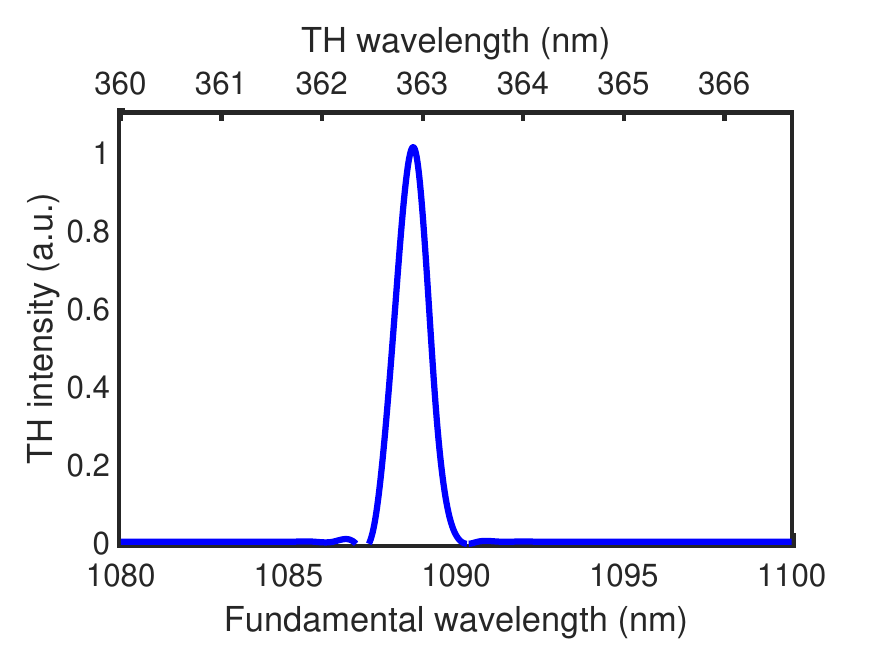}}
    \subfloat[]{\includegraphics[width=0.33\columnwidth]{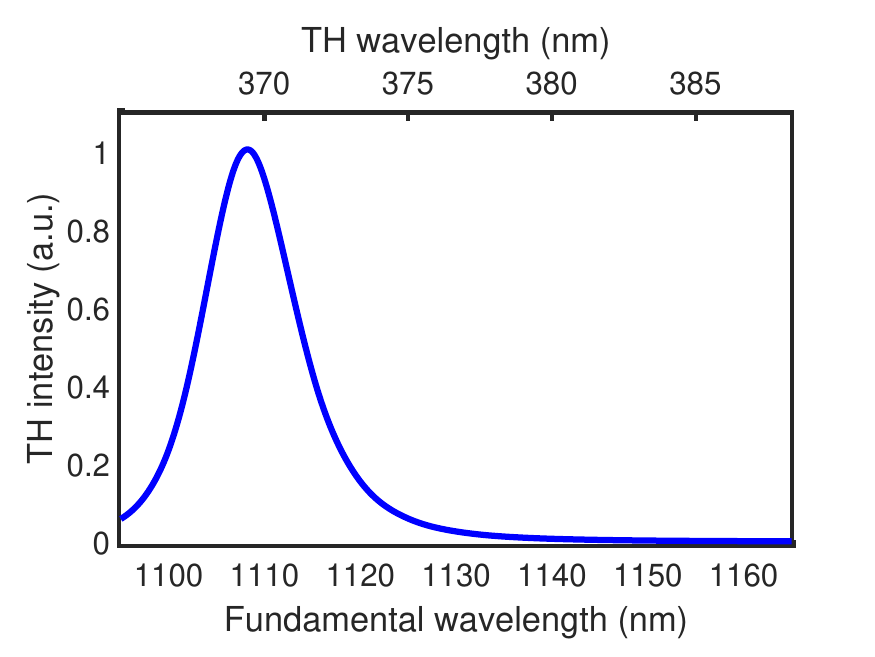}}
    \caption{Simulated TH intensities at the eight resonances, computed at $\delta = 0.14$, peak pump intensity $1.6\,\mathrm{GW/cm^{2}}$, and effective pump-beam area $314\,\mathrm{\mu m^{2}}$. The simulated TH intensities correspond to excitation at the resonances (a) $M_{1}$, (b) $M_{2}$, (c) $M_{3}$, (d) $M_{4}$, (e) $M_{5}$, (f) $M_{6}$, (g) $M_{7}$, and (h) $M_{8}$.}
    \label{fig_M_wavelength_variation}
\end{figure}

The generated TH signal is evaluated in transmission. The TH conversion efficiency is defined as $\eta = P_{3\omega}/P_{\omega}$, where $P_{3\omega}$ and $P_{\omega}$ represent the TH and pump powers, respectively. The simulated TH spectra as a function of pump wavelength across the multiband resonances are shown in Figure~\ref{fig_M_wavelength_variation}, computed at $\delta = 0.14$ with a peak pump intensity of $1.6\,\mathrm{GW/cm^{2}}$. The corresponding conversion efficiencies are presented in Figure~S4 of the supplementary material. As expected, both the TH signal and the conversion efficiency reach their maxima near the resonance wavelengths, where the electromagnetic field localization is strongest.
\begin{figure}[t!]
    \centering
    \subfloat[]{\includegraphics[width=0.7\columnwidth]{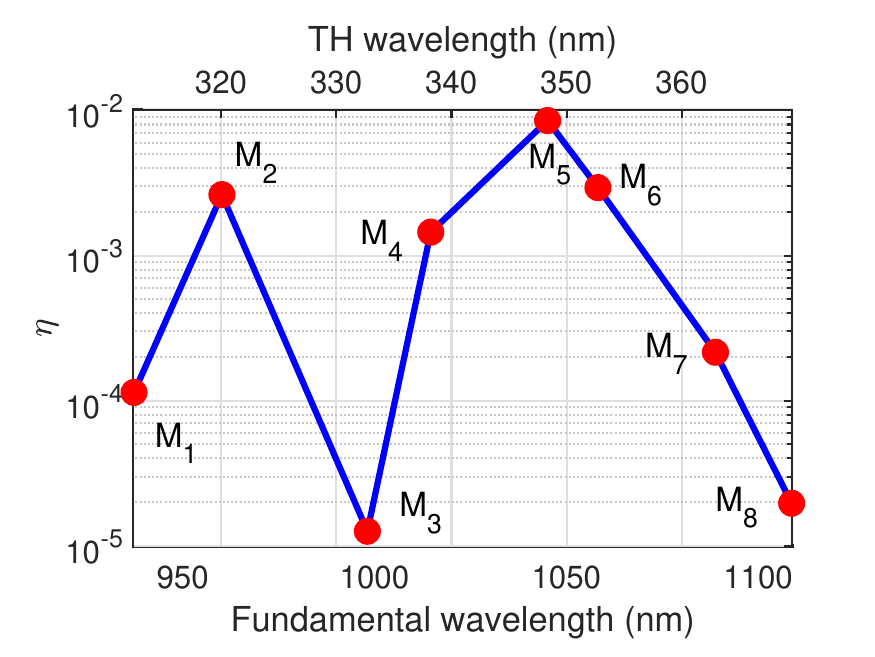}}\\[2mm]
    \subfloat[]{\includegraphics[width=0.85\columnwidth]{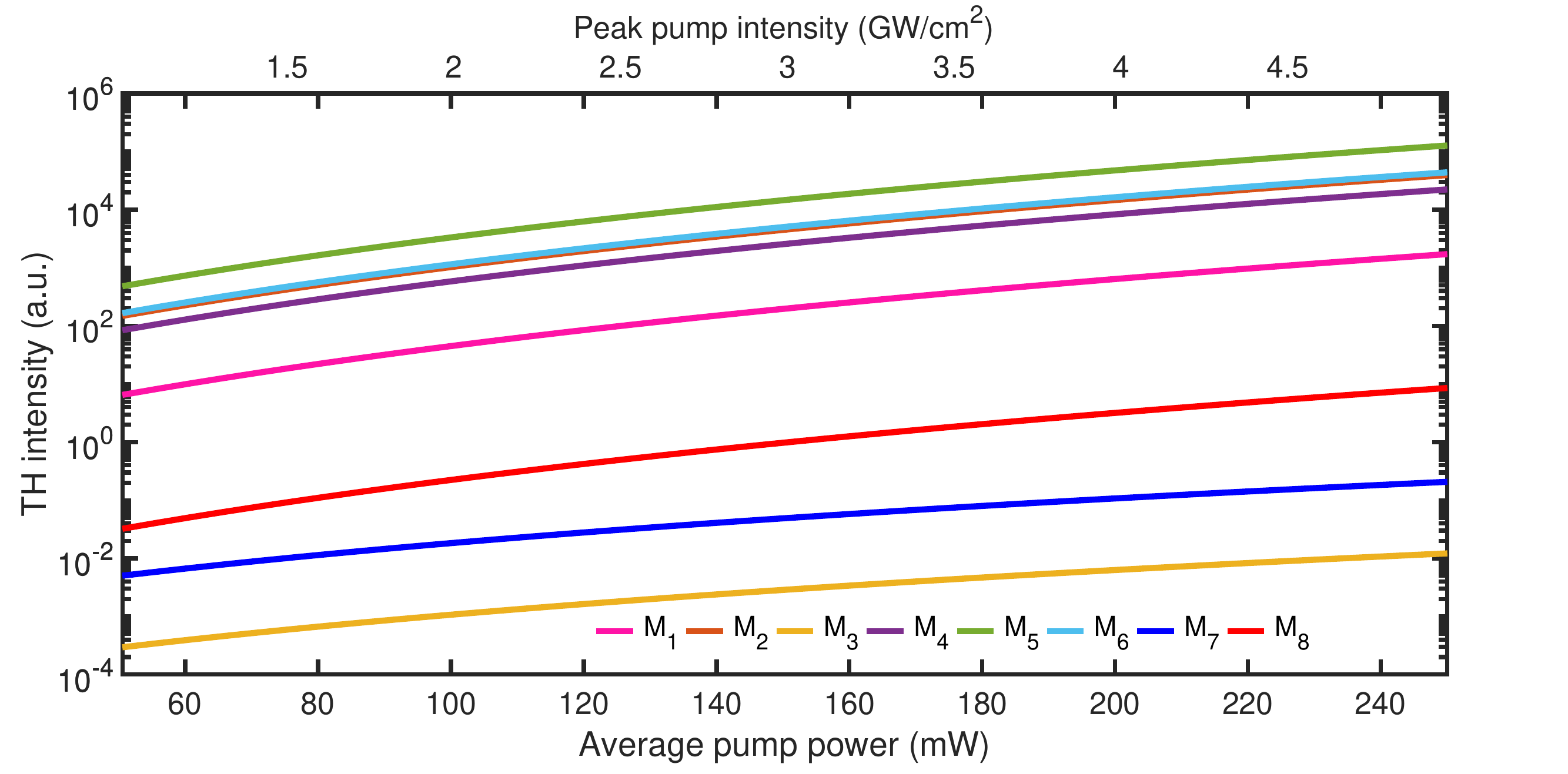}}
    \caption{(a) Simulated conversion efficiency at each resonance for $\delta = 0.14$ and a peak pump intensity of $1.6\,\mathrm{GW/cm^{2}}$. (b) Simulated TH intensity as a function of input pump power at each resonance for $\delta = 0.14$.}
    \label{fig_sim_eff_power_variation}
\end{figure}

A pronounced variation in conversion efficiency is observed among different resonances. This behavior can be attributed to differences in modal contributions and in the corresponding $Q$ factors. Lower $Q$ factors result in increased radiative leakage, thereby reducing local field enhancement and, consequently, the TH conversion efficiency. Figure~\ref{fig_sim_eff_power_variation}(a) summarizes the simulated TH conversion efficiencies of modes $M_1$--$M_8$ at $\delta=0.14$ and a peak pump intensity of $1.6\,\mathrm{GW/cm^{2}}$. It is observed that the TH efficiency does not vary monotonically with the $Q$ factor. Although $M_1$, $M_3$, and $M_7$ exhibit relatively high $Q$ factors, $M_5$, with an intermediate $Q$ factor of $1\,002$, provides the highest conversion efficiency of $8.5\times10^{-3}$. According to temporal coupled-mode theory~\cite{koshelev2019nonlinear}, the total $Q$ factor satisfies $Q^{-1}=Q_{\mathrm{r}}^{-1}+Q_{\mathrm{nr}}^{-1}$, where $Q_{\mathrm{r}}$ and $Q_{\mathrm{nr}}$ are the radiative and non-radiative $Q$ factors, respectively. Maximum field enhancement generally occurs near the critical-coupling condition $Q_{\mathrm{r}}=Q_{\mathrm{nr}}$, rather than at the largest total $Q$ factor. 

The high efficiency of $M_5$ is consistent with a more favorable balance between radiative coupling and intrinsic (non-radiative) loss, together with its strong field localization within the nonlinear a-Si volume, although the radiative and non-radiative contributions are not separated in the present analysis. Differences in effective mode volume, multipolar composition, nonlinear modal overlap, and TH outcoupling further contribute to the mode-dependent efficiencies. It should be noted, however, that in practical implementations the efficiency may be slightly reduced due to TPA, FCA, and fabrication imperfections. Figure~\ref{fig_sim_eff_power_variation} (b) illustrates the dependence of TH intensity on the input pump power for all resonant wavelengths. In each case, the TH intensity increases with pump power, with the strongest response observed at $M_{5}$, consistent with its highest conversion efficiency. To further confirm the nonlinear order of the process, a log--log plot of TH power $P_{3\omega}$ versus peak pump intensity $I_{\omega}$ is generated for $M_{8}$ (see Figure~S5 in the supplementary material). The simulation data are fitted to the power-law relation $P_{3\omega} = a\,(I_{\omega})^b$, where $a$ is a constant prefactor and $b$ is the fitted exponent. The fit yields $a = 0.07082$ and $b = 3.0$, confirming the ideal cubic dependence of the generated TH power on the incident pump intensity and, consequently, its third-order nonlinear origin.





\section{Experimental results}
\begin{figure}[t!]
    \centering

    \subfloat[]{%
        \includegraphics[width=0.45\columnwidth]
        {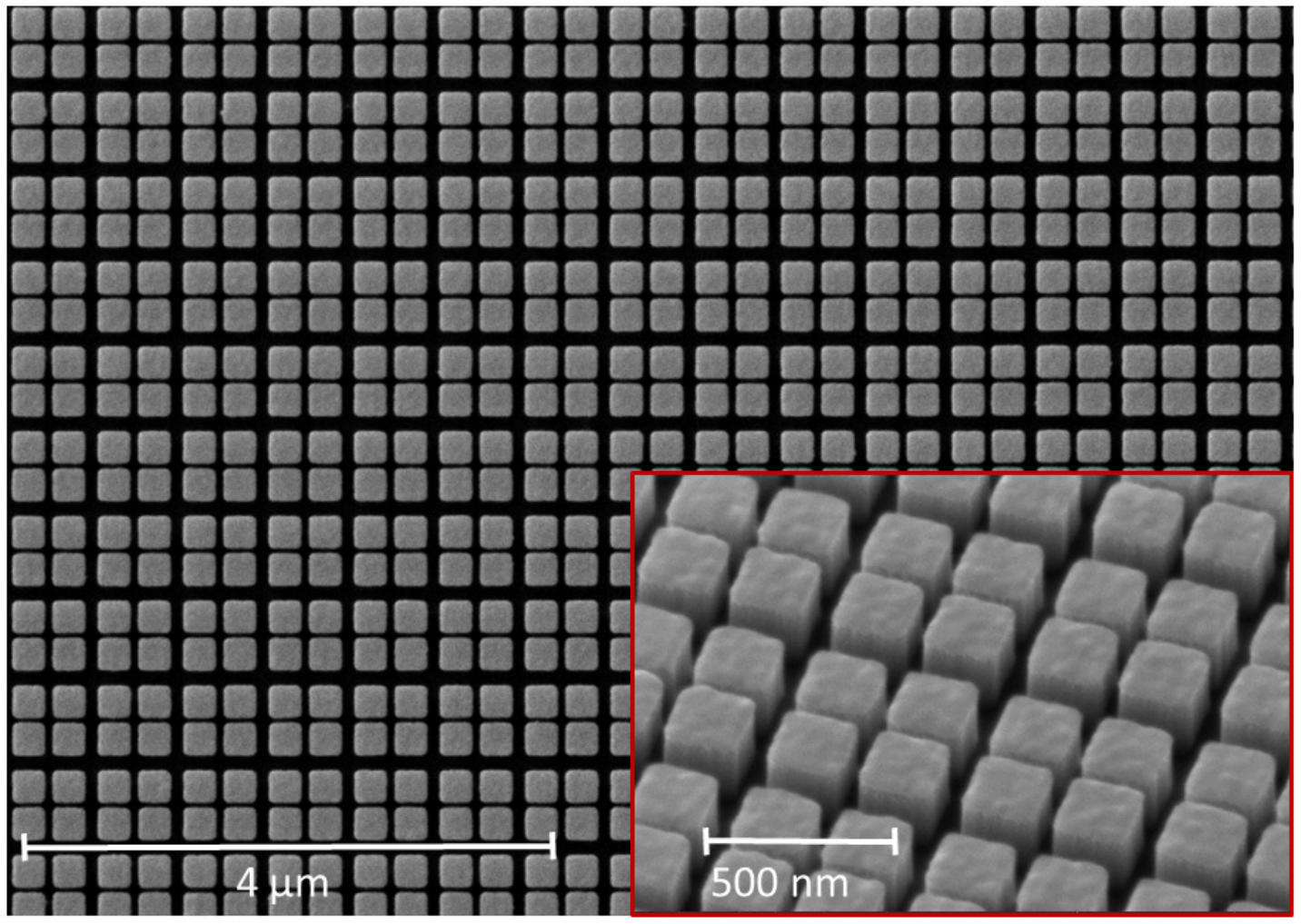}}%
    \subfloat[]{%
        \includegraphics[width=0.37\columnwidth]
        {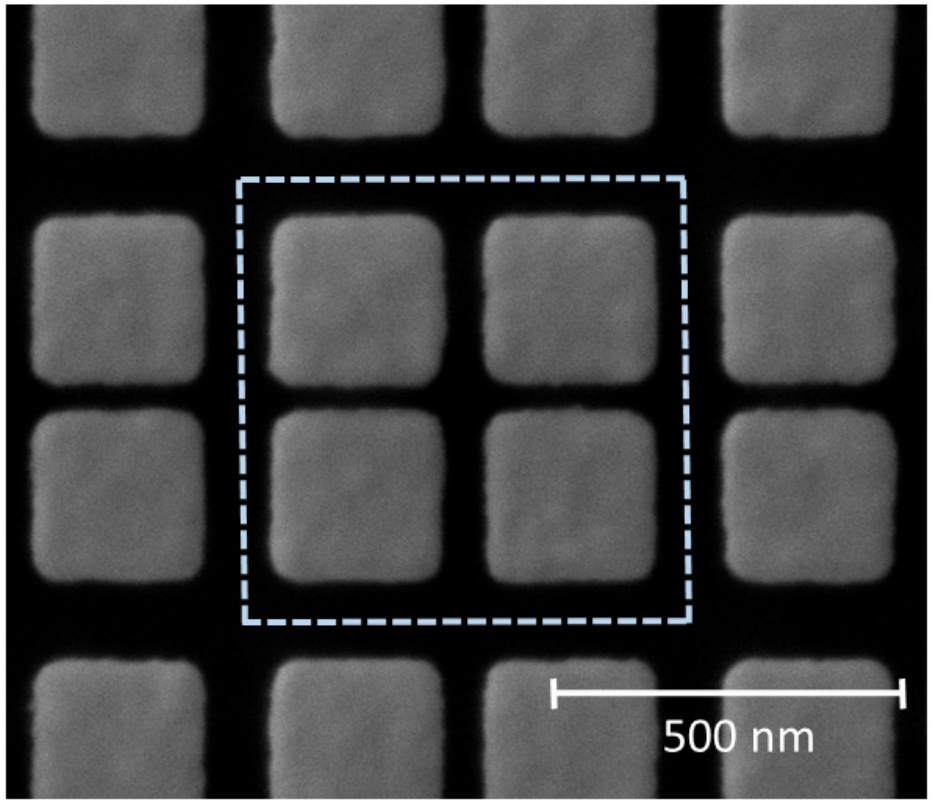}}\\

    \subfloat[]{%
        \includegraphics[width=0.45\columnwidth]
        {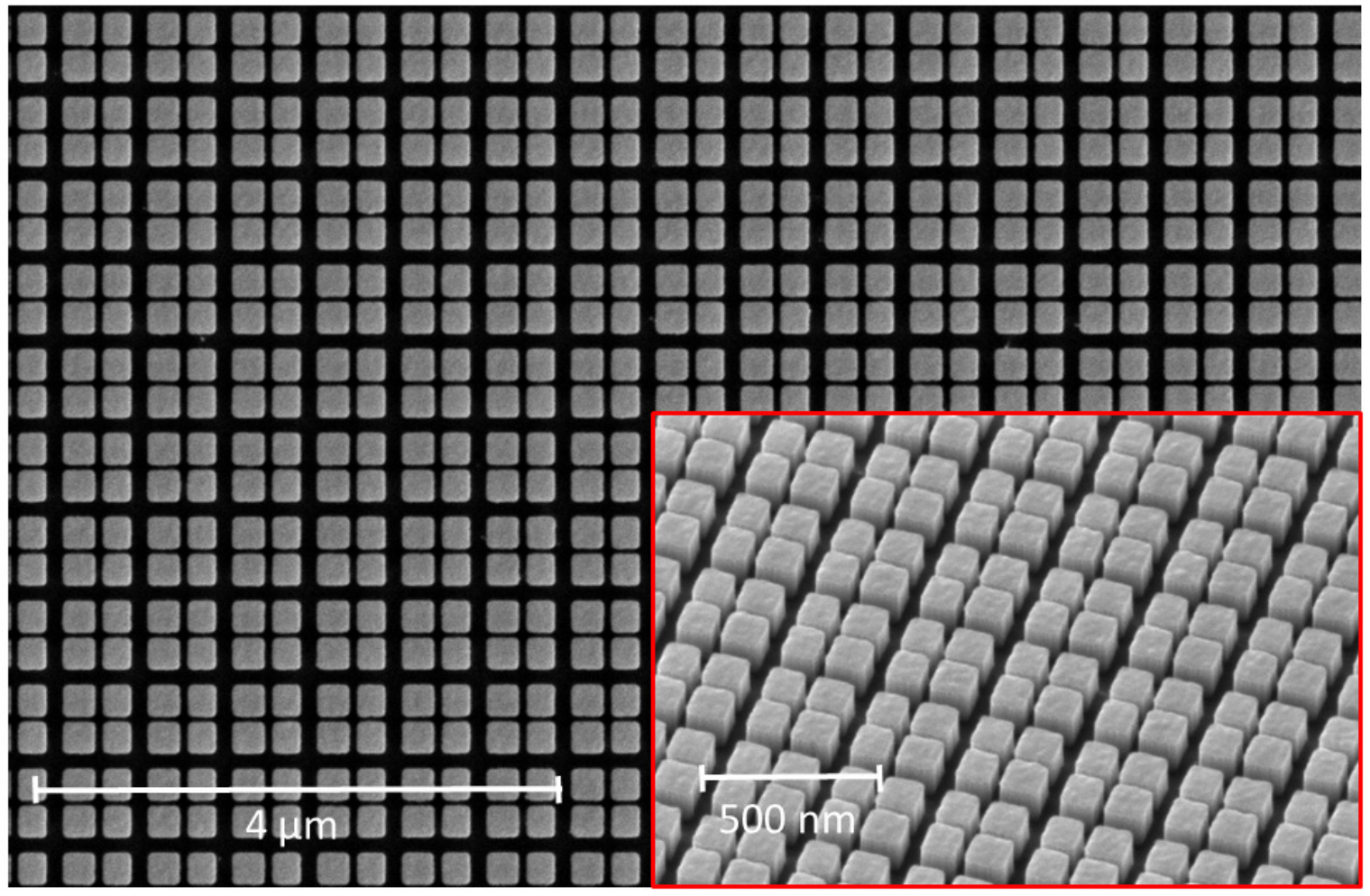}}%
    \subfloat[]{%
        \includegraphics[width=0.355\columnwidth]
        {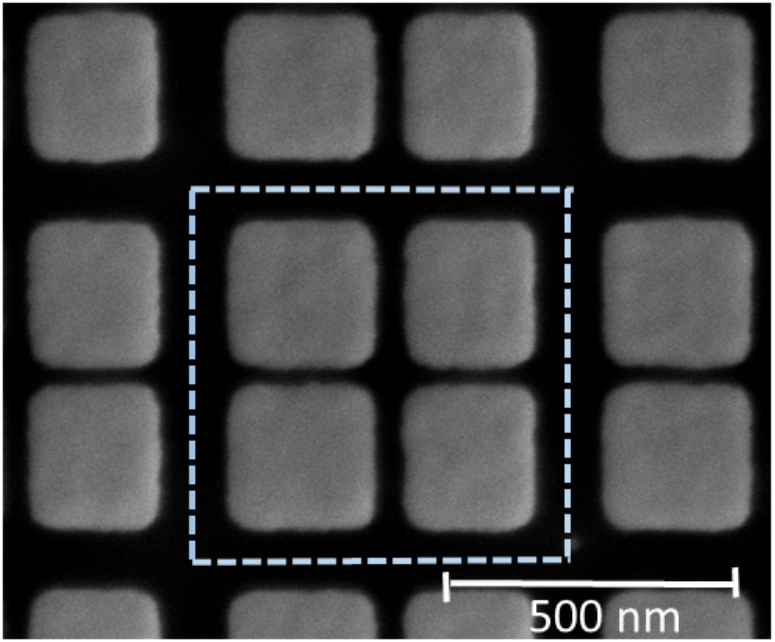}}\\

    \makebox[\columnwidth][c]{%
        \subfloat[]{%
            \includegraphics[width=0.5\columnwidth]
            {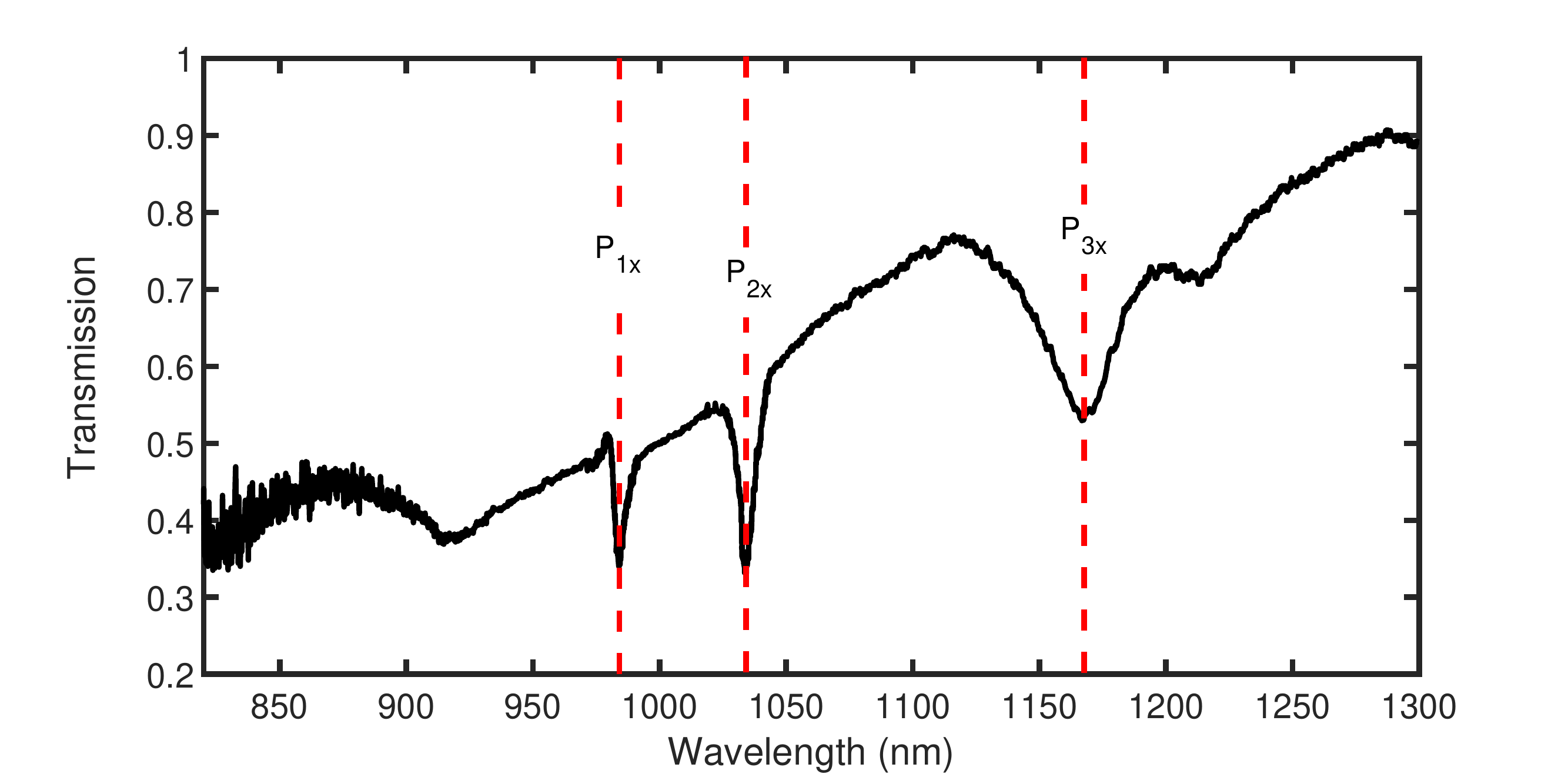}}%
        \hspace{-0.025\columnwidth}%
        \subfloat[]{%
            \includegraphics[width=0.5\columnwidth]
            {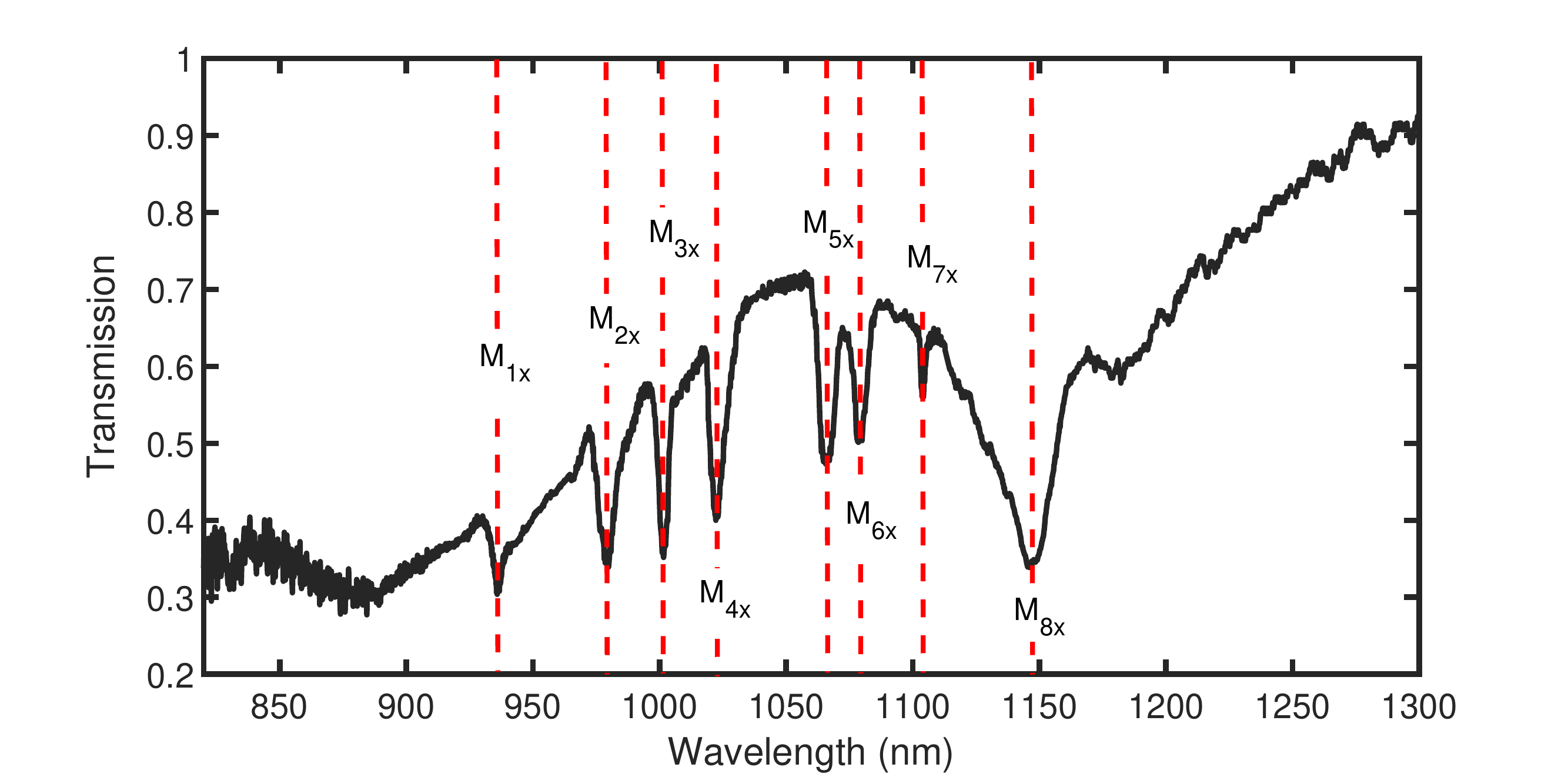}}%
    }

    \caption{SEM images and measured transmission spectra of the fabricated metasurfaces. (a) Top-view SEM image of the symmetric metasurface. (b) Enlarged SEM image of the symmetric unit cell. (c) Top-view SEM image of the asymmetric metasurface. (d) Enlarged SEM image of the asymmetric unit cell. Measured transmission spectra of the (e) symmetric and (f) asymmetric metasurfaces.}

    \label{fig_sym_asym_image_spectra}
\end{figure}

To experimentally demonstrate the multiresonances and enhanced TH generation, the designed metasurface structures are fabricated in both symmetric and asymmetric configurations. a-Si is first deposited on a glass substrate using plasma-enhanced chemical vapor deposition (PECVD). The metasurface patterns are subsequently defined by electron-beam lithography (EBL) using a JEOL JBX-6300FS system on a $200\,\mathrm{nm}$-thick layer of AR-P 6200 (CSAR 62) positive electron-beam resist spin-coated onto the a-Si film. Following resist development, a $20\,\mathrm{nm}$-thick chromium (Cr) layer is deposited as a hard mask. The pattern is then transferred into the a-Si layer through a lift-off process followed by reactive ion etching (RIE). Finally, symmetric and asymmetric metasurfaces with a footprint of $400\,\mathrm{\mu m} \times 400\,\mathrm{\mu m}$ are realized on the glass substrate. Detailed fabrication procedures are provided in Figure S6 in the supplementary material. The top-view scanning electron microscope (SEM) images of the fabricated symmetric and asymmetric structures are shown in Figures~\ref{fig_sym_asym_image_spectra}(a) and~\ref{fig_sym_asym_image_spectra}(c), respectively. Figures~\ref{fig_sym_asym_image_spectra}(b) and~\ref{fig_sym_asym_image_spectra}(d) show enlarged top-view SEM images of the corresponding unit cells for the symmetric and asymmetric structures, respectively.

\begin{figure}[t!]
    \centering
     \subfloat[]{\includegraphics[width=0.6\columnwidth]{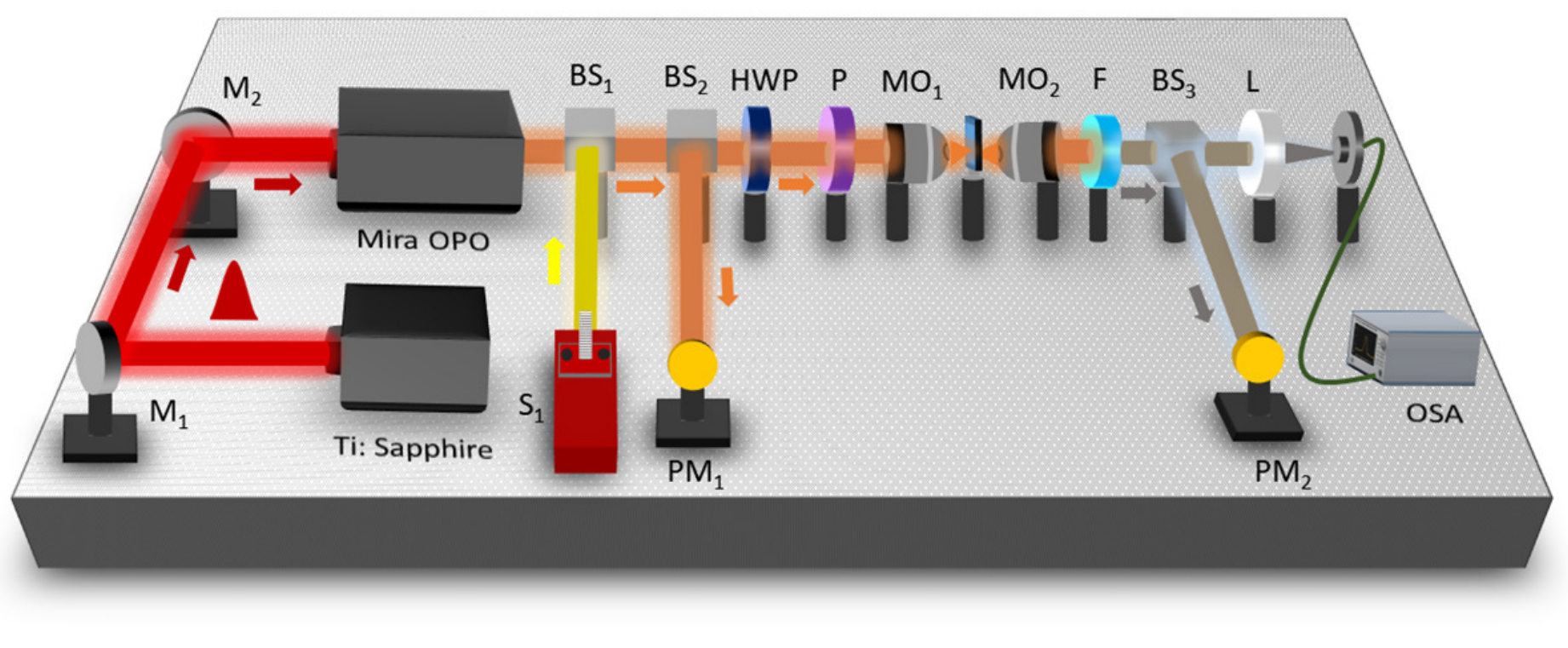}}\\
    \subfloat[]{\includegraphics[width=0.4\columnwidth]{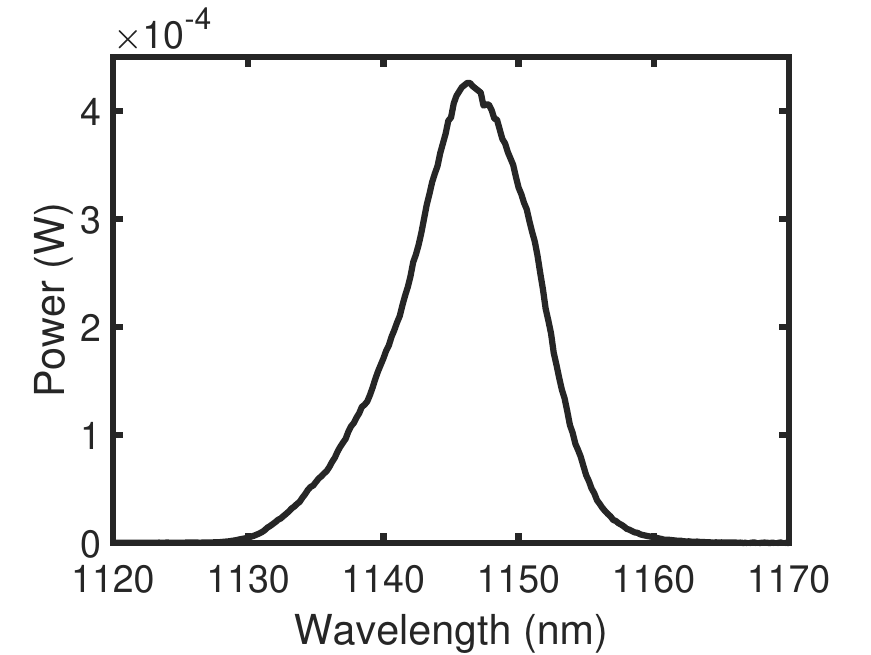}}
   \subfloat[]{\includegraphics[width=0.4\columnwidth]{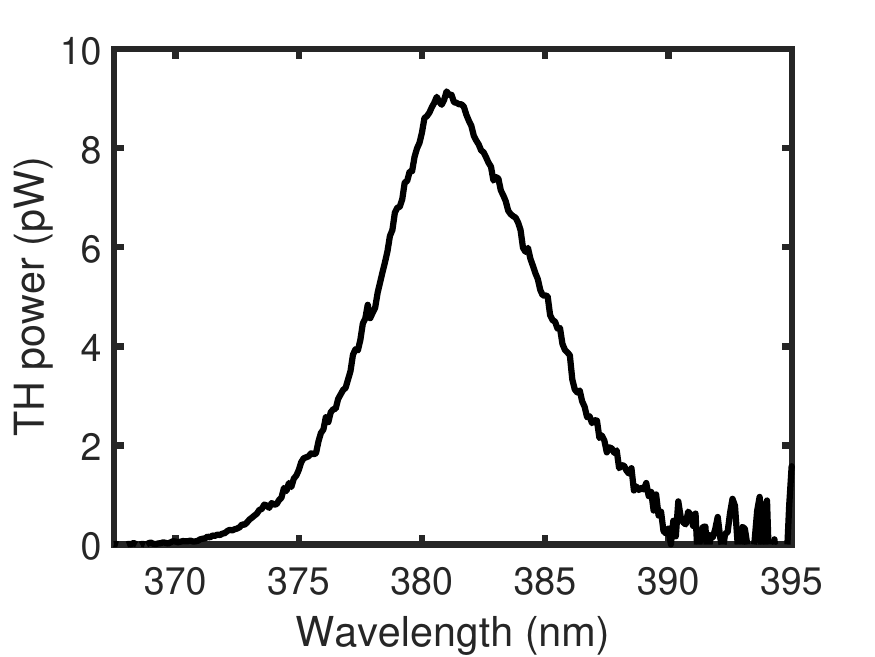}}\\
    \subfloat[]{\includegraphics[width=0.45\columnwidth]{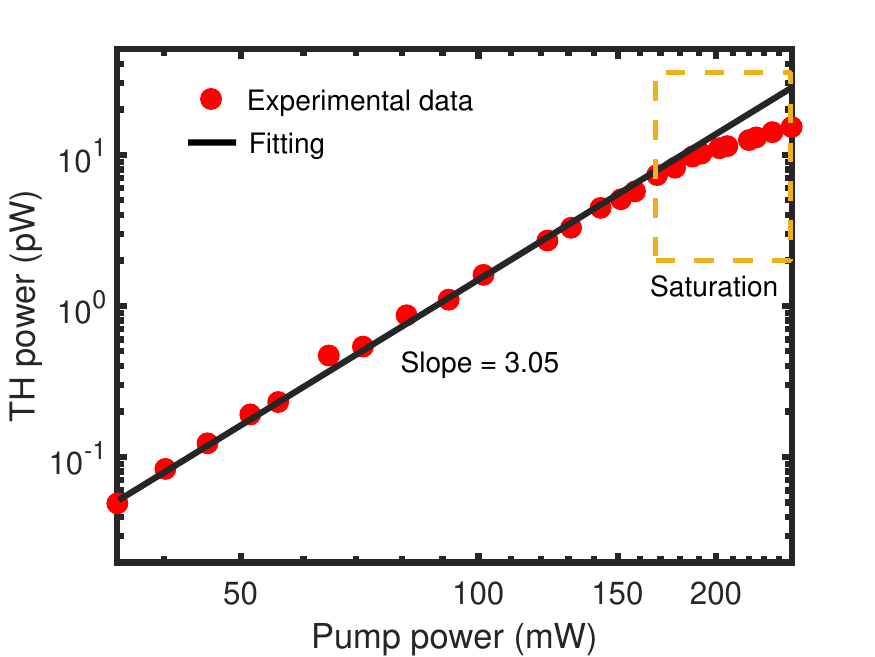}}
   \subfloat[]{\includegraphics[width=0.35\columnwidth]{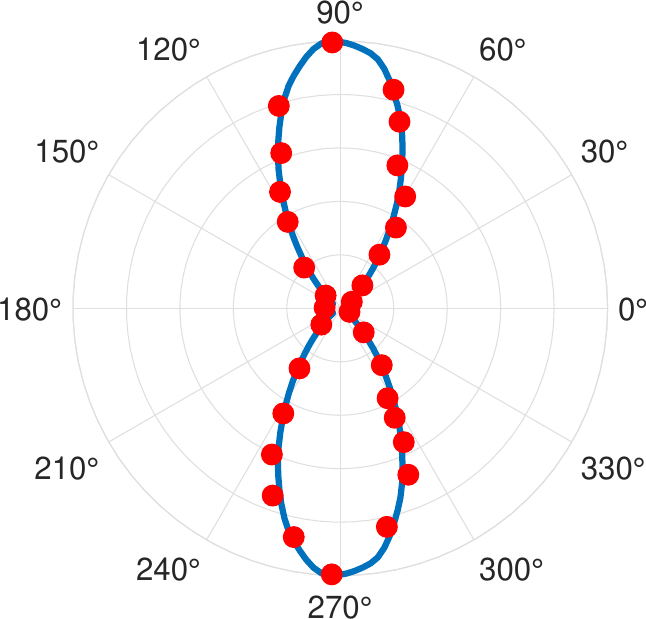}} 
    \caption{(a) Schematic of the linear and nonlinear characterization setup. $\mathrm{M_{1}}$, $\mathrm{M_{2}}$: mirror; $\mathrm{BS_{1}}$--$\mathrm{BS_{3}}$: beam splitter; $\mathrm{S_{1}}$: halogen source; HWP: half-wave plate; $\mathrm{PM_{1}}$, $\mathrm{PM_{2}}$: power meter; P: polarizer; $\mathrm{MO_{1}}$, $\mathrm{MO_{2}}$: microscope objective; F: short-pass filter; L: lens; OSA: optical spectrum analyzer. (b) Spectrum of the fundamental pump laser when pumping at $M_{8\mathrm{x}}$. (c) Measured TH spectrum of the Si metasurface. (d) Measured TH power as a function of the incident pump power. (e) Measured TH power as a function of the incident pump polarization angle.}
    \label{fig_THG_exp_results}
\end{figure}
The SEM images confirm high structural uniformity, minimal surface roughness, and well-defined vertical sidewalls. For optical characterization, a broadband halogen light source (Thorlabs SLS201L) is used, and the incident polarization is controlled using a broadband polarizer. The input beam is focused at normal incidence onto the metasurface using microscope objectives. The transmitted light is collected and analyzed using an optical spectrum analyzer (Yokogawa AQ6370B), and all measured spectra are normalized to the transmission of a bare glass substrate. Figure~\ref{fig_sym_asym_image_spectra}(e) presents the measured transmission spectrum of the symmetric structure under a $y$-polarized incident wave. Three resonance peaks are observed at $984.1\,\mathrm{nm}$ ($P_{1\mathrm{x}}$), $1034.2\,\mathrm{nm}$ ($P_{2\mathrm{x}}$), and $1167.5\,\mathrm{nm}$ ($P_{3\mathrm{x}}$). In contrast, the asymmetric structure exhibits multiple resonance peaks, as shown in Figure~\ref{fig_sym_asym_image_spectra}(f), located at $935.8\,\mathrm{nm}$ ($M_{1\mathrm{x}}$), $979.6\,\mathrm{nm}$ ($M_{2\mathrm{x}}$), $1000.8\,\mathrm{nm}$ ($M_{3\mathrm{x}}$), $1022.6\,\mathrm{nm}$ ($M_{4\mathrm{x}}$), $1066.2\,\mathrm{nm}$ ($M_{5\mathrm{x}}$), $1079.5\,\mathrm{nm}$ ($M_{6\mathrm{x}}$), $1103.9\,\mathrm{nm}$ ($M_{7\mathrm{x}}$), and $1146.4\,\mathrm{nm}$ ($M_{8\mathrm{x}}$). Both the symmetric and asymmetric measured spectra show good agreement with the corresponding numerical simulations presented in Figure~\ref{fig2}. The $Q$ factors of the experimentally observed resonances are extracted by fitting the measured transmission spectrum with the same Fano line-shape model employed for the simulated spectra. The corresponding experimental data and fitting curves are presented in Figure~S7 of the supplementary material. The experimental $Q$ factors of modes $M_{1\mathrm{x}}$--$M_{8\mathrm{x}}$ are found to be $271$, $148$, $260$, $161$, $156$, $195$, $550$, and $51$, respectively, compared with the corresponding simulated values of $8\,731$, $1\,509$, $15\,084$, $2\,942$, $1\,002$, $635$, $10\,167$, and $52$. The lower experimental values are primarily attributed to fabrication-induced dimensional variations, sidewall roughness and tapering, structural nonuniformity, material absorption, and additional scattering losses. Finite illumination numerical aperture, angular averaging, and the spectral resolution of the measurement system further broaden the measured resonances. These effects are particularly significant for the ultranarrow high-$Q$ modes $M_{1}$, $M_{3}$, and $M_{7}$. In contrast, the intrinsically broad $M_{8}$ mode exhibits close agreement between simulation and experiment. 

Next, the TH response of the metasurfaces is characterized. The excitation source is an optical parametric oscillator (Mira OPO S2 fs, S04471) pumped by a Ti:sapphire laser Coherent Mira-HP-F operating at $830\,\mathrm{nm}$ with a pulse width of $\sim 150\,\mathrm{fs}$ and a repetition rate of $76\,\mathrm{MHz}$ before the OPO~\cite{almalawi2020enhanced, alreshidi2024enhanced, alamoudi2025carrier}. Mira-HP-F is pumped by Coherent Verdi with a power of $18\,\mathrm{W}$. The OPO is tuned to a central wavelength of $1146.4\,\mathrm{nm}$, with a repetition rate of $200\,\mathrm{kHz}$ and a pulse duration of $200\,\mathrm{fs}$. A schematic of the combined linear and nonlinear optical setup is presented in Figure~\ref{fig_THG_exp_results} (a). The laser beam first passes through a half-wave plate and a polarizer, enabling precise control of the input polarization and pump power. The beam is then focused onto the metasurface using a near-infrared microscope objective ($20\times$, $\mathrm{NA} = 0.4$). The generated TH signal is filtered using a short-pass filter to suppress the fundamental wavelength and is subsequently analyzed with an optical spectrum analyzer (OSA). In addition, a power meter placed after the filter is used to measure the TH power. For the experimental demonstration of TH generation, ultrafast pump pulses are incident on the metasurface at the $M_{8\mathrm{x}}$ resonance. The measured spectra of the pump wave at peak intensity of $3.25\,\mathrm{GW/cm^{2}}$ and the corresponding TH signals are shown in Figures~\ref{fig_THG_exp_results}(b) and (c), respectively.
It is observed that the bandwidth of the TH signal is narrower than that of the fundamental pump laser. The dependence of the TH power on the pump power at $M_{8\mathrm{x}}$ is presented in Figure~\ref{fig_THG_exp_results}(d). The plot confirms a cubic dependence with a fitted slope of $3.05$, thereby validating the third-order nonlinear origin of TH generation. The slight deviation of the experimental slope from the simulated value (see Figure~S5 in the supplementary material) can be attributed to fabrication imperfections and measurement uncertainties. At a peak pump intensity of $3.25\,\mathrm{GW/cm^{2}}$, an experimental conversion efficiency of $\eta=1.2 \times 10^{-6}$ is obtained at $M_{8\mathrm{x}}$. The measured TH conversion efficiency is substantially lower than the simulated value of approximately $10^{-5}$. Since the measured and simulated $Q$ factors of $M_{8\mathrm{x}}$ are nearly identical ($51$ and $52$, respectively), resonance broadening is not the dominant cause of this discrepancy. At the TH wavelength, multiple diffraction orders are allowed, and the simulation accounts for the total generated power distributed among all propagating channels. In contrast, the finite numerical aperture of the experimental collection optics captures only a fraction of the emitted TH power, excluding higher-order, large-angle, and backward-emitted components. Further differences arise from imperfect spatial, spectral, and angular coupling of the focused pump to the QBIC mode and uncertainties in the effective peak intensity. Because the TH conversion efficiency scales quadratically with pump intensity, such uncertainties significantly affect the inferred efficiency. Nonlinear absorption, free-carrier generation, and thermal effects in a-Si may additionally suppress the measured TH response at high pump intensities. 
 \begin{table*}[ht!]
\centering
\caption{Comparison of QBIC metasurfaces for enhanced TH generation.}
\label{tab:table1}
\renewcommand{\arraystretch}{1.35}
\setlength{\tabcolsep}{2.5pt}
\scriptsize

\begin{tabularx}{\textwidth}{
@{}
l
|>{\raggedright\arraybackslash}p{2.1cm}
|>{\centering\arraybackslash}p{1.1cm}
|>{\centering\arraybackslash}p{1.8cm}
|>{\centering\arraybackslash}p{1.8cm}
|>{\centering\arraybackslash}p{1.7cm}
|>{\centering\arraybackslash}p{2.1cm}
|>{\raggedright\arraybackslash}X
@{}
}
\hline

\textbf{Ref.} &
\textbf{Structure type} &
\textbf{\# high-$Q$ bands} &
\textbf{Pump wavelength (nm)} &
\textbf{Peak pump intensity ($\mathrm{GW/\,cm^{2}}$)} &
\textbf{TH wavelength (nm)} &
\textbf{Max. TH efficiency} &
\textbf{Key functionalities} \\
\hline

\cite{liu2019high} &
Si nanoblock array &
1 &
1588 &
N.R. &
529 &
$1.4\times10^{-8}\,\mathrm{W^{-2}}$ (normalized) &
High-$Q$ QBIC-assisted resonantly enhanced TH and SH generation (Linear and nonlinear experiment)\\
\hline

\cite{xu2019dynamic} &
Si nanodisk with an off-centred hole &
1 &
1345 &
1 &
448 &
$5\times 10^{-6}$ &
$>500$-fold TH enhancement and nonlinear image control (Linear and nonlinear experiment)\\
\hline

\cite{https://doi.org/10.1002/andp.202200263} &
Chiral dielectric metasurface &
2 &
1074.9 \& 1277.9 &
5.3 &
358 \& 426 &
$\sim 10^{-4}$ &
Dual-band TH generation combined with nonlinear circular dichroism (Numerical)\\
\hline

\cite{10.1021/acs.nanolett.5c00146} &
Si nanodisk metasurface on glass &
1 &
1539 &
5.85 &
513 &
$1.03 \times  10^{-5}$ &
Polarization-independent TH generation from a doubly degenerate QBIC (Linear and nonlinear experiment) \\
\hline

\cite{https://doi.org/10.1002/admt.202500297} &
Si nanodisk metasurface on glass &
1 &
2380 &
0.82 &
793 &
$1.6 \times  10^{-2}$ $\mathrm{W^{-2}}$ &
Frequency up conversion leveraging QBIC in elliptical dimer (Linear and nonlinear experiment) \\
\hline

\cite{TaoZhangJinPanQinHongFangShu+2024+3373+3384} &
Hybrid dielectric–GST phase-change metasurface  &
2 &
2450 \& 2904 &
0.001 (CW wave)  &
816.6 \& 968 &
$2.2 \times  10^{-5}$  &
Dynamic wavelength and polarization control using phase-change material (Numerical) \\
\hline

\cite{qin2024polarization} &
Rhomboidal a-Si on glass substrate &
2 &
1001 \& 1055 &
0.03  &
333.6 \& 351.6 &
$3.74 \times  10^{-3}$  &
QBIC-induced dual-band TH generation (Numerical) \\
\hline

\textbf{This work} &
\textbf{Four a-Si nanoblocks on glass} &
\textbf{8} &
Span over 937--1125
 &
\textbf{1.6 (Sim.) 3.25 (Exp.)} &
\textbf{Span over 312--375} &
\textbf{$8.5 \times 10^{-3}$ (Max. Sim.)} \newline
\textbf{$1.2 \times 10^{-6}$ at 1146.4 nm pump (Exp.)}
 &
\textbf{Highest number of distinct QBIC-assisted TH channels; broad near-UV coverage using a simple single-layer geometry (Linear and nonlinear experiment)} \\
\hline

\end{tabularx}

\vspace{1.5mm}
\begin{minipage}{\textwidth}

\end{minipage}
\end{table*}
Notably, at higher pump power levels, the TH power exhibits saturation-like behavior. This deviation is primarily attributed to pump-induced thermal and nonlinear effects in the silicon nanoresonators. At higher pump powers, optical absorption can induce local heating and free-carrier generation, leading to changes in the refractive index and increased optical losses. These effects can shift and broaden the resonant mode, thereby reducing the local-field enhancement at the pump wavelength and limiting the efficiency of the TH process~\cite{hahnel2023multi,tang2024realizing}. Furthermore, the polarization dependence of the TH signal is illustrated in Figure~\ref{fig_THG_exp_results}(e). The polarization state of the pump plays a crucial role in modulating the TH intensity. The maximum TH power is obtained when the pump polarization is along the $y$-axis, while the minimum occurs when the polarization is along the $x$-axis. This behavior is expected: the $y$-polarized pump resonantly excites the multiresonant modes and produces the strong near-field enhancement that drives the nonlinear process, whereas $x$-polarized excitation couples poorly to these modes, yielding a much weaker TH signal. Overall, the measured nonlinear response is consistent with the theoretical predictions.

Table~\ref{tab:table1} compares the TH generation performance of the proposed metasurface with that of previously reported high-$Q$ dielectric metasurfaces. Most previous studies demonstrated TH generation enhancement through a single QBIC resonance, whereas only a few achieved dual-band operation using chiral, phase-change, or polarization-dependent configurations. In contrast, the proposed single-layer metasurface, comprising four a-Si nanoblocks, supports eight distinct high-$Q$ QBIC resonances across the pump-wavelength range of $937$--$1125\,\mathrm{nm}$, thereby providing multiple TH generation channels spanning approximately $312$--$375\,\mathrm{nm}$. The key advantage of the present design lies not only in its strong conversion efficiency but also in its uniquely rich multiband QBIC spectrum, broad near-UV frequency-conversion range, simple, fabrication-friendly architecture, and experimental validation of both the linear and nonlinear optical responses. The availability of multiple high-$Q$ resonances further provides flexibility for multichannel nonlinear frequency conversion, sensing, spectroscopy, and wavelength-selective photonic applications.

\section{Conclusion}
In conclusion, an all-dielectric metasurface supporting multiple high-$Q$ Fano resonances in the near-IR regime has been designed, fabricated, and experimentally characterized. The resonances originate from symmetry-protected QBICs, activated by introducing a controlled in-plane structural asymmetry within the unit cell. Their physical origin is established through multipolar decomposition and field distributions at the resonant wavelengths, and the spectra are quantitatively fitted with a Fano line-shape model to confirm high-$Q$ factors across multiple bands. The dependence of the resonances on the geometrical parameters is also examined, providing design rules for spectral tuning. Full-wave nonlinear simulations show that the strong field localization at these high-$Q$ modes enables efficient TH generation across multiple bands, with a maximum theoretical conversion efficiency of $8.5 \times 10^{-3}$ at the $M_{5}$ resonance. 

The metasurface is then fabricated in both symmetric and asymmetric configurations; the measured linear transmission spectra agree well with simulations, and TH generation is measured at the $M_{8\mathrm{x}}$ resonance, yielding a conversion efficiency of $1.2 \times 10^{-6}$ at a peak pump intensity of $3.25\,\mathrm{GW/cm^{2}}$. Experimentally, TH generation is demonstrated at the longest-wavelength resonance, $M_{8\mathrm{x}}$, using a pump wavelength of $1\,146.4\,\mathrm{nm}$. Although the $M_5$ mode exhibits the highest simulated TH conversion efficiency, quantitative characterization of the remaining TH channels spanning $312$--$375\,\mathrm{nm}$ was not feasible because of limitations in the available excitation and calibrated near-UV detection systems. These experimental constraints do not negate the predicted multiband TH generation enhancement. Future studies employing a broadly tunable ultrafast source and a calibrated near-UV detection system will enable quantitative characterization of the remaining TH channels and further establish the potential of the proposed metasurface for multiband nonlinear photonic applications. Beyond its relevance to Fano-resonant photonic devices, the demonstrated multiband response provides a promising platform for applications including multiwavelength lasers, multiband harmonic generation, and single-photon sources for quantum photonics.

\section*{Acknowledgement}
The authors acknowledge the KAUST Nanofabrication Core Lab for providing the fabrication facilities.

\setcounter{figure}{0}
\renewcommand{\thefigure}{S\arabic{figure}}
\setcounter{section}{0}
\renewcommand{\thesection}{S\arabic{section}}
\setcounter{equation}{0}
\renewcommand{\theequation}{S\arabic{equation}}

\clearpage\newpage

\begin{center}
     {\LARGE Supplementary Material: \\[4pt] Enhanced Third-Harmonic Generation in a Bound State in the Continuum Assisted Multiband All-Dielectric Metasurface \par}
\end{center}

\section{Multipolar decomposition}
The scattering response of the metasurface is analyzed using a Cartesian multipole decomposition, in which the total scattering cross section is written as the sum of the contributions from the electric dipole (ED), toroidal dipole (TD), magnetic dipole (MD), electric quadrupole (EQ), and magnetic quadrupole (MQ) moments~\cite{alaee2018electromagnetic, baryshnikova2019optical, hinamoto2021menp},
\begin{equation*}
\begin{aligned}
C_{\mathrm{sca}} &= C_{\mathrm{sca}}^{\mathrm{ED}} + C_{\mathrm{sca}}^{\mathrm{TD}} + C_{\mathrm{sca}}^{\mathrm{MD}} + C_{\mathrm{sca}}^{\mathrm{EQ}} + C_{\mathrm{sca}}^{\mathrm{MQ}} + \cdots\\
&= \frac{k^4}{6\pi\varepsilon_{0}^2 |\mathbf{E}_{\mathrm{inc}}|^2} \Biggl[ \sum_{\alpha} \Bigl(|p_{\alpha} + ik\,T_{\alpha}|^2 + \frac{|m_{\alpha}|^2}{c}\Bigr) + \frac{1}{120} \sum_{\alpha\beta} \Bigl(|kQ_{\alpha\beta}^{e}|^2 + \Bigl|\frac{kQ_{\alpha\beta}^{m}}{c}\Bigr|^2\Bigr) + \cdots\Biggr],
\end{aligned}
\end{equation*}
where $p_{\alpha}$ and $T_{\alpha}$ are the electric and toroidal dipole moments, $m_{\alpha}$ is the magnetic dipole moment, and $Q_{\alpha\beta}^{e}$ and $Q_{\alpha\beta}^{m}$ are the electric and magnetic quadrupole moments, respectively. Here, $\alpha,\beta \in \{x,y,z\}$ are Cartesian indices, $\mathbf{J}$ is the induced current density with component $J_{\alpha}$, $\mathbf{r}$ is the position vector with component $r_{\alpha}$ and magnitude $r$, $\omega$ is the angular frequency, $\delta_{\alpha\beta}$ is the Kronecker delta, $|\mathbf{E}_{\mathrm{inc}}|$ is the amplitude of the incident plane-wave field, $k$ is the wavenumber, $c$ is the speed of light, and $\varepsilon_0$ is the vacuum permittivity. The corresponding multipole moments are given by~\cite{alaee2018electromagnetic}
\begin{equation*}
p_{\alpha} \approx -\frac{1}{i\omega} \int J_{\alpha}\,\mathrm{d}^3\mathbf{r},
\end{equation*}
\begin{equation*}
T_{\alpha} \approx \frac{1}{10c} \int \Bigl\{(\mathbf{r}\cdot\mathbf{J}) r_{\alpha} - 2r^{2} J_{\alpha}\Bigr\}\,\mathrm{d}^3\mathbf{r},
\end{equation*}
\begin{equation*}
m_{\alpha} \approx \frac{1}{2}\int (\mathbf{r}\times \mathbf{J})_{\alpha}\,\mathrm{d}^3\mathbf{r},
\end{equation*}
\begin{equation*}
\begin{aligned}
Q_{\alpha\beta}^{e} &\approx -\frac{1}{i\omega} \Biggl[\int \Bigl\{3(r_{\beta}J_{\alpha} + r_{\alpha}J_{\beta}) - 2(\mathbf{r} \cdot \mathbf{J})\delta_{\alpha\beta}\Bigr\}\,\mathrm{d}^3\mathbf{r} \\
&\quad +\frac{k^2}{4} \int \Bigl\{4r_{\alpha}r_{\beta}(\mathbf{r}\cdot\mathbf{J})-5r^{2}(r_{\alpha} J_{\beta} + r_{\beta}J_{\alpha}) + 2r^{2}(\mathbf{r} \cdot \mathbf{J}) \delta_{\alpha\beta}\Bigr\}\,\mathrm{d}^3\mathbf{r}\Biggr],
\end{aligned}
\end{equation*}
\begin{equation*}
Q_{\alpha\beta}^{m} \approx \int \Bigl\{r_{\alpha}(\mathbf{r}\times \mathbf{J})_{\beta} + r_{\beta} (\mathbf{r} \times \mathbf{J})_{\alpha}\Bigr\}\,\mathrm{d}^3\mathbf{r}.
\end{equation*}
Here, $p_{\alpha}$ denotes the leading electric dipole term, and the toroidal dipole moment $T_{\alpha}$, retained beyond the long-wavelength approximation, is treated as a separate scattering channel; the two contribute jointly through the combination $p_{\alpha} + ik\,T_{\alpha}$ in the cross section.

\section{Fano fitting: Simulated transmission spectra}
\begin{figure}[!ht]
    \centering
    \subfloat[]{\includegraphics[width=0.325\columnwidth]{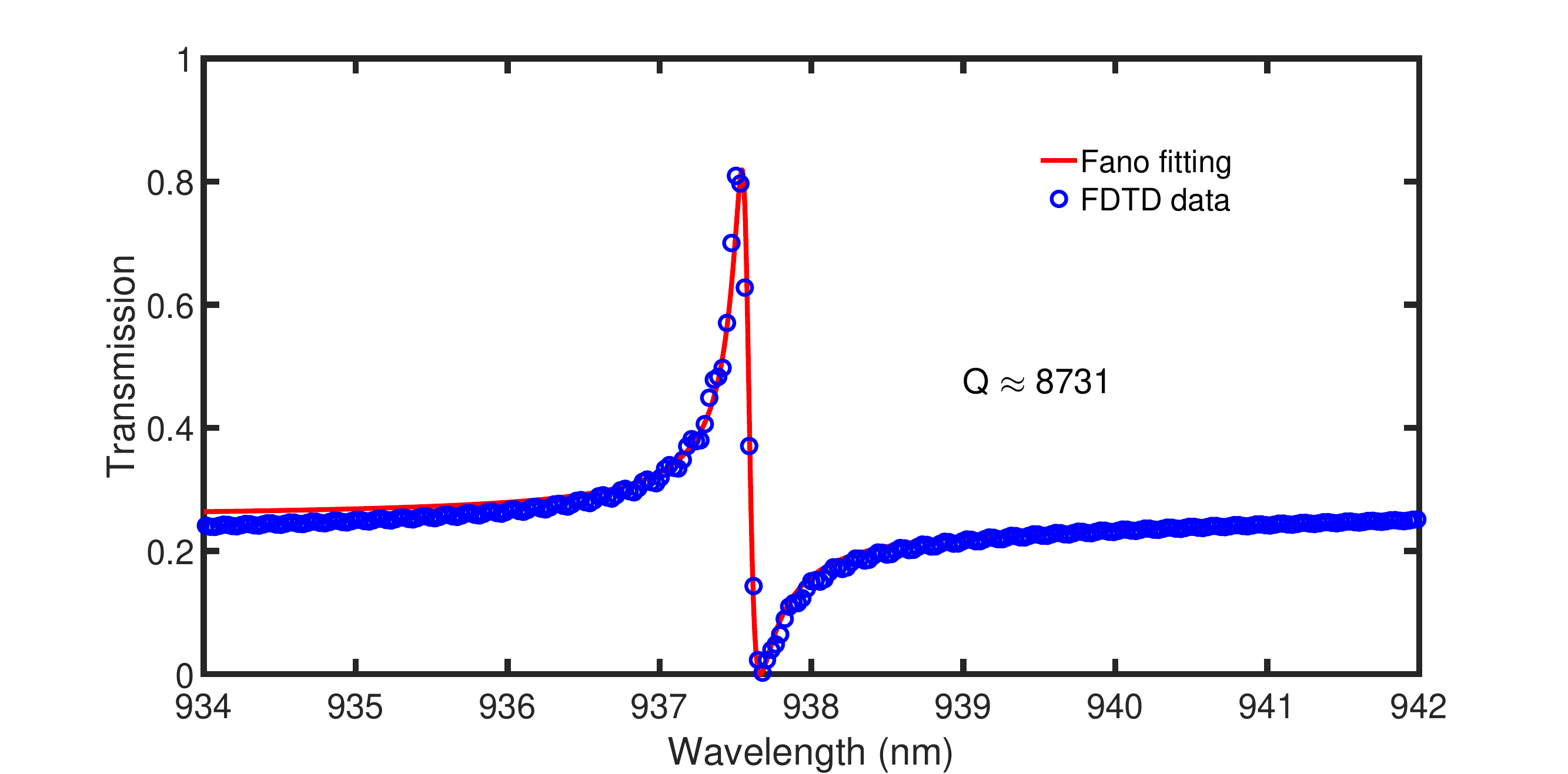}}
    \subfloat[]{\includegraphics[width=0.325\columnwidth]{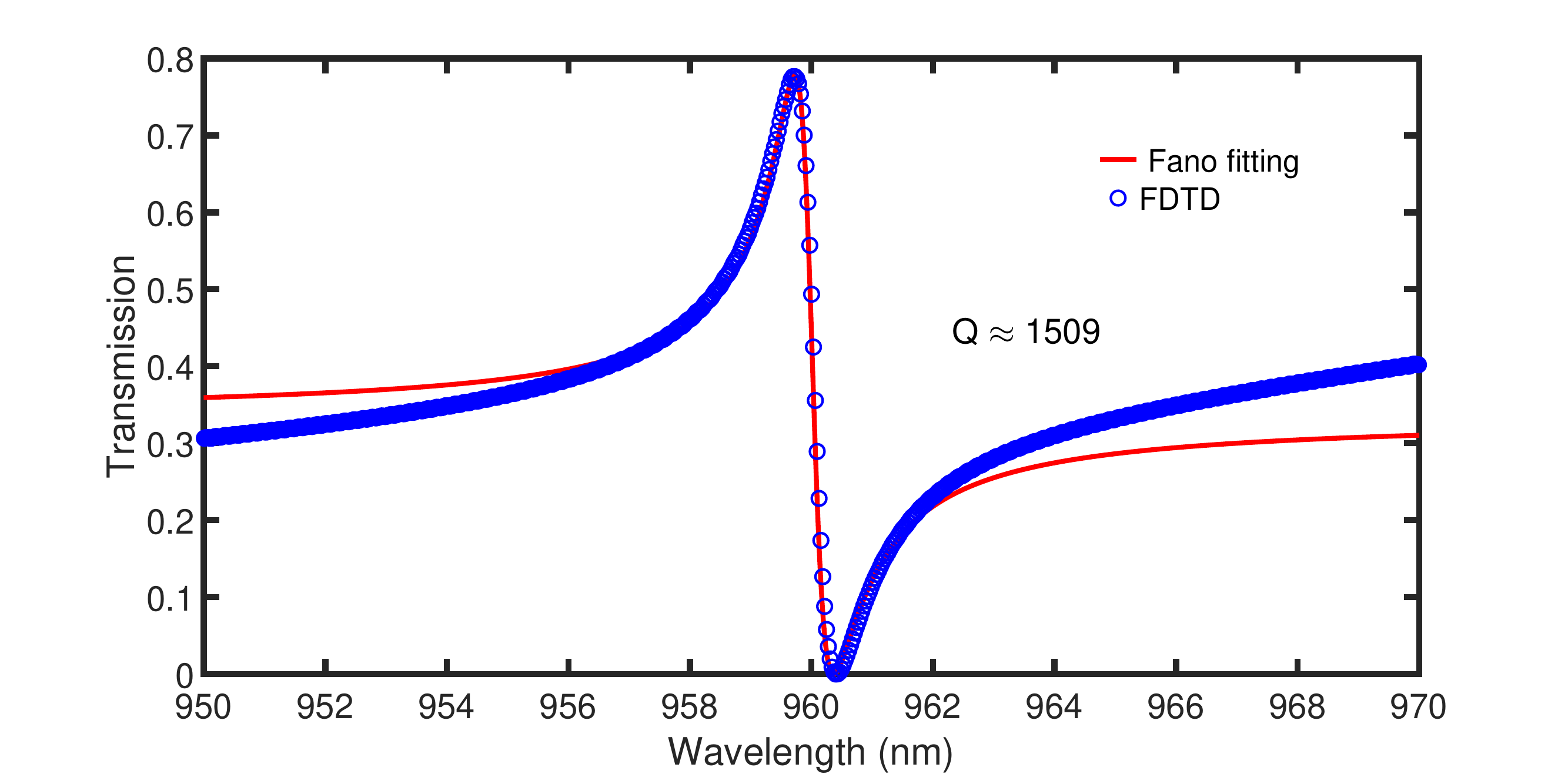}}
    \subfloat[]{\includegraphics[width=0.325\columnwidth]{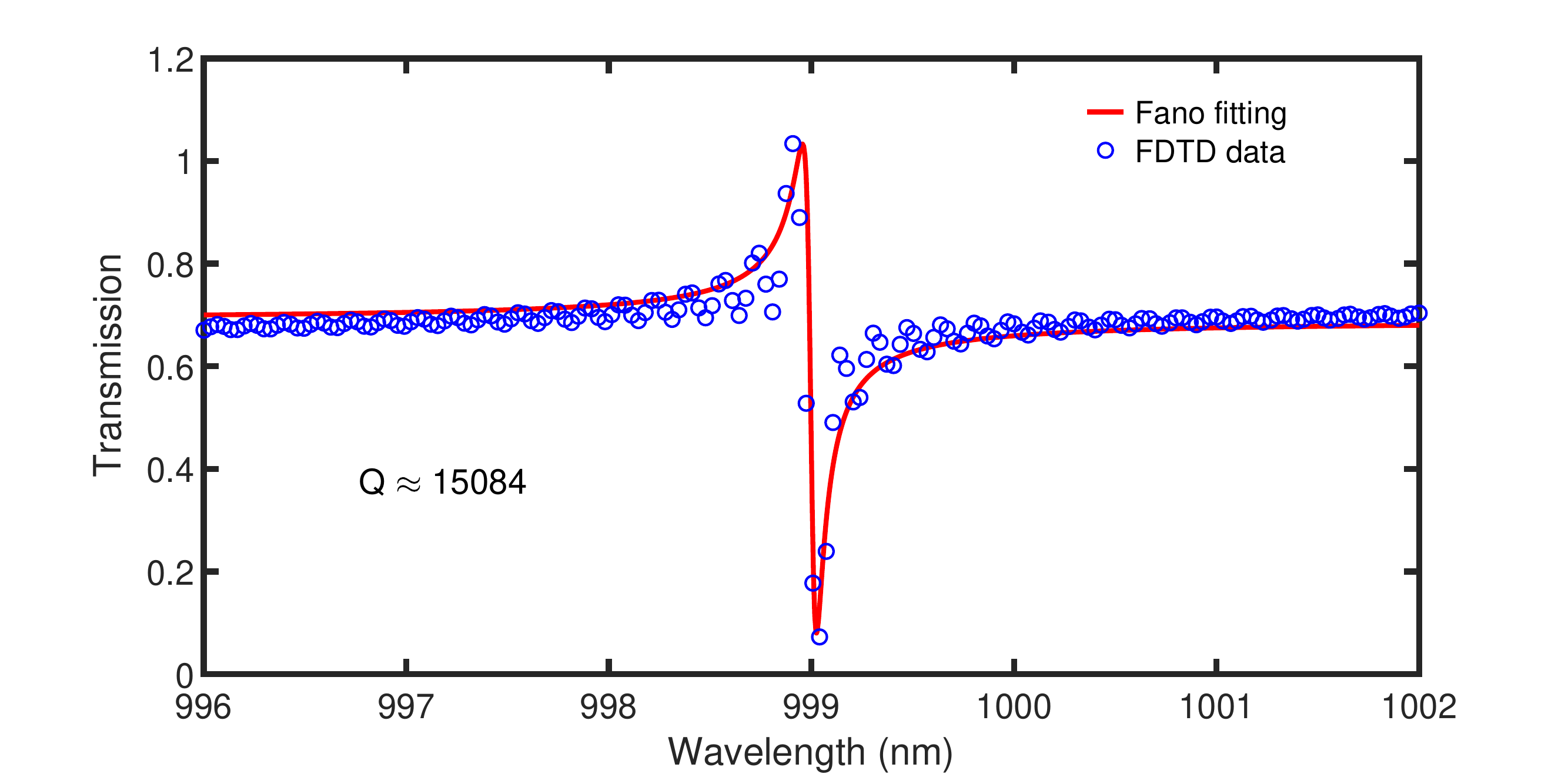}}\\
    \subfloat[]{\includegraphics[width=0.325\columnwidth]{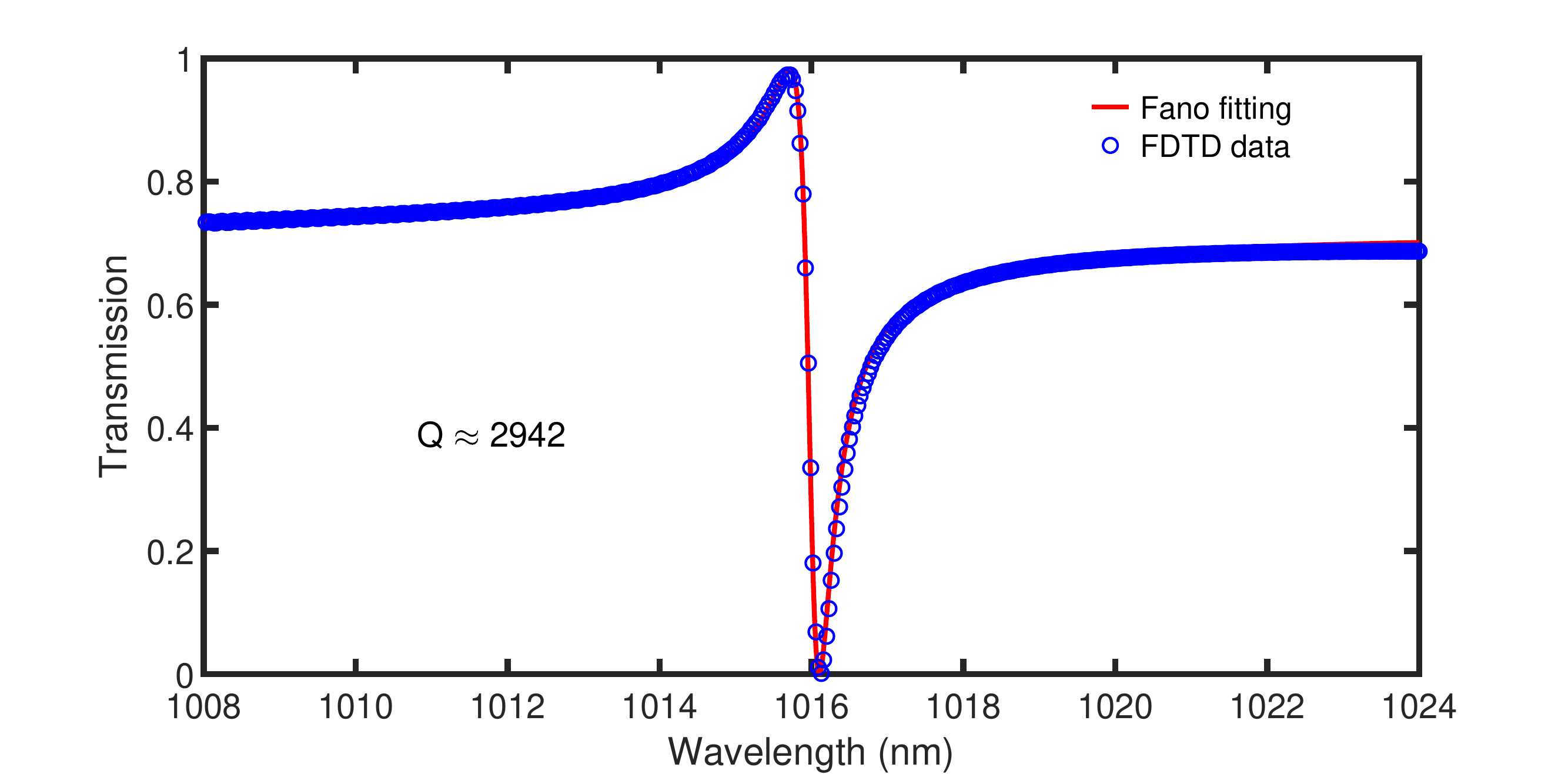}}
    \subfloat[]{\includegraphics[width=0.325\columnwidth]{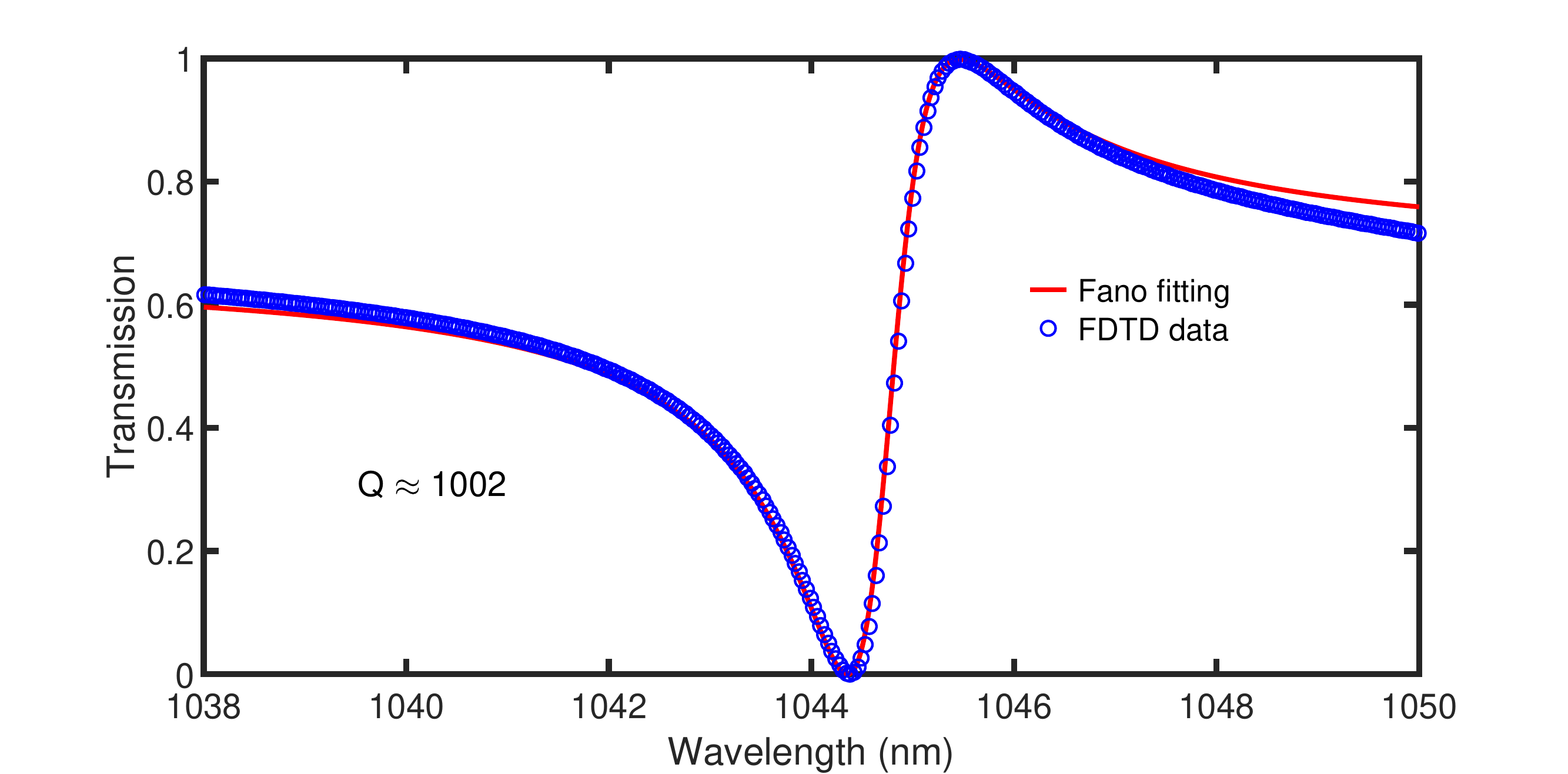}}
    \subfloat[]{\includegraphics[width=0.325\columnwidth]{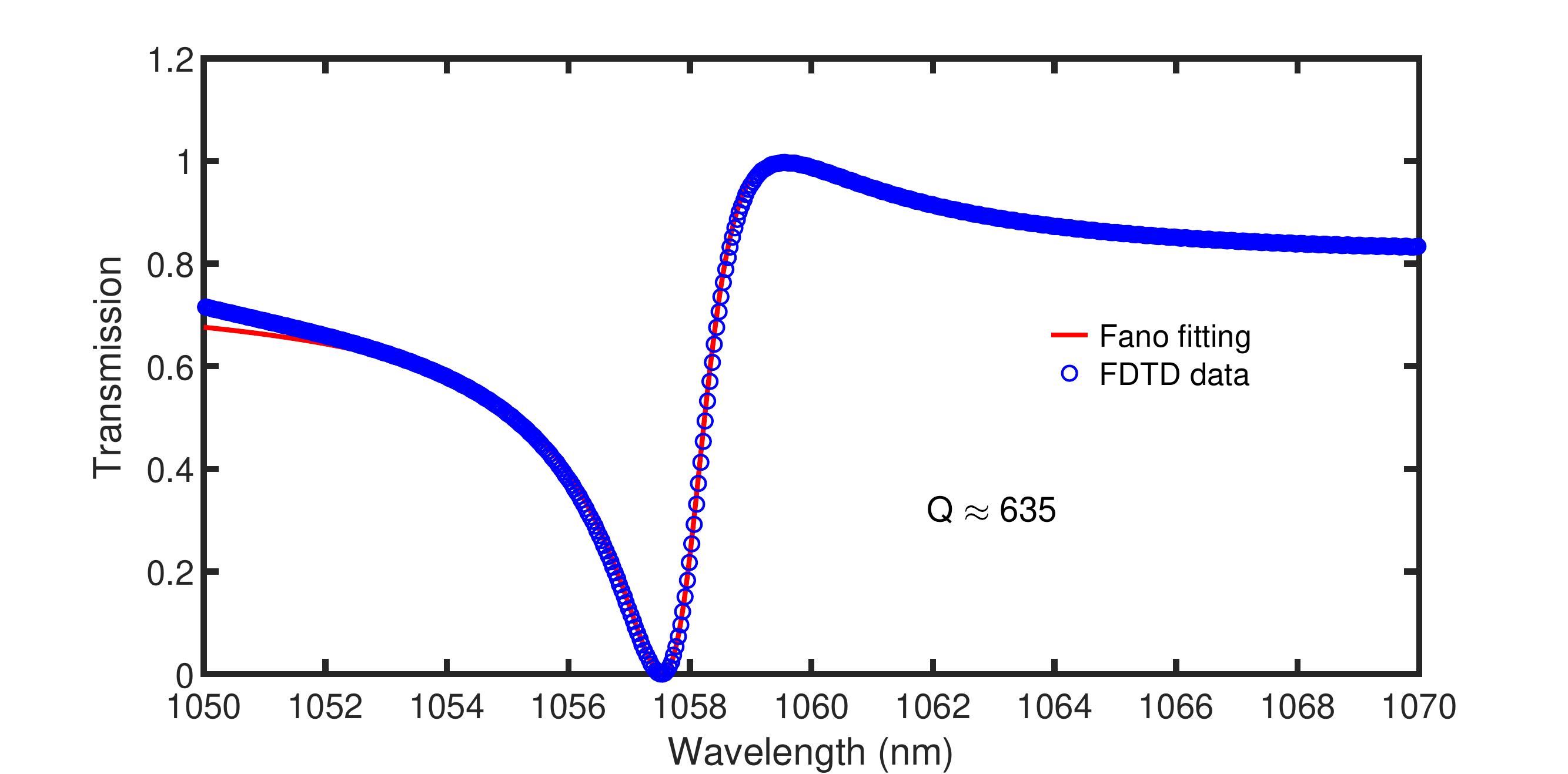}}\\
    \subfloat[]{\includegraphics[width=0.325\columnwidth]{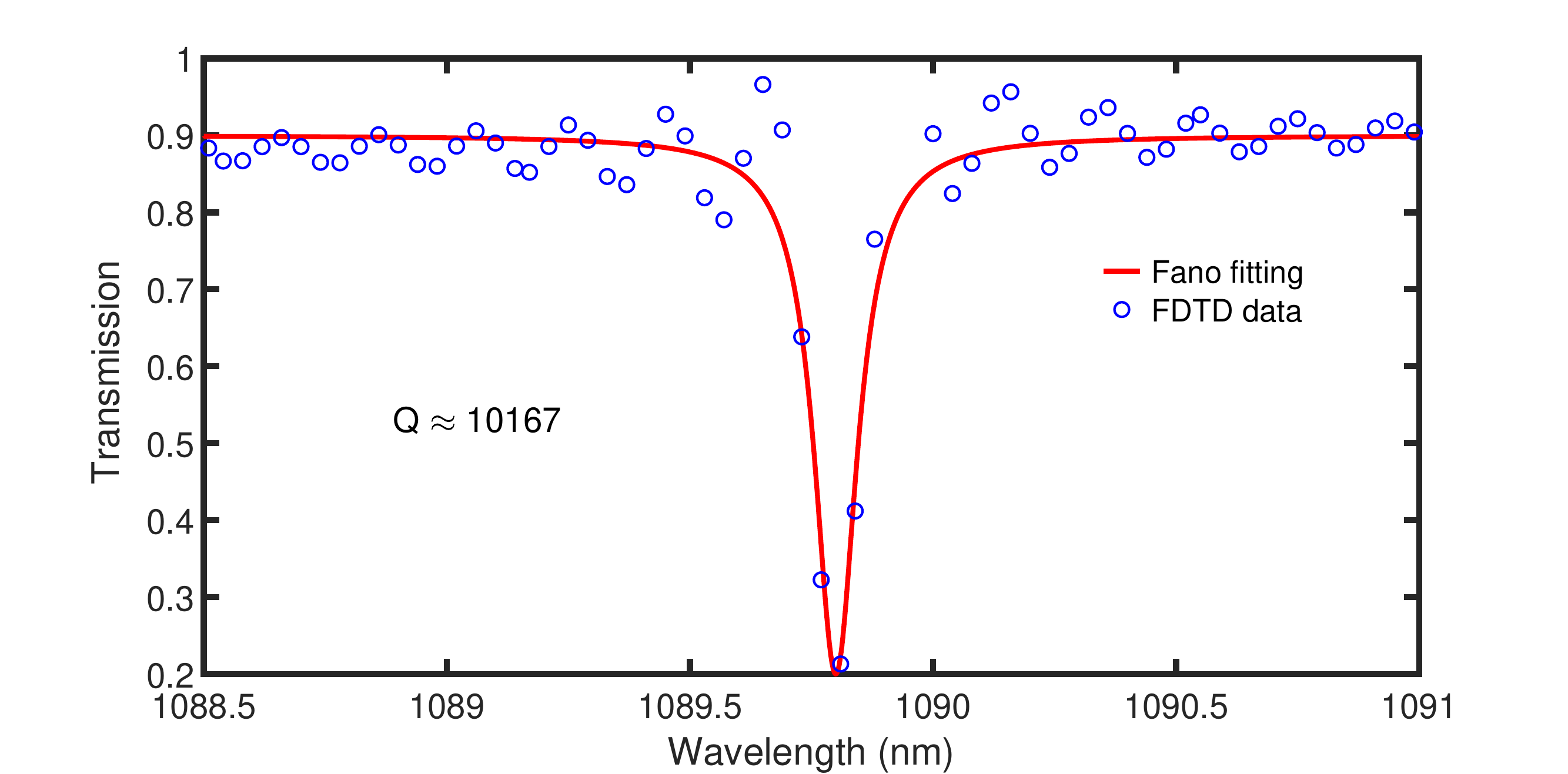}}
    \subfloat[]{\includegraphics[width=0.325\columnwidth]{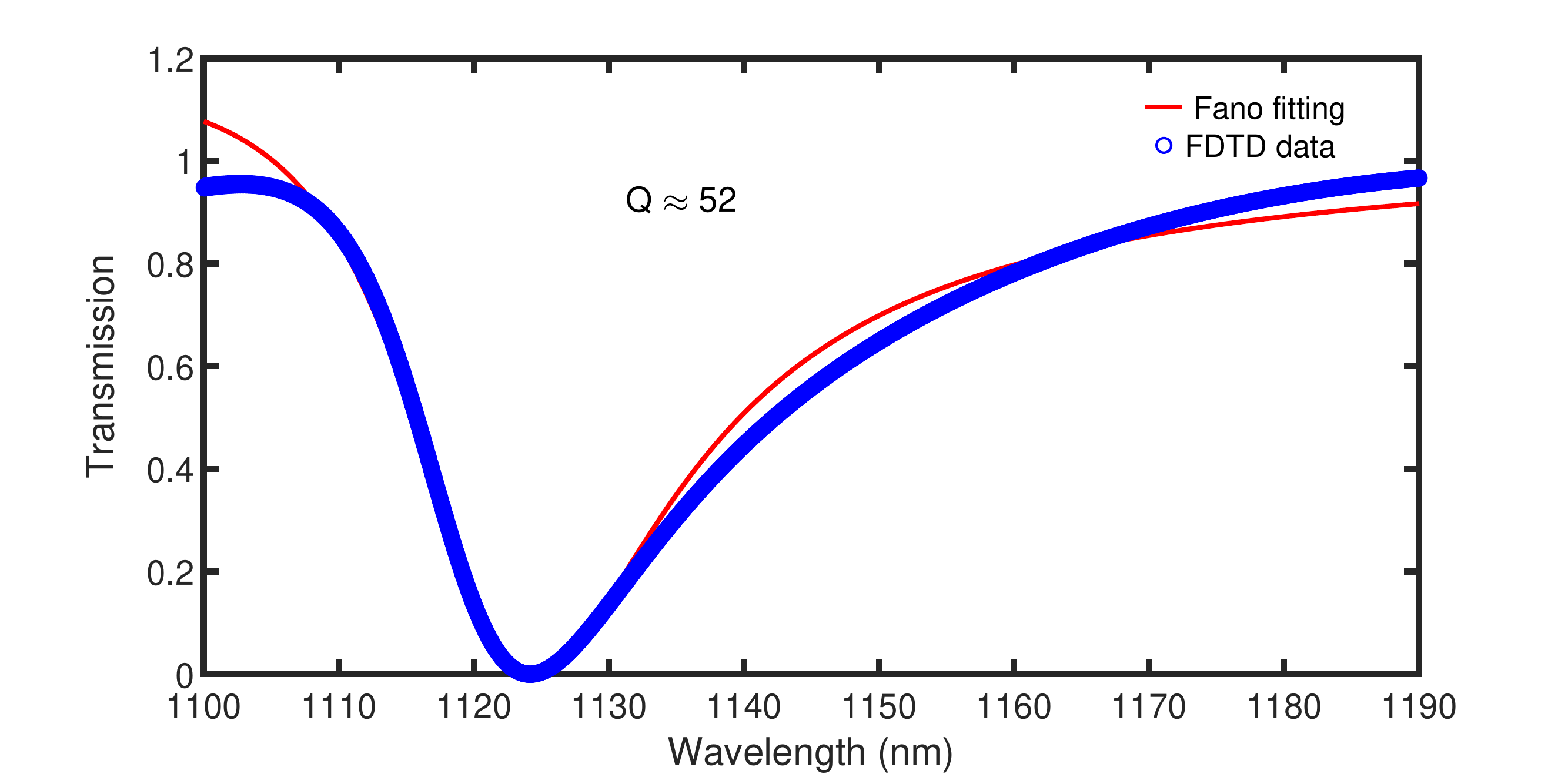}}
    \caption{Fano fitting of the simulated transmission spectra at $\delta = 0.14$ for the resonances (a) $M_{1}\; (Q=8\,731)$, (b) $M_{2}\; (Q=1\,509)$, (c) $M_{3}\; (Q=15\,084)$, (d) $M_{4}\; (Q=2\,942)$, (e) $M_{5}\; (Q=1\,002)$, (f) $M_{6}\; (Q=635)$, (g) $M_{7}\; (Q=10\,167)$, and (h) $M_{8}\; (Q=52)$.}
    \label{fig_fano_fitting}
\end{figure}

The simulated transmission spectra at each resonance are fitted with the Fano line-shape model to extract the corresponding $Q$ factors. Figure~\ref{fig_fano_fitting} shows the fitted spectra and provides the extracted $Q$ factors for the eight resonances $M_1$--$M_8$ at $\delta = 0.14$

\newpage\clearpage
\section{Field distribution}
\begin{figure}[ht!]
    \centering
    \subfloat[]{\includegraphics[width=0.33\columnwidth]{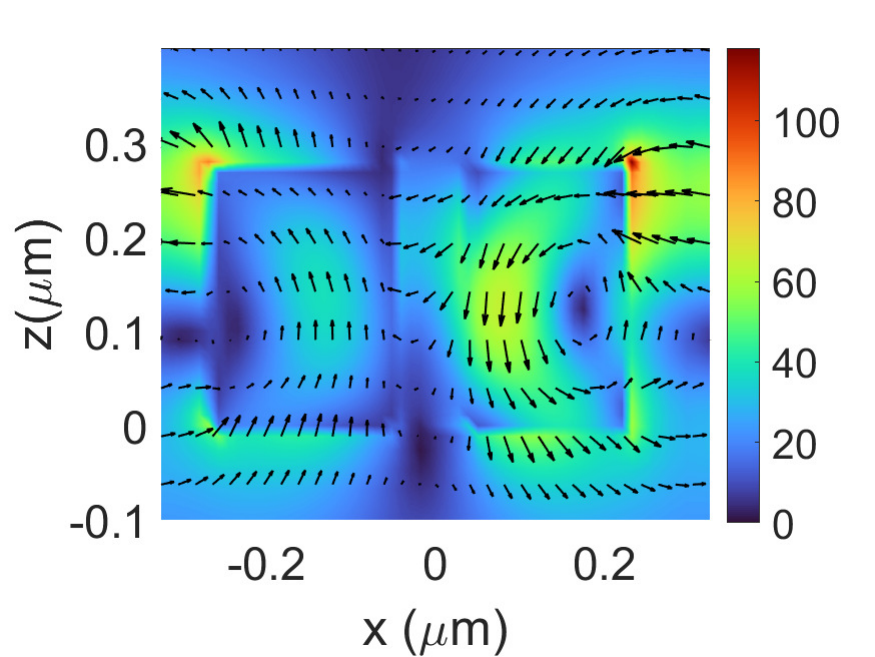}}
    \subfloat[]{\includegraphics[width=0.33\columnwidth]{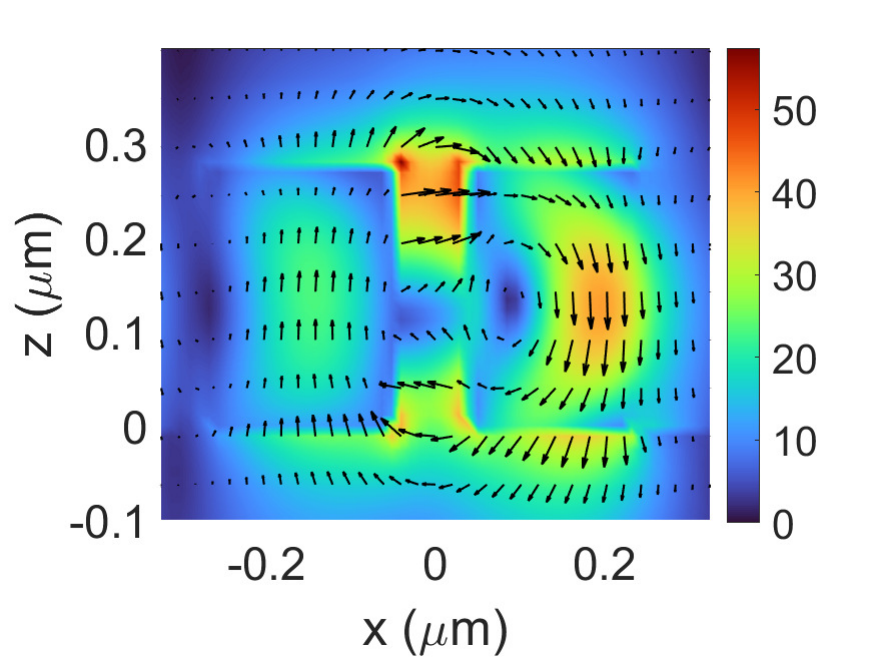}} 
    \subfloat[]{\includegraphics[width=0.33\columnwidth]{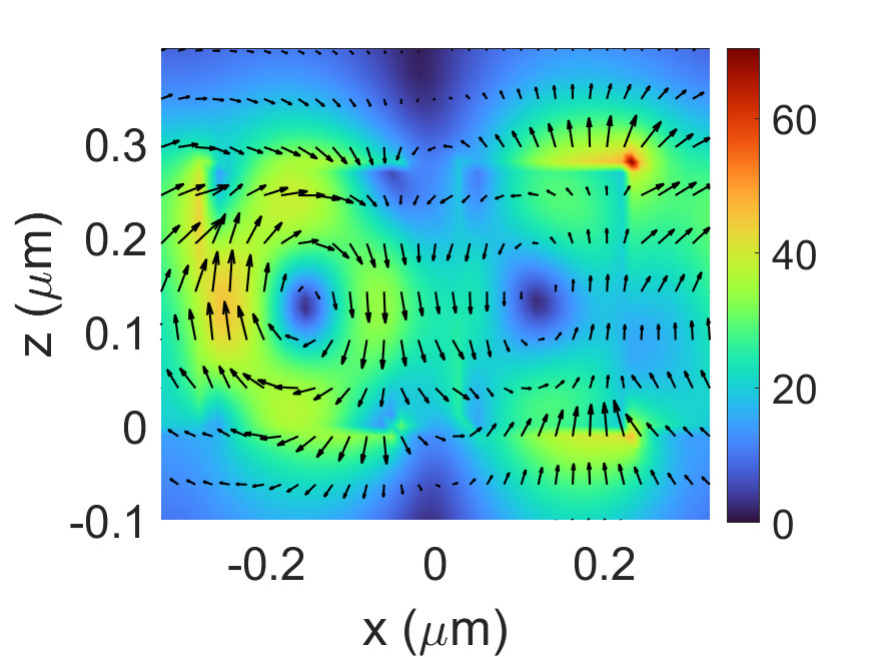}}\\
    \subfloat[]{\includegraphics[width=0.33\columnwidth]{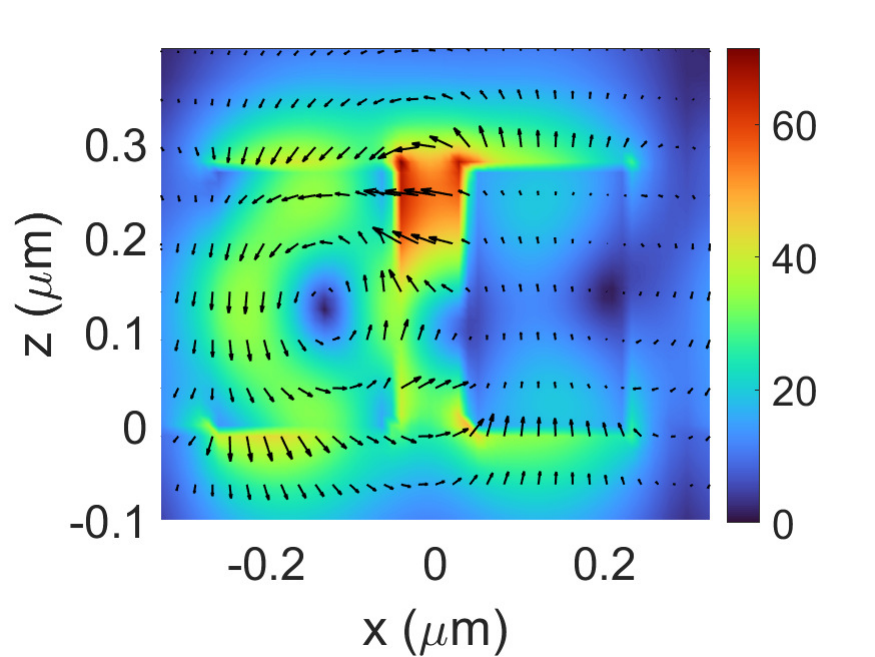}}
    \subfloat[]{\includegraphics[width=0.33\columnwidth]{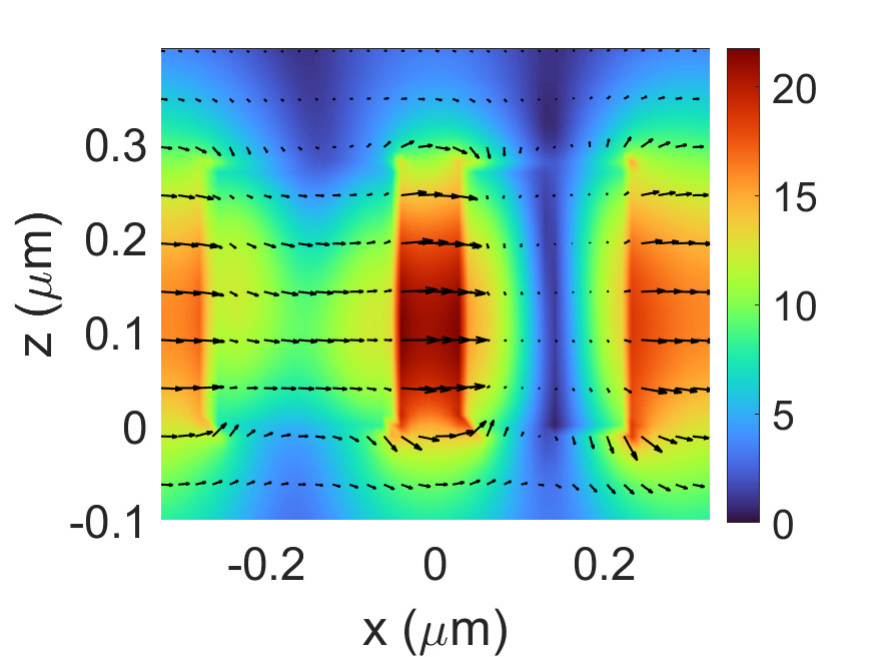}}
    \subfloat[]{\includegraphics[width=0.33\columnwidth]{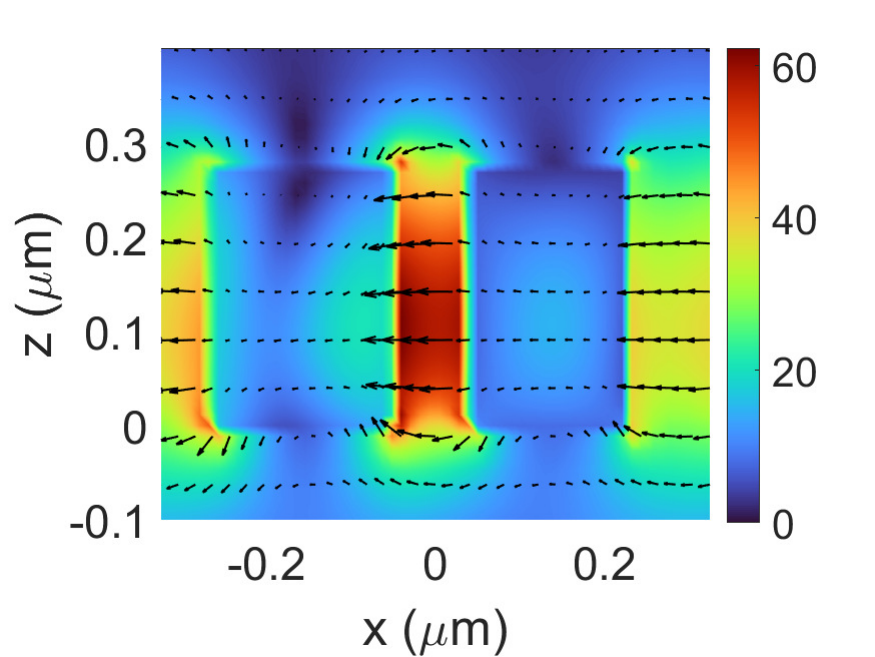}}\\
    \subfloat[]{\includegraphics[width=0.33\columnwidth]{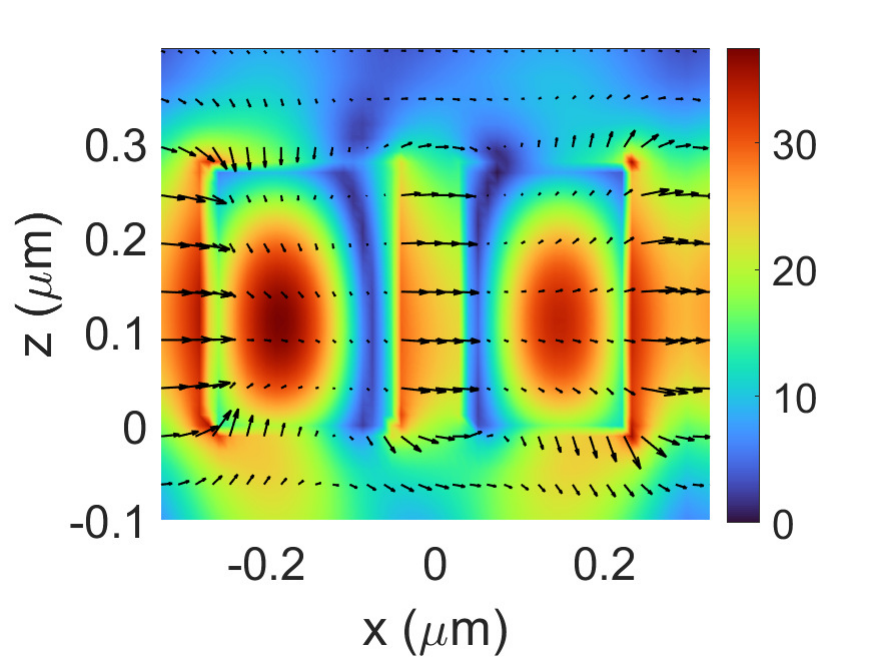}}
    \subfloat[]{\includegraphics[width=0.33\columnwidth]{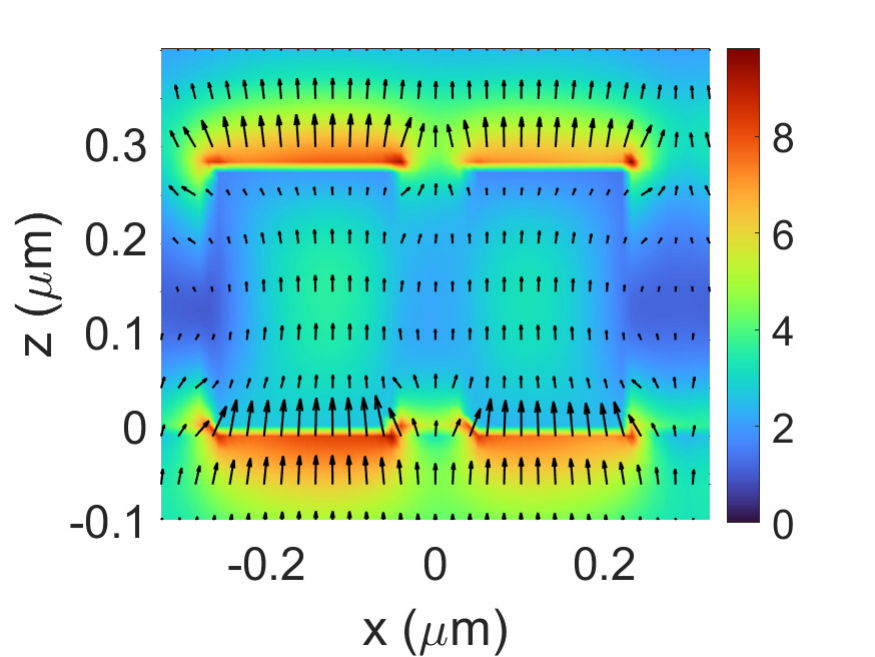}}
    \caption{Simulated electric field distributions in the $xz$-plane at $\delta = 0.14$ for the resonances (a) $M_{1}$, (b) $M_{2}$, (c) $M_{3}$, (d) $M_{4}$, (e) $M_{5}$, (f) $M_{6}$, (g) $M_{7}$, and (h) $M_{8}$.}
    \label{fig_Efield_xz}
\end{figure}

\begin{figure}[ht!]
    \centering
    \subfloat[]{\includegraphics[width=0.33\columnwidth]{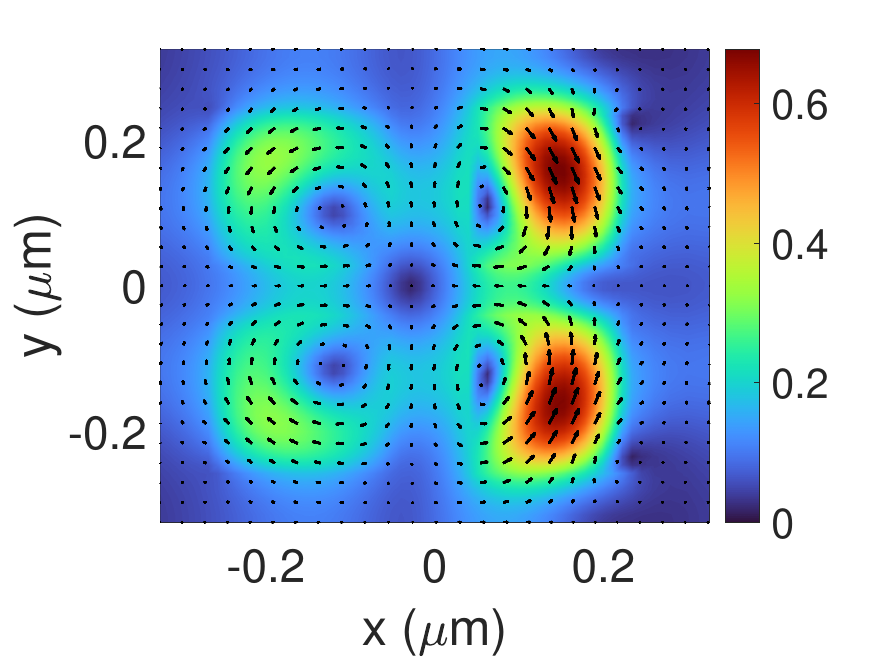}}
    \subfloat[]{\includegraphics[width=0.33\columnwidth]{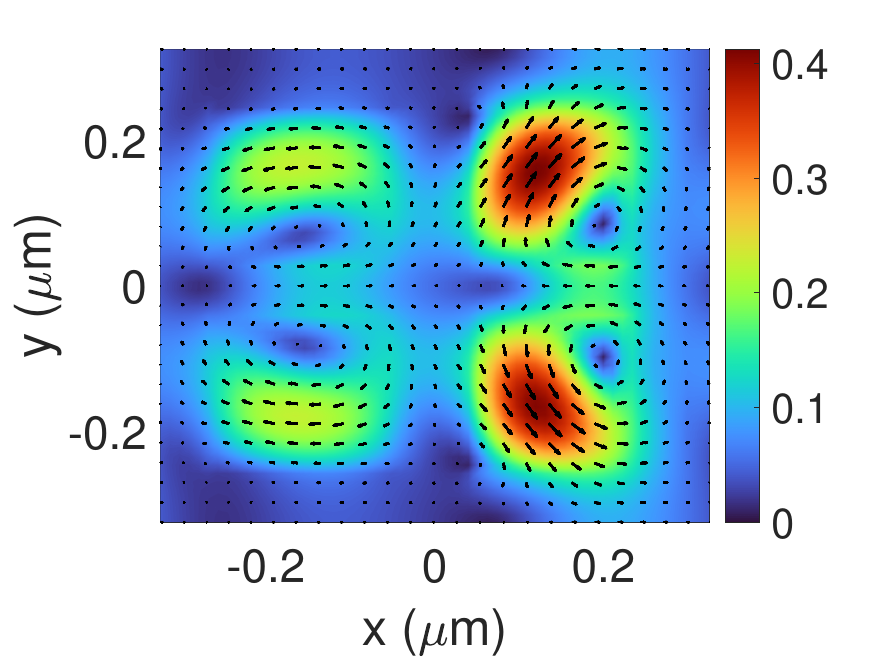}} 
    \subfloat[]{\includegraphics[width=0.33\columnwidth]{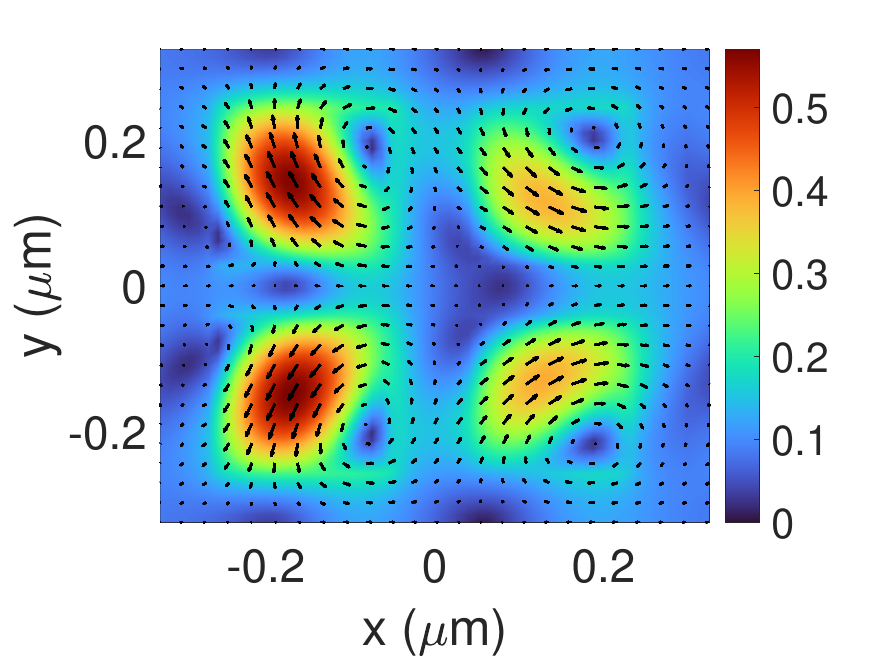}}\\
    \subfloat[]{\includegraphics[width=0.33\columnwidth]{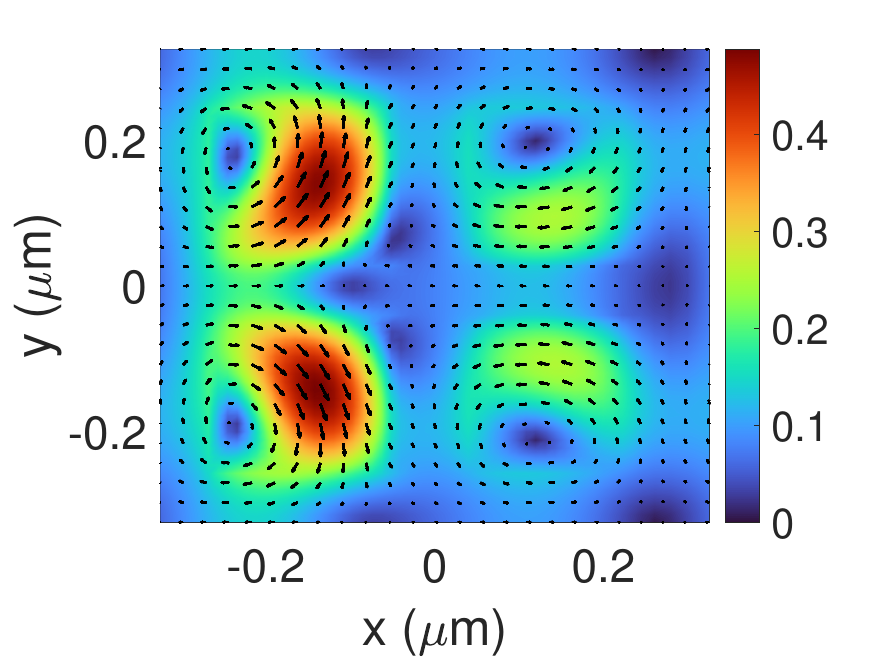}}
    \subfloat[]{\includegraphics[width=0.33\columnwidth]{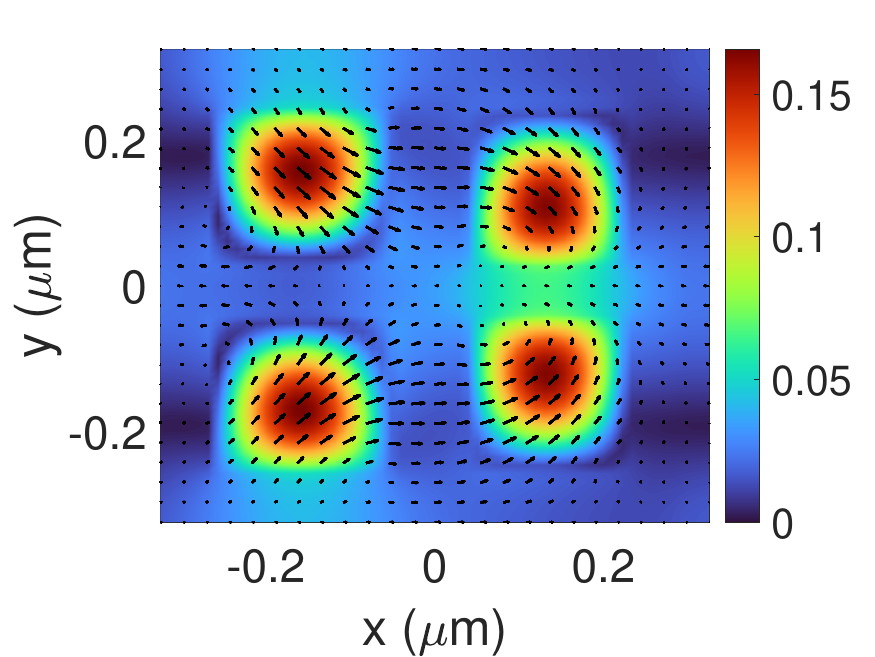}}
    \subfloat[]{\includegraphics[width=0.33\columnwidth]{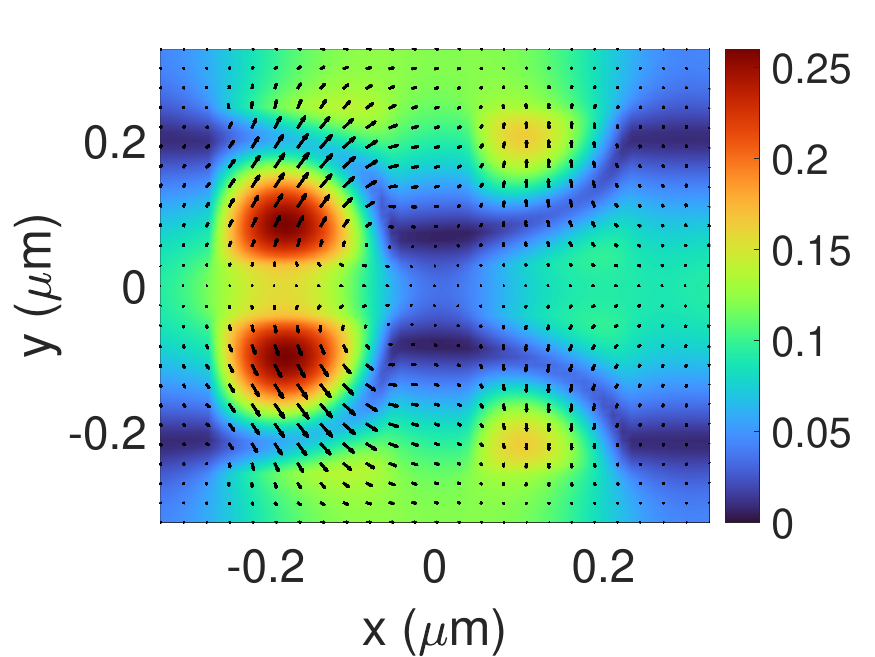}}\\
    \subfloat[]{\includegraphics[width=0.33\columnwidth]{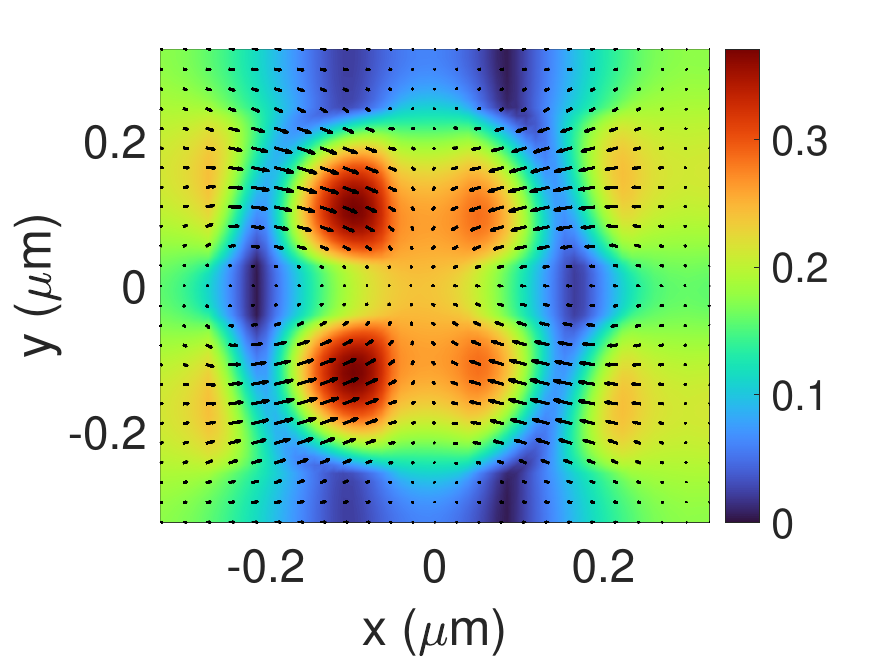}}
    \subfloat[]{\includegraphics[width=0.33\columnwidth]{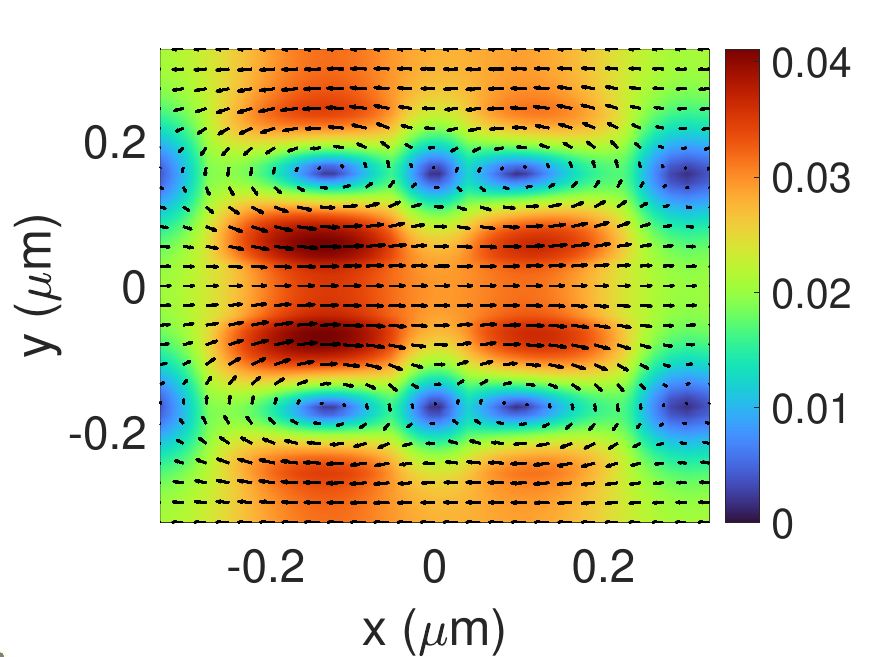}}
    \caption{Simulated magnetic field distributions in the $xy$-plane at $\delta = 0.14$ for the resonances (a) $M_{1}$, (b) $M_{2}$, (c) $M_{3}$, (d) $M_{4}$, (e) $M_{5}$, (f) $M_{6}$, (g) $M_{7}$, and (h) $M_{8}$.}
    \label{fig_Hfield_xy}
\end{figure}

The electric and magnetic field distributions at each resonance are examined to identify the modes underlying the observed resonances. Figure~\ref{fig_Efield_xz} shows the simulated electric field distributions in the $xz$-plane, and Figure~\ref{fig_Hfield_xy} shows the magnetic field distributions in the $xy$-plane, all at $\delta = 0.14$ for the eight resonances $M_1$--$M_8$.

\newpage\clearpage
\section{TH conversion efficiency}
\begin{figure}[!ht]
    \centering
    \subfloat[]{\includegraphics[width=0.33\columnwidth]{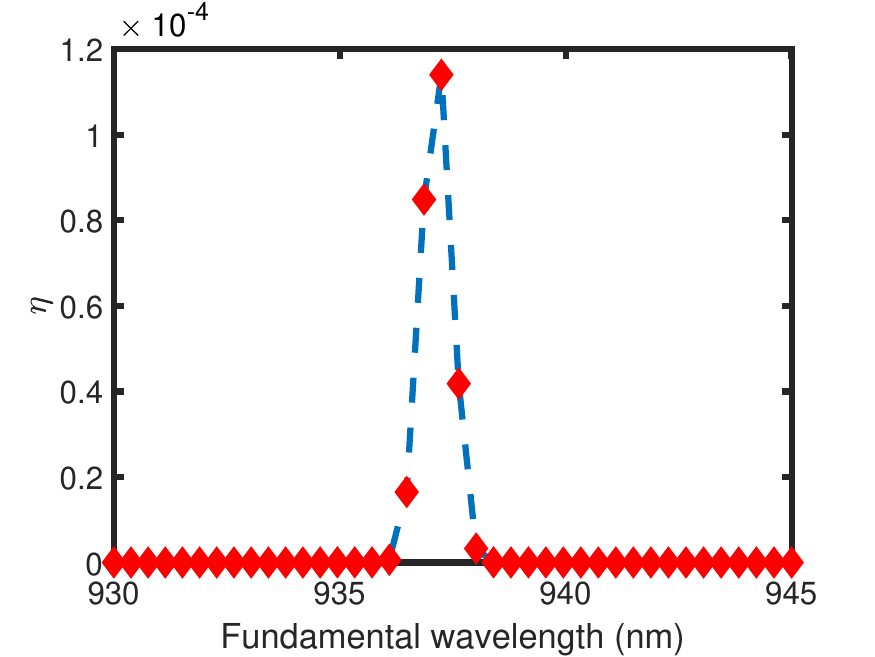}}
    \subfloat[]{\includegraphics[width=0.34\columnwidth]{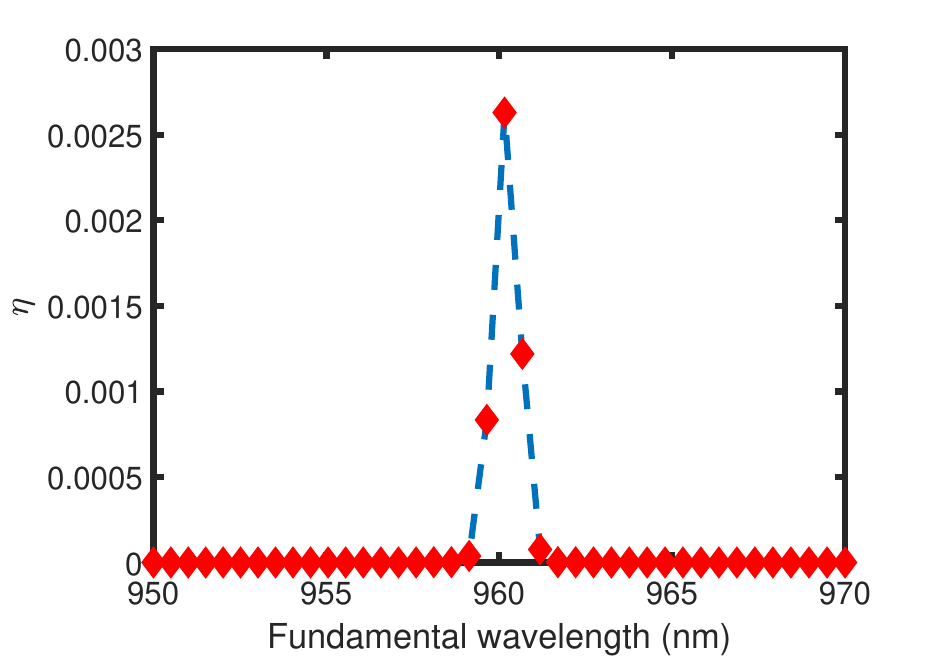}} 
    \subfloat[]{\includegraphics[width=0.33\columnwidth]{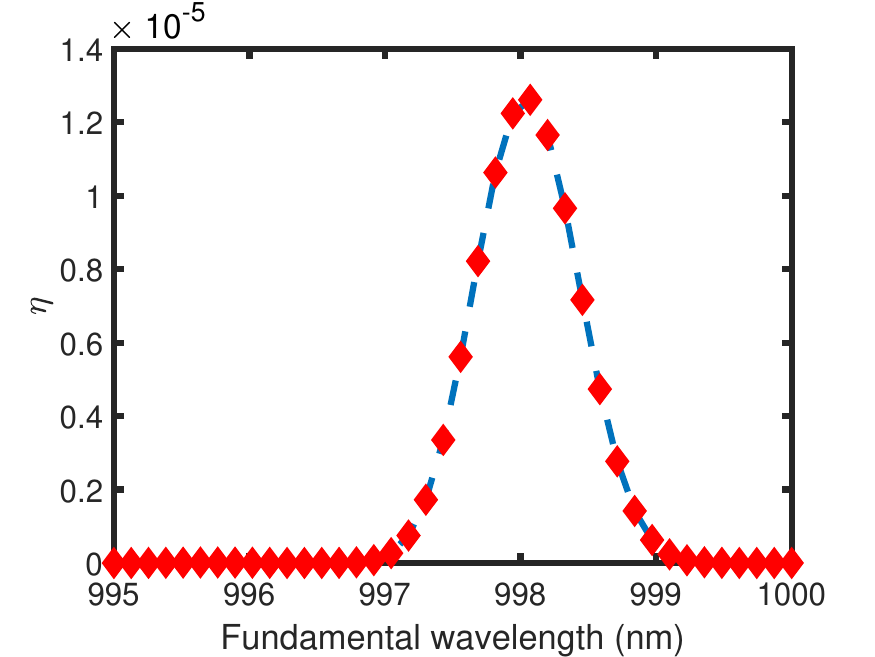}}\vspace{-0.2cm}\\
    \subfloat[]{\includegraphics[width=0.33\columnwidth]{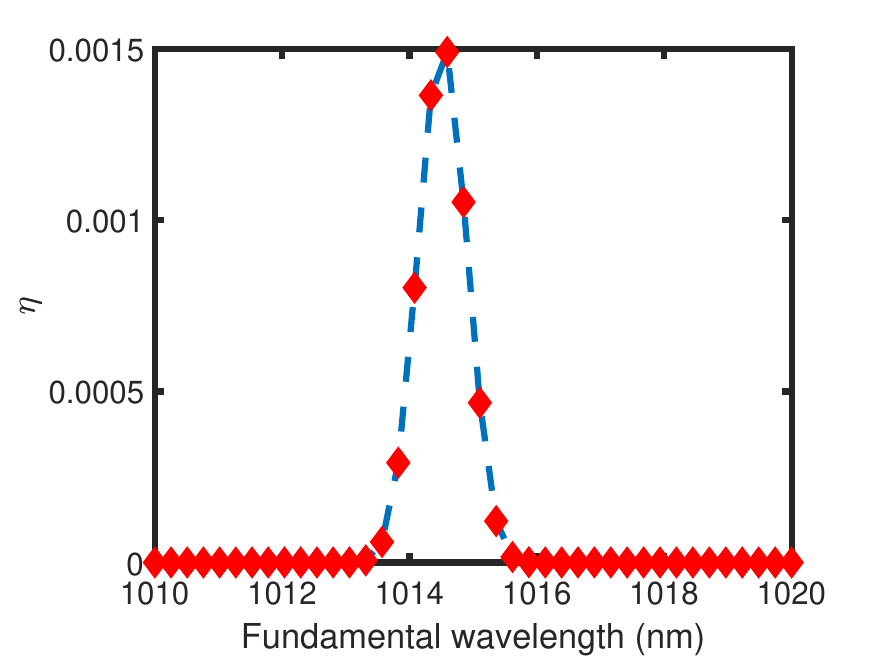}}
    \subfloat[]{\includegraphics[width=0.33\columnwidth]{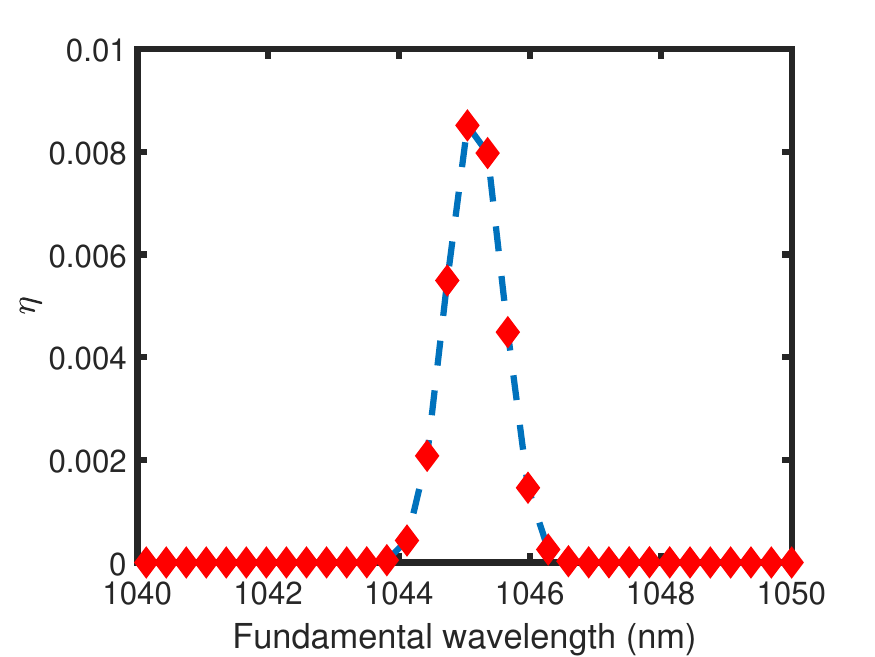}}
    \subfloat[]{\includegraphics[width=0.33\columnwidth]{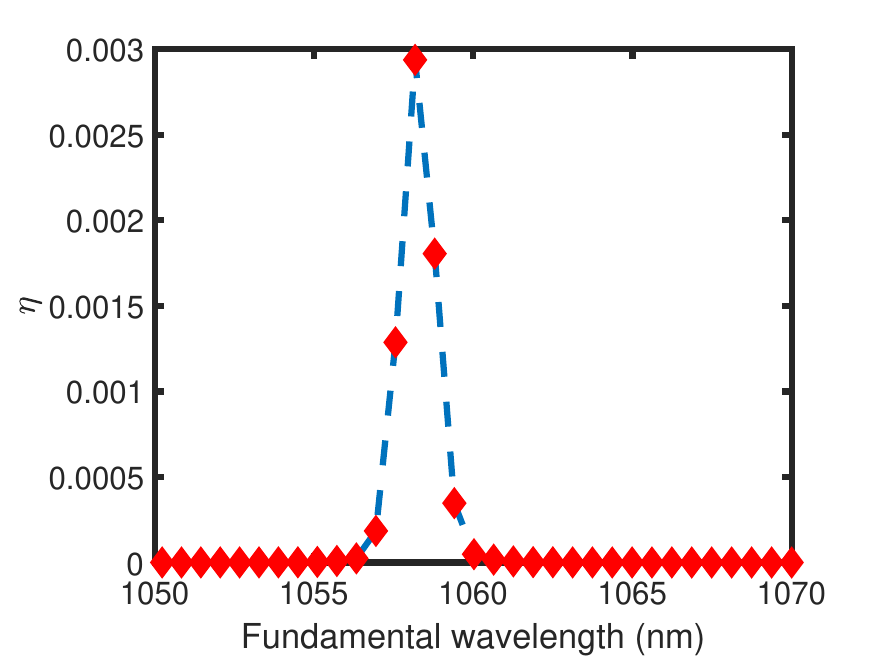}}\vspace{-0.2cm}\\
    \subfloat[]{\includegraphics[width=0.33\columnwidth]{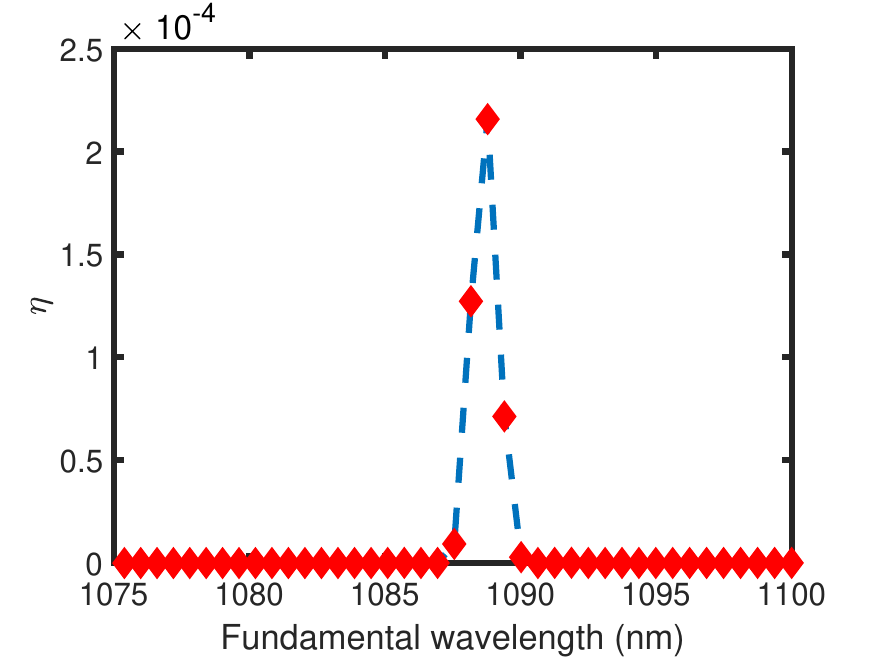}}
    \subfloat[]{\includegraphics[width=0.33\columnwidth]{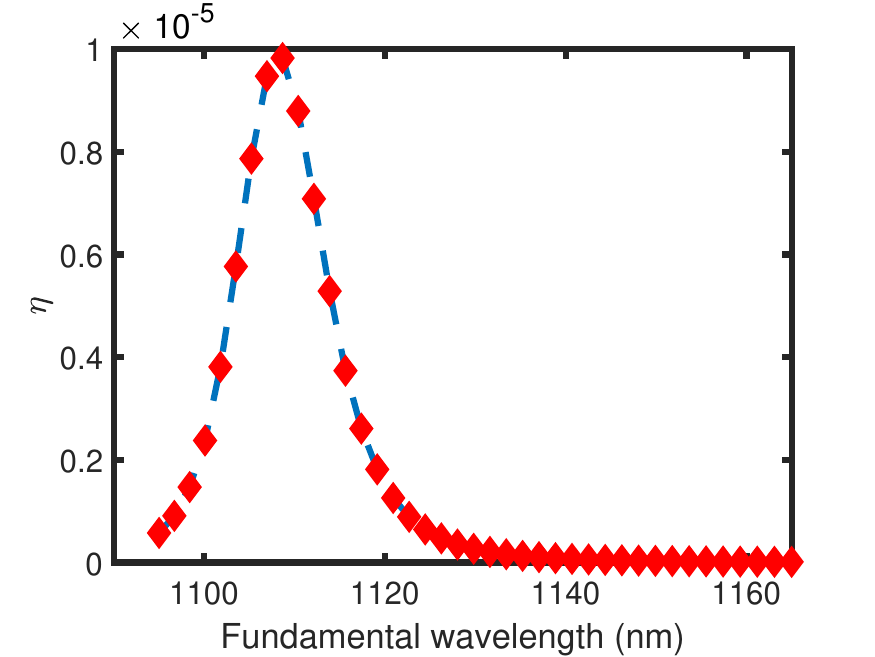}}
    \caption{Simulated conversion efficiency as a function of pump wavelength across the multiband resonances, computed at $\delta = 0.14$, peak pump intensity $1.6\,\mathrm{GW/cm^{2}}$, and effective pump-beam area $314\,\mathrm{\mu m^{2}}$. The plotted conversion efficiencies correspond to excitation at the resonances (a) $M_{1}$, (b) $M_{2}$, (c) $M_{3}$, (d) $M_{4}$, (e) $M_{5}$, (f) $M_{6}$, (g) $M_{7}$, and (h) $M_{8}$.}
    \label{fig_M_eta_variation}
\end{figure}

Nonlinear simulations are performed to evaluate the TH response at each resonance. Figure~\ref{fig_M_eta_variation} shows the simulated conversion efficiencies of the TH spectra for the eight resonances $M_1$--$M_8$ at $\delta = 0.14$.

\newpage\clearpage
\section{Nonlinear fitting}
\begin{figure}[ht!]
    \centering
   {\includegraphics[width=0.6\columnwidth]{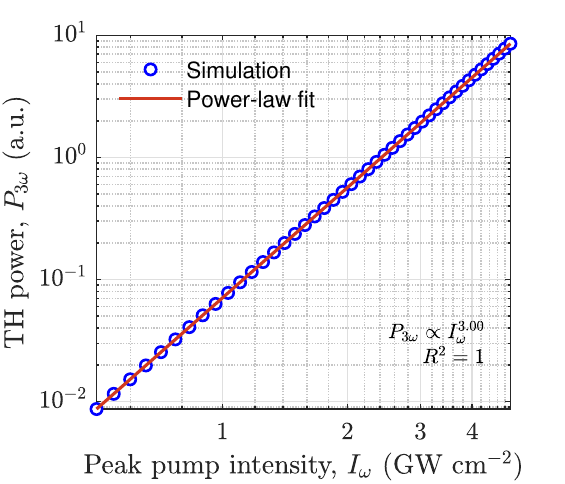}}
      \caption{Simulated TH power $P_{3\omega}$ as a function of the peak pump intensity $I_{\omega}$ at the $M_{8}$ resonance, plotted on a log--log scale. The solid line represents the power-law fit $P_{3\omega}=a\,(I_{\omega})^{b}$, yielding $a=0.07082$, $b=3.00$, and $R^{2}=1$. The cubic dependence confirms the third-order nonlinear origin of the generated TH signal.
}
    \label{fig_slope}
\end{figure}

The third-order nature of TH generation is confirmed by the power dependence of the simulated TH power. Figure~\ref{fig_slope} shows a log--log plot of the simulated TH power versus average input pump power for the $M_8$ resonance.

\newpage\clearpage
\section{Device fabrication}
\begin{figure}[!ht]
    \centering
   {\includegraphics[width=0.8\columnwidth]{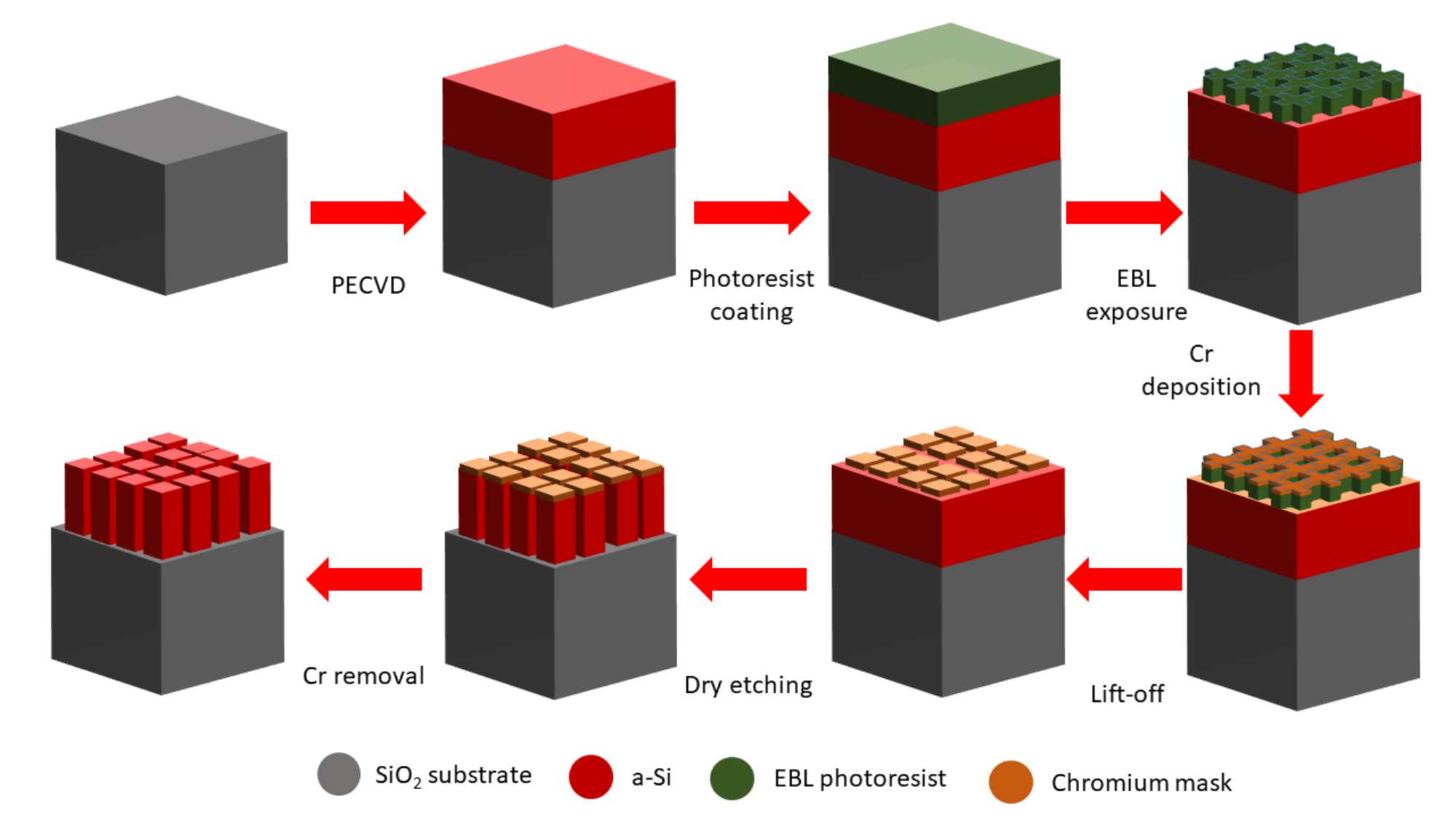}}
    \caption{Fabrication steps of the all-dielectric metasurface. An amorphous silicon (a-Si) layer is deposited onto a glass substrate using plasma-enhanced chemical vapor deposition (PECVD). The designed metasurface patterns are then defined in a thin layer of positive photoresist (AR-P 6200) by EBL. A $20\,\mathrm{nm}$-thick chromium (Cr) film is deposited by electron-beam evaporation to serve as a hard etching mask, and lift-off leaves the mask defining the pattern. Finally, reactive ion etching (RIE) transfers the pattern into the a-Si layer, and the chromium mask is subsequently removed.}
    \label{fig_fabrication}
\end{figure}

The metasurface is fabricated using standard thin-film deposition and electron-beam lithography (EBL). Figure~\ref{fig_fabrication} shows the main fabrication steps.
\section{Fano fitting: Measured transmission spectra}
\begin{figure}[!ht]
    \centering
    \subfloat[]{\includegraphics[width=0.33\columnwidth]{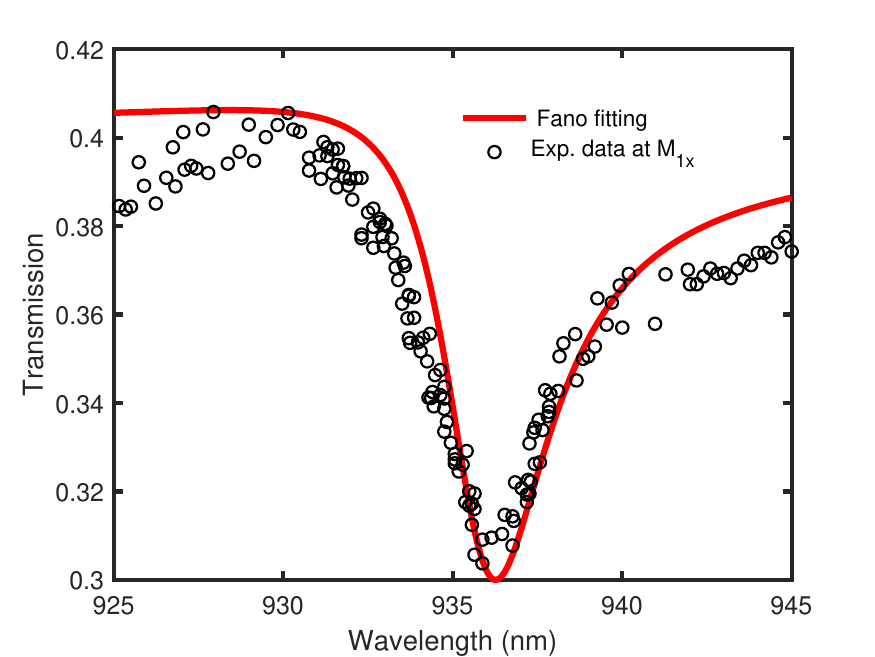}}
    \subfloat[]{\includegraphics[width=0.33\columnwidth]{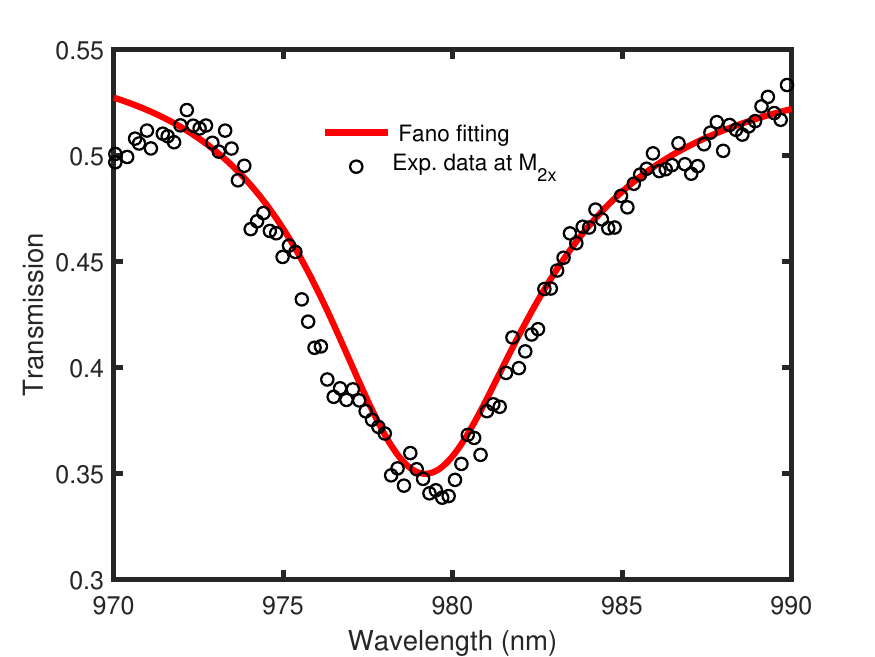}}
    \subfloat[]{\includegraphics[width=0.33\columnwidth]{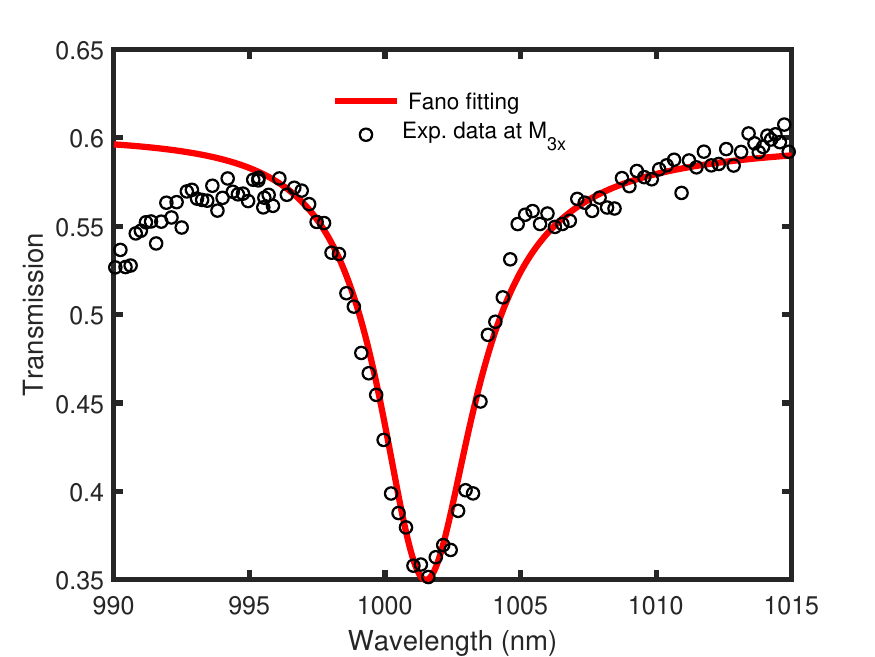}}\\
    \subfloat[]{\includegraphics[width=0.33\columnwidth]{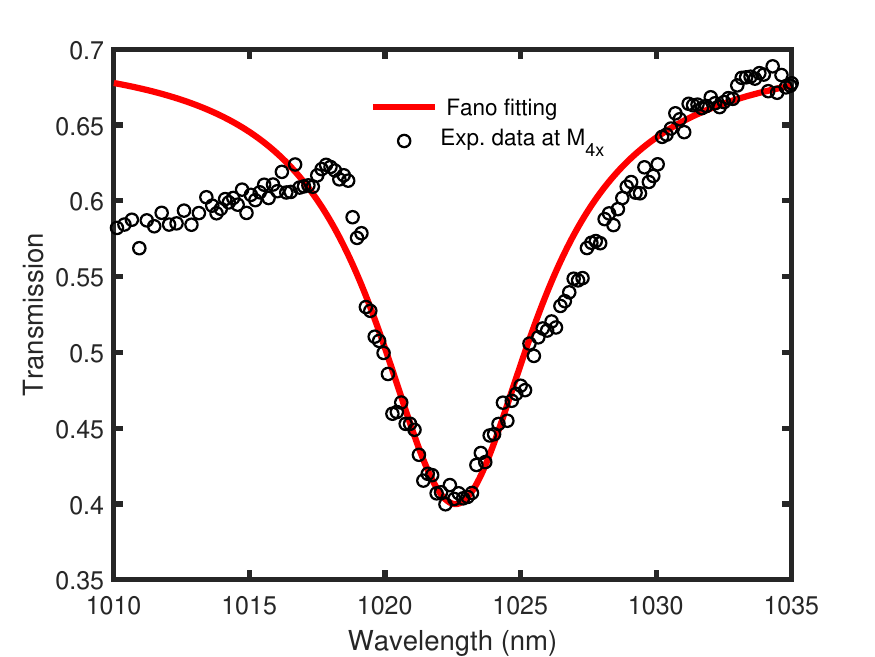}}
    \subfloat[]{\includegraphics[width=0.33\columnwidth]{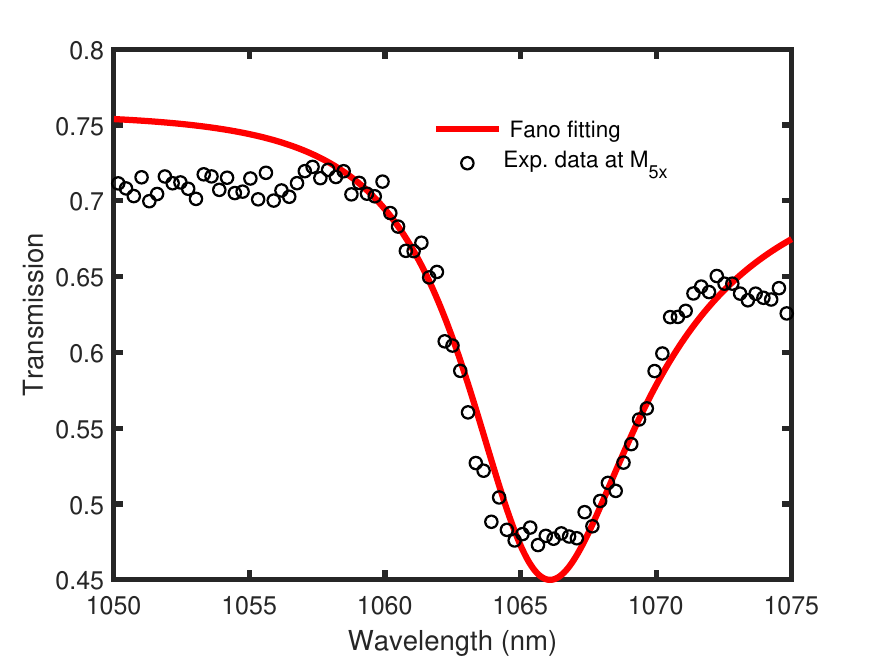}}
    \subfloat[]{\includegraphics[width=0.33\columnwidth]{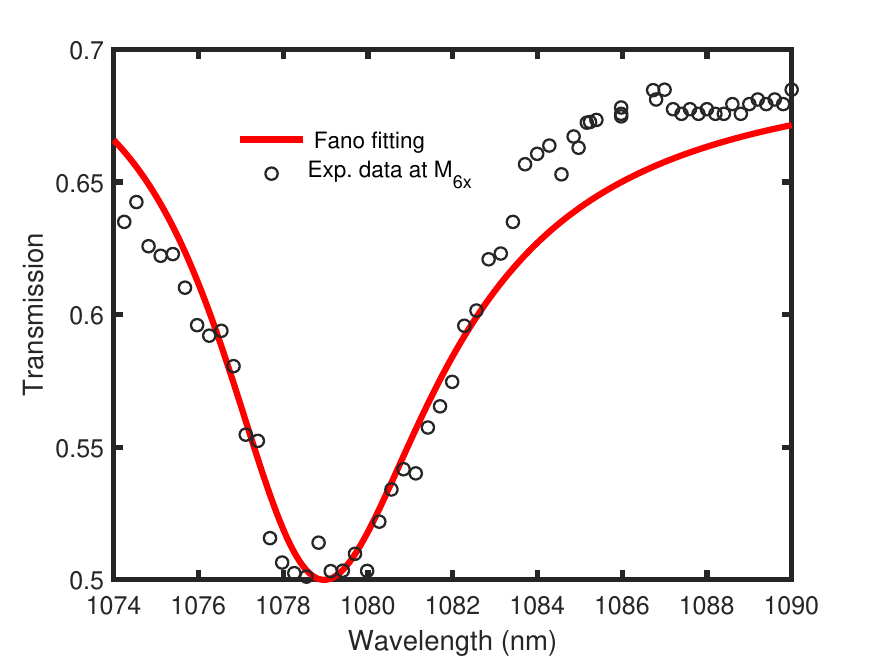}}\\
    \subfloat[]{\includegraphics[width=0.33\columnwidth]{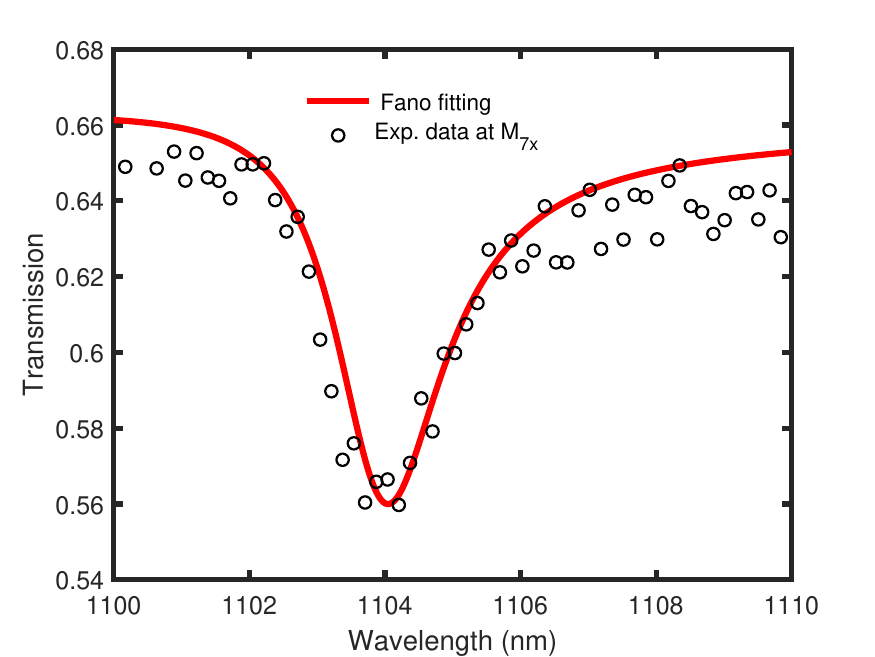}}
    \subfloat[]{\includegraphics[width=0.33\columnwidth]{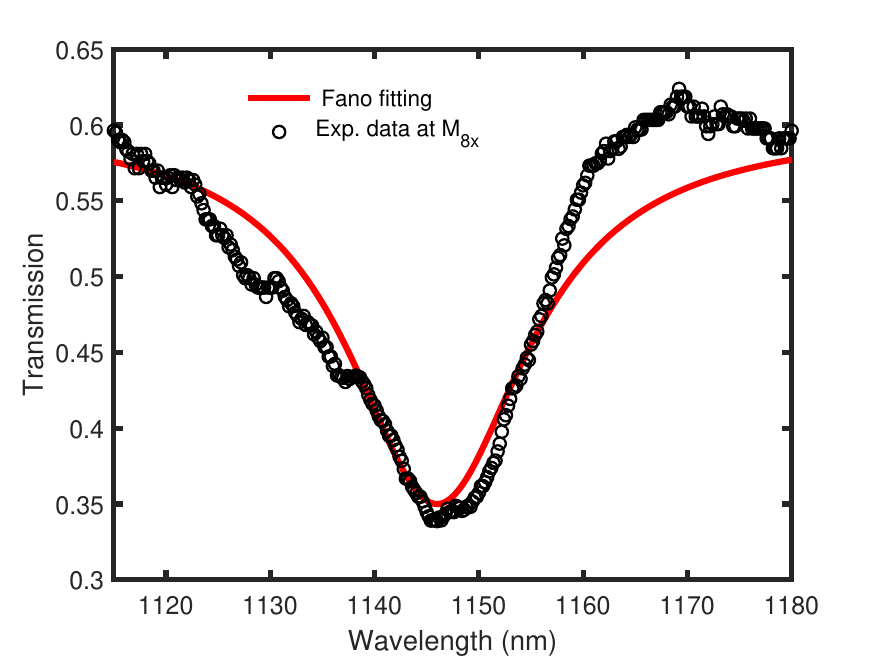}}
    \caption{Fano fitting of the measured transmission spectra at $\delta = 0.14$ for the resonances (a) $M_{1\mathrm{x}}\; (Q=271)$, (b) $M_{2\mathrm{x}}\; (Q=148)$, (c) $M_{3\mathrm{x}}\; (Q=260)$, (d) $M_{4\mathrm{x}}\; (Q=161)$, (e) $M_{5\mathrm{x}}\; (Q=156)$, (f) $M_{6\mathrm{x}}\; (Q=195)$, (g) $M_{7\mathrm{x}}\; (Q=550)$, and (h) $M_{8\mathrm{x}}\; (Q=51)$.}
    \label{fig_fano_fitting_ex}
\end{figure}

The measured transmission spectra at each resonance are fitted with the Fano line-shape model to extract the corresponding $Q$ factors. Figure~\ref{fig_fano_fitting_ex} shows the fitted spectra and provides the extracted $Q$ factors for the eight resonances $M_{1\mathrm{x}}$--$M_{8\mathrm{x}}$ at $\delta = 0.14$.

\newpage\clearpage
\bibliographystyle{unsrt}
\bibliography{references_revised_arXiv_combined}

\end{document}